\documentclass[twocolumn,times]{aastex63}

\accepted{ApJ}

\shorttitle{Dynamics of Narrow Line Regions}
\shortauthors{Meena et al.}

\usepackage{float}
\usepackage{color}
\usepackage{comment}
\usepackage{graphics}
\usepackage{epstopdf}
\usepackage{microtype}
\usepackage{subfigure}
\usepackage[normalem]{ulem}
\usepackage{natbib}
\usepackage{amsmath}
\usepackage{natbib}
\defcitealias{Meena2021}{Paper I}

\newcommand{\othree}{[O~III]~}
\newcommand{\halpha}{H$\alpha$~}
\newcommand{\hbeta}{H$\beta$~}

\newcommand{\llothree}{$\lambda \lambda$4959, 5007~\AA}
\newcommand{\kms}{km~s$^{-1}$}

\definecolor{malachite}{rgb}{0.01, 0.8, 0.24}
\usepackage{soul,xcolor}
\setstcolor{red}

\begin{document}

\title{Investigating the Narrow Line Region Dynamics in Nearby Active Galaxies}

\correspondingauthor{Beena Meena}
\email{bmeena@stsci.edu}

\author[0000-0001-8658-2723]{Beena Meena}
\affil{Department of Physics and Astronomy,
Georgia State University,
25 Park Place, Suite 605,
Atlanta, GA 30303, USA}
\affil{Space Telescope Science Institute, 3700 San Martin Drive, Baltimore, MD 21218, USA}

\author[0000-0002-6465-3639]{D. Michael Crenshaw}
\affil{Department of Physics and Astronomy,
Georgia State University,
25 Park Place, Suite 605,
Atlanta, GA 30303, USA}

\author[0000-0003-2450-3246]{Henrique R. Schmitt}
\affil{Naval Research Laboratory,
Washington, DC 20375, USA}

\author[0000-0002-4917-7873]{Mitchell Revalski}
\affil{Space Telescope Science Institute, 3700 San Martin Drive, Baltimore, MD 21218, USA}

\author[0000-0003-3401-3590]{Zo Chapman}
\affil{Department of Physics and Astronomy,
Georgia State University,
25 Park Place, Suite 605,
Atlanta, GA 30303, USA}

\author[0000-0002-3365-8875]{Travis C. Fischer}
\affiliation{AURA for ESA, Space Telescope Science Institute, 3700 San Martin Drive, Baltimore, MD 21218, USA}

\author[0000-0002-6928-9848]{Steven B. Kraemer}
\affil{Institute for Astrophysics and Computational Sciences,
Department of Physics,
The Catholic University of America,
Washington, DC 20064, USA}

\author[0000-0002-4262-4845]{Justin H. Robinson}
\affil{Department of Chemistry and Physics, Troy University, Troy, AL 36081, USA}

\author[0000-0001-7238-7062]{Julia Falcone}
\affil{Department of Physics and Astronomy,
Georgia State University,
25 Park Place, Suite 605,
Atlanta, GA 30303, USA}

\author[0000-0001-5862-2150]{Garrett E. Polack}
\affil{Department of Physics and Astronomy,
Georgia State University,
25 Park Place, Suite 605,
Atlanta, GA 30303, USA}

\begin{abstract}
We present dynamical models of the narrow line region (NLR) outflows in the nearby Seyfert galaxies Mrk~3, Mrk~78, NGC~1068, and NGC~4151 using observations from the Hubble Space Telescope and Apache Point Observatory. We employ long-slit spectroscopy to map the spatially-resolved outflow and rotational velocities of the ionized gas. We also perform surface brightness decompositions of host galaxy images to constrain the enclosed stellar mass distributions as functions of distance from the supermassive black holes (SMBHs). Assuming that the NLR gas is accelerated by AGN radiation pressure, and subsequently decelerated by the host galaxy and SMBH gravitational potentials, we derive outflow velocity profiles where the gas is launched in situ at multiple distances from the SMBH. We find a strong correlation between the turnover (from acceleration to deceleration) radii from our models, with the turnovers seen in the observed velocities and spatially-resolved mass outflow rates for the AGN with bolometric luminosities $>$ 10$^{44}$ erg sec$^{-1}$. This consistency indicates that radiation pressure is the dominant driving mechanism behind the NLR outflows in these moderate-luminosity AGN, with a force multiplier $\sim$500 yielding the best agreement between the modeled and observed turnover radii. However, in \cite{Meena2021} we found that this trend may not hold at lower luminosities, where our modeled turnover distance for NGC 4051 is much smaller than in the observed kinematics. This result may indicate that either additional force(s) are responsible for accelerating the NLR outflows in low-luminosity AGN, or higher spatial resolution observations are required to quantify their turnover radii.
\end{abstract}


\section{INTRODUCTION}\label{sec:intro}

\subsection{AGN-Driven Outflows} \label{subsec:nlr_outflows}

Feedback from Active Galactic Nuclei (AGN) is an essential ingredient for constraining galaxy evolution scenarios in large-scale cosmological simulations \citep{Sijacki2007,Booth2009,Steinborn2015} and for explaining the correlations between supermassive black holes (SMBHs) and their host galaxies. Specifically, the empirical scaling relationships between the SMBH mass and bulge luminosity ($M_\mathrm{BH}$-$L_{bulge}$, \citealp{Magorrian1998, Marconi2003}), and stellar velocity dispersion ($M_\mathrm{BH}$-$\sigma_{\star}$, \citealp{Ferrarese2000, Gebhardt2000, McConnell2013}) indicate that SMBHs may co-evolve with their host galaxies. AGN feedback may exist in two forms: (1) narrow beams of relativistic jets, which often extend beyond their host galaxies and impact the interstellar medium (ISM) and/or intergalactic medium, and (2) wider angle ``winds'' or ``outflows'' produced by the interaction between the accretion disk radiation and gas surrounding the AGN, including the ISM of the host galaxies. These AGN-driven winds are multi-phase and multi-scale in nature and can be observed across the electromagnetic spectrum \citep{Cicone2018}.
Highly ionized winds including ultra-fast outflows (UFOs, \citealp{Tombesi2013}), broad absorption line (BAL) outflows (\citealp{Miller2020}), and ``warm'' or ``intrinsic'' absorbers (WA) are observed in X-ray and/or UV absorption (\citealp{Crenshaw2003}) over a broad range of distances, including very close to the nucleus. In contrast, large-scale outflows of molecular and/or neutral gas may extend beyond the host galaxy disk (\citealp{Veilleux2020}).

Ionized gas outflows traced by optical emission lines are often found in the narrow line regions (NLRs, \citealp{Veilleux1987,Pogge1988,Osterbrock_Ferland2006}) and extended NLRs (ENLRs, \citealp{Unger1987}) of active galaxies. With the high-resolution capabilities of current telescopes, the NLRs of nearby galaxies can be spatially-resolved from within a few parsecs (pcs) of the SMBH, and extend several kpcs into the host galaxy bulges and/or disks \citep{Kang2018}.
Due to their larger physical extents compared to UFOs, and their higher velocities compared to their molecular/neutral gas counterparts, NLR outflows offer excellent laboratories to investigate the interaction between AGN and their host galaxies.

The kinematics and physical conditions of the ionized gas in the NLR provide information regarding the size, mass, and energetics of the outflows, and reveal their dynamic impact on black hole feeding and nuclear/circumnuclear star-formation. In general, the velocity distribution of the ionized gas is mapped using the often asymmetric and/or multicomponent profiles found in strong emission lines such as [O~III] \llothree, \halpha$\lambda$6563\AA, and [S~III] $\lambda$9533 \AA, which allow us to determine the outflow sizes, geometries, and kinematics.

The overall strength and feedback efficiency of the outflows can be quantified by measuring the ionized gas mass outflow rates ($\dot{M}_{out} = \Delta M v(r)/\Delta r$) and kinetic luminosities ($L_{K.E.} = (1/2)\dot{M}_{out} v(r)^{2}$), where  $\Delta M$ is the mass and $v(r)$ is the velocity of the ionized gas in each spatial bin $\Delta r$ \citep{Harrison2018, Revalski2021}. Many of these studies have found the mass outflow rates exceed the black hole accretion rate \citep{Crenshaw2015}, indicating mass depletion of the fueling reservoirs and/or that most of the fueling flows do not reach the inner accretion disk. Similarly, one can measure the total energy deposition into the ISM, and assess whether the NLR outflows are sufficiently energetic to evacuate the nuclear/circumnuclear star-forming gas, thus providing significant negative feedback. Despite the progress that has been made to characterize the properties of outflows in AGN with a wide range of mass, luminosities and redshifts, we do not have a clear consensus on their origins and driving mechanisms \citep{Wylezalek2018, Laha2021}.

There are three primary mechanisms by which the central engine (AGN) can directly accelerate and produce gas outflows: magnetic fields, thermal expansion, and radiation pressure \citep{Crenshaw2003, Proga2007}. Magnetohydrodynamical (MHD) acceleration  \citep{Emmering1992,Bottorff1997, Bottorff2000a} has been proposed to explain the X-ray warm absorbers and/or ultra-fast outflows (UFOs) via accretion-disk winds in the absence of line-driving \citep{Konigl1994,Fukumura2010,Fukumura2015,Fukumura2018, Kraemer2018}. Magnetic fields are also thought to be responsible for the propagation of jets \citep{Blandford1982}. However, it remains to be proven that magnetic acceleration can drive wide-angle winds to the large distances of the NLR clouds from the SMBH. Thermal expansion \citep{Begelman1983} from the hot corona above the accretion disc has been suggested to launch the warm absorbers \citep{Krolik1995,Woods1996,Krolik2001,Mizumoto2019}. \cite{Everett2007} developed an isothermal Parker wind model to test and explain spatially-resolved velocities in the NLR of NGC~4151, although they found that isothermal winds cannot be sustained at larger radii due to adiabatic cooling and are therefore unable to produce the observed NLR velocities. Radiation pressure \citep{Castor1975, Abbott1982} exerted on the surrounding gas by a strong continuum source (the accretion disk) can accelerate the gas via bound-bound (line-driving), bound-free (continuum opacity), and free-free (Thomson scattering) transitions \citep{Arav1994,Murray1995,Proga1998,Proga2000, Chelouche2001}.
Beyond the dust sublimation radius, the dust may provide additional opacity and hence contribute to gas acceleration induced by AGN radiation \citep{Dopita2002,Fabian2006,Ishibashi2015, Trindade2021}.

Outflows that are radiatively launched close to the SMBH and accretion disk can achieve high velocities and escape the gravitational potential of the SMBH \citep{Crenshaw2003}.
However, at NLR distances, one must also consider the galaxy's gravitational potential acting on the gas. \cite{Das2007} proposed a dynamical model for the NLR in NGC~1068 consisting of radiative acceleration of the gas and deceleration due to gravity from the SMBH and the galaxy, as well as drag forces. This model assumed a single point of acceleration, primarily from a launch distance of few parsecs from the SMBH, but was unable to reproduce the observed velocity distribution of the NLR clouds.

A number of secondary driving mechanisms have been proposed that rely on the impact of gas accelerated by one or more primary mechanisms.
These secondary processes include momentum or shock driving from accretion disk winds/UFOs \citep{Pounds2011, Pounds2013, Tombesi2013, Mou2017, Veilleux2017}, expansion of radio plasma or hot cocoons around jets \citep{Greene2014, Mukherjee2016, May2017, Venturi2021, May2020}, mechanical driving by AGN winds \citep{Fischer2019} or jets \citep{May2018}, and entrainment by X-ray winds \citep{Trindade2021}. It is likely that more than one process may play a role in accelerating the gas in different environments or spatial scales. However, none of these processes alone (or in combination) have been shown to fully reproduce the radial variations observed in the NLR outflow velocities and mass outflow rates.

Most early mass outflow rate studies used a global technique to calculate the integrated mass outflow rates and energetics, which assumes the NLR outflows are produced and driven at the nucleus \citep{Storchi-Bergmann2010, Muller2011, Kakkad2016, Karouzos2016, Bae2017}.  However, using spatially-resolved measurements of the nearby Seyfert 1 galaxy NGC~4151, \cite{Crenshaw2015} found that the mass outflow rates increase as a function of distance and peak at $\sim$70 pc from the SMBH before steadily decreasing. This study provided evidence of ``in-situ'' acceleration of the ionized gas outflows, as compared to the nuclear outflow origin model.

\cite{Fischer2017} revised the radiative-gravity formalism (initially proposed by \citealp{Das2007}) by considering the in-situ production of outflows and found that gas can be radiatively driven at hundreds of pc from the SMBH in the host disk of the Seyfert~2 galaxy Mrk 573. This in-situ acceleration in Mrk 573 was also substantiated by \cite{Revalski2018a} based on spatially-resolved mass outflow rate measurements performed using similar analysis as to those for NGC~4151 \citep{Crenshaw2015}.

More recently, in \citealt{Meena2021} (hereafter \citetalias{Meena2021}), we employed a similar dynamical model of radiative acceleration and gravitational deceleration for the outflows in the Narrow-Line Seyfert~1 galaxy NGC 4051. We measured the spatially-resolved \othree ionized gas kinematics using ground-based spectroscopy and developed an improved kinematic model of biconical outflow, with larger spatial extent \& turnover radius and wider opening angle than what was previously presented in \cite{Fischer2013}.
We also determined a radial mass distribution using 2D surface brightness decomposition of the host galaxy adopted from \cite{Bentz2009} and \cite{Bentz2018}. We then utilized the radiative-gravity model \citep{Das2007,Fischer2017} and numerically solved for the launch radii of observed outflows. Unlike the large-scale in-situ acceleration discussed for the higher luminosity AGN Mrk 573 \citep{Fischer2017}, we found that most of the observed large-scale ($\sim$1 kpc) outflows were radiatively launched within $\sim$2 pc of its low-mass, low-luminosity AGN. Our results in \citetalias{Meena2021} suggest that the in-situ acceleration increases with the bolometric luminosity of the AGN, and that higher luminosity AGN may be more efficient in driving outflows at larger distances.

Similar to NGC 4051 \citepalias{Meena2021}, in this work we measure velocity profiles for the ionized gas NLR outflows and determine their launch distances in four nearby Seyfert galaxies. This methodology assumes the outflow kinematics are shaped exclusively by the forces of radiative acceleration and gravitational deceleration acting on the gas. Furthermore, we compare our model with the previous biconical outflow kinematic models of \cite{Fischer2013}, and the spatially-resolved mass outflow rates measured by \citealt{Revalski2021}, to determine if the NLR dynamics can be entirely explained by the combination of these two forces.

\subsection{Sample Selection and Overview}\label{subsec:targets}

\setlength{\tabcolsep}{0.1in}
\begin{deluxetable*}{lccccccccc}[ht!]
\tablecaption{General Characteristics of the AGN Sample  \label{tab:sample}}
\tablehead{
\colhead{Galaxy} & \colhead{AGN} & \colhead{Redshift} & \colhead{Distance} & \colhead{Scale} & \colhead{log($L_{bol}$)} & \colhead{log($M_{BH}$)} & \colhead{$L_{bol}/L_{Edd}$} & \colhead{Refs.}   & \colhead{Models \vspace{-0.5em}}\\
\colhead{Name} & \colhead{Type} & \colhead{(21 cm)} & \colhead{(Mpc)} & \colhead{(pc/\arcsec)} & \colhead{(ergs s$^{-1}$)} & \colhead{($M_{\odot}$)} & \colhead{} & \colhead{Col 6,7} & \colhead{Refs.\vspace{-0.5em}}\\
\colhead{(1)} & \colhead{(2)} & \colhead{(3)} &\colhead{(4)} & \colhead{(5)} & \colhead{(6)} & \colhead{(7)} & \colhead{(8)} & \colhead{(9)} & \colhead{(10)}
}
\startdata
Mrk 3 & 2 & 0.0135 & 56.6 & 274.5 & 45.4 & 8.7 & 0.04 & $\star$,1 & $\star$\\
Mrk 78 & 2 & 0.0372 & 154.2 & 747.4 & 45.9 & 7.9 & 0.6 & $\star$,1 & $\star$\\
NGC 1068 & 2 & 0.0038 & 14.4$^a$ & 54.0 & 45.0 & 7.2 & 0.50 & $\star$, 1 & $\star$, 10\\
NGC 4151 & 1 & 0.0033 & 15.8$^b$ & 76.7 & 44.7$^d$ & 7.4 & 0.2 & $\star$,2 & $\star$\\ \hline
Mrk 34 & 2 & 0.0505 & 207.9 & 1007.7 & 46.2 & 7.5 & 3.98 & 3, 4 &  11\\
Mrk 573 & 2 & 0.0172 & 72.0 & 349.1 & 45.5 & 7.3 & 0.75 & 5, 1 & 12\\
NGC 4051 & 1 & 0.00234 & 16.6$^c$ & 80.5 & 42.9 & 5.9 & 0.053 & 6, 2 & 6\\ 
NGC 5643 & 2 & 0.0040 & 16.9 & 81.9 & 43.9 & 6.43 & 0.25 & 7, 8 & 7\\
2MASX J0423 & 2 & 0.0461 & 195 & 940 & 45.6 & ... & ... & 9 & 9\\
\enddata
\tablecomments{The active galaxies sample and associated AGN properties. Columns are: (1) Galaxy name, (2) Type of AGN (1 or 2), (3) Redshifts using HI 21 cm observations, retrieved from NASA/IPAC Extragalactic Database, (4) Distance using redshift and the Hubble constant H$_{0}$ = 70 km s$^{-1}$ Mpc$^{-1}$, except for $^a$NGC~1068 \citep{Bland-Hawthorn1997} using Tully-Fisher distance \citep{Tully1988} to be consistent with literature measurements of the rotation curves as discussed in \S \ref{subsec:rotation} and Figure~\ref{fig:rotation}, $^b$NGC~4151 \citep{Yuan2020} and $^c$NGC~4051 \citep{Yuan2021} using Cepheid measurements, (5) Transverse scale on the plane of the sky, calculated using distance in Column 3, (6) Bolometric luminosities, corrected for galactic extinction, $^d$ see \S \ref{subsec:assum} for the variations in different $L_{bol}$ measurements of NGC~4151, (7) Black hole mass, (8) Eddington ratio, where $L_{Edd}$ = 1.26$\times$10$^{38}$ ($M/M_{\odot}$) erg s$^{-1}$, (9) References for values adopted in column 6 and 7, (10) References for the derived radiative driving models for each AGN. The numbered references are: 
1. \cite{WooUrry2002}, 2. \cite{Bentz2015}, 3. \cite{Revalski2018b}, 4. \cite{Oh2011}, 5. \cite{Revalski2018a}, 6. \citetalias{Meena2021}, 7. \cite{Garcia2021}, 8. \cite{Goulding2010}, 9. \cite{Fischer2019}, 10. \cite{Das2007}, 11. \cite{Trindade2021}, 12. \cite{Fischer2017} and $\star$ is this work. The uncertainties in bolometric luminosities are $\pm$0.3 dex for all targets except Mrk 573 (uncertainties = $\pm$0.6 dex, \cite{Kraemer2009}) based on uncertainties in the \othree to bolometric luminosity conversion \cite{Heckman2004} and from scattering in $L_{14-195~\mathrm{keV}}$ \cite{Ricci2017} for NGC 5643 in their respective references.} 
\vspace{-2em}
\end{deluxetable*}

One of our goals is to compare dynamical outflow models with the observed spatially-resolved properties of the NLR outflows. We selected the four Seyfert galaxies Mrk~3, Mrk~78, NGC~1068, and NGC~4151 for this analysis, as they have existing spatially-resolved mass outflow rate measurements presented in \cite{Revalski2021} (including results from \citealp{Crenshaw2015, Revalski2018a, Revalski2018b}).
We include 5 additional AGN in our sample that have been studied with a similar radiative driving model. The full AGN sample and their general characteristics are provided in Table~\ref{tab:sample}. In this work we determine the bolometric luminosity ($L_{bol}$) based on the \othree luminosity of the NLR in each target, which represents the weighted value of the luminosity of the AGN over the light travel time of the NLR.
We determined the \othree luminosities using the continuum-subtracted [O~III] emission from narrow-band and nearby continuum images.
We calculated $L_{bol}$ = 3500 $\times$ L$_\mathrm{[O III]}$ based on the calibration of \citet{Heckman2004}.
Some of the physical properties of the four targets studied in this sample are discussed below.

Mrk~3 (UGC~3426) is an S0 classified galaxy, which hosts a large bulge, a nuclear stellar disk \citep{Gnilka2020} and a type 2 AGN that is being fed externally through a tidal interaction with gas-rich spiral galaxy UGC 3422 at a distance of $\sim$100 kpc to the NW \citep{Noordermeer2005,Gnilka2020}. Mrk~78 is a potential post-merger system \citep{Whittle2004} with a type 2 AGN. It is classified as an SB galaxy with a large bulge similar to Mrk~3. Large-scale ionized gas kinematics using ground-based spectroscopy also shows distinct rotation indicating an outer disk \citep{Revalski2021}. Both Mrk~3 and Mrk~78 show no sign of ongoing star-formation from Baldwin-Phillips-Terlevich (BPT) diagnostics \citep{Gnilka2020, Revalski2021}. Similarly, there are no spiral arms visible in existing images of these galaxies (\citealt{Schmitt2000, Gnilka2020, Revalski2021}).

NGC~1068 (M~77) is morphologically classified as a (R)SA(rs)b galaxy \citep{deVaucouleurs1991} for its outer rings structure and pronounced spiral arms. There is also evidence of a nuclear stellar cluster \citep{Thatte1997}. It is the closest and one of the most studied Seyfert galaxies, with a variety of multi-wavelength observations that show multi-phase outflows driven by its bright type 2 AGN \citep{Cecil2002,Garcia-Burillo2014,May2017,Riffel2014,Mizumoto2019,Lamastra2019,Saito2022}.
NGC~4151 is a barred Seyfert (De Vaucouleurs classification: (R')SAB(rs)ab ) galaxy with a type 1 AGN as indicated by broad Balmer emission (including \hbeta and \halpha) from its nucleus \citep{Shapovalova2010}.
NGC~4151 also shows extended outflows in X-ray emission \citep{Kraemer2020, Trindade2022}. The AGN of NGC~4151 is highly variable in both continuum and line emissions \citep{Ulrich1991,Kraemer2006, Shapovalova2008}. Based on UV continuum flux averaged over $\sim$20 years, \cite{Kraemer2020} derived an AGN bolometric luminosity of $L_{bol}$ = 1.4 $\times$ 10$^{44}$ erg s$^{-1}$. However, we calculated $L_{bol}$ $\approx$ 5 $\times$ 10$^{44}$ erg s$^{-1}$ from L$_\mathrm{[O III]}$ \citep{Heckman2004}, representing a much longer time average. We discuss the implications of adopting different luminosities in \S \ref{subsec:corr_TO}. 

We have previously investigated these four targets and generated biconical outflow models of their kinematics (Mrk~3: \citealp{Ruiz2001}, Mrk~78: \citealp{Fischer2011}, NGC~1068: \citealp{Das2006} and NGC~4151: \citealp{Das2005}) using some of the long-slit observations provided in \S \ref{sec:obs}. In this work, we employ improved spectral fitting methods (\S \ref{subsec:app_spec_fit}) to compare the gas kinematics with those of the previous measurements (\S \ref{app_kinematics}). Using additional ground-based spectra, we also derive spatially-resolved rotation curves (\S \ref{subsec:rotation}).
We perform 2D surface brightness decomposition of the host galaxies and constrain the mass-to-light ratios ($M/L$) using space- and ground-based imaging (\S \ref{subsec:galfit}) and determine host galaxy enclosed mass profiles as a function of distance from the SMBHs (\S \ref{subsec:mass_vel})
We then develop our dynamical models using radiative-gravity formalism (\S \ref{subsec:radiative_driving}) and determine the launch distances of the observed outflowing knots of emission. We present the scale of outflow launch distances for our sample across a wide range of AGN luminosities (\S \ref{subsec:corr_lum}) and compare our model results with those of the outflow model geometries provided by \cite{Fischer2013} to establish the validity of our models (\S \ref{subsec:corr_TO}).
Additionally, we will also discuss the feedback implications of ionized gas outflows by connecting our modeled outflow launch distances with the morphological features found in various molecular gas phases (\S \ref{subsec:outflow_origins}). Finally, we discuss whether radiation pressure and gravity are the primary forces that shape the kinematics NLR outflows (\S \ref{subsec:outflow_mech}) and provide a summary (\S \ref{sec:conclusion}).

\section{OBSERVATIONS}\label{sec:obs}

\subsection{Hubble Space Telescope (HST)}\label{subsec:hstobs}

We utilized Hubble Space Telescope (HST) observations of our targets available from the Mikulski Archive (MAST) at the Space Telescope Science Institute. For this study, we used spectroscopic observations from the HST Space Telescope Imaging Spectrograph (STIS), and imaging from the Faint Object Camera (FOC), Advanced Camera for Surveys (ACS), and Wide Field and Planetary Camera 2 (WFPC2). Details of the observations are given in Table \ref{tab:hst-obs}.

\begin{deluxetable*}{lhcccccccccccc}
\setlength{\tabcolsep}{0.03in}
\renewcommand{\arraystretch}{1.05}
\tabletypesize{\footnotesize}
\tablecaption{HST~Observations of the Sample \label{tab:hst-obs}}
\tablehead{
\colhead{Target} & \nocolhead{Observing} & \colhead{Instrument} & \colhead{Grating/} & \colhead{Slit} & \colhead{Proposal} & \colhead{Observation} & \colhead{Date}  & \colhead{Exposure} & \colhead{Spectral} & \colhead{Wavelength} & \colhead{Spatial} & \colhead{Position} & \colhead {Spatial} \vspace{-2ex}\\
\colhead{Name} & \nocolhead{Facility} & \colhead{Name} & \colhead{Filter} & \colhead{Name} & \colhead{ID} & \colhead{ID} & \colhead{(UT)}& \colhead{Time} & \colhead{Dispersion} & \colhead{Range} & \colhead{Scale} & \colhead{Angle} & \colhead {Offset} \vspace{-2ex}\\
\colhead{} & \nocolhead{} & \colhead{}  & \colhead{} & \colhead{} & \colhead{} & \colhead{} & \colhead{}  & \colhead{(s)}& \colhead{(\AA~pix$^{-1}$)} & \colhead{(\AA)} & \colhead{($\arcsec$~pix$^{-1}$)} & \colhead{(deg)} & \colhead {($\arcsec$)}
}
\startdata
Mrk 3 & HST & STIS & G430L & 1 & 8480 & O5KS01010 & 2000-08-22 & 1080 & 2.73 & 2900-5700 & 0.051 & 71.8 & 0.0\\
Mrk 3 & HST & FOC & F502M & ... & 5140 & X2580103T & 1994-03-20 & 750 & ... & 4645-5389 & 0.014 & ... & ...\\
Mrk 3 & HST & FOC & F550M & ... & 5140 & X2580104T & 1994-03-20 & 1196 & ... & 5303-5726 & 0.014 & ... & ...\\
\hline
Mrk 78 & HST & STIS & G430M & 1 & 7404 & O4DJ02020 & 1998-02-28 & 1727 & 0.28 & 4950-5236 & 0.051 &  88.05 & 0.125\\
Mrk 78 & HST & STIS & G430M & 2 & 7404 & O4DJ02050 & 1998-02-28 & 1938 & 0.28 & 4950-5236 & 0.051 &  88.05 & -0.27\\
Mrk 78 & HST & STIS & G430M & 3 & 7404 & O4DJ02080 & 1998-02-28 & 2052 & 0.28 & 4950-5236 & 0.051 &  88.05 & -0.55\\
Mrk 78 & HST & STIS & G430M & 4 & 7404 &  O4DJ04020 & 1998-03-01 & 1800 & 0.28 & 4950-5236 & 0.051 &  61.56 & -0.05\\
Mrk 78 & HST & FOC & F502M & ... & 5140 &  X2580303T & 1994-03-19 & 800 & ... & 4645-5389 & 0.014 &  ... & ...\\
Mrk 78 & HST & FOC & F550M & ... & 5140 &  X2580304T & 1994-03-19 & 1196 & ... & 5303-5726 & 0.014 &  ... & ...\\
Mrk 78 & HST & ACS/WFC & F814W & ... & 15444 & JDRW5M010 & 2019-01-06 & 674 & ... & 7077-9588 & 0.05 &  ... & ...\\
\hline
NGC 1068 & HST & STIS & G430M & 1 & 7353 & O56502010 & 1999-10-02 & 2585 & 0.28 & 4818-5104 & 0.051 & 38.05 & -0.6\\
NGC 1068 & HST & STIS & G430M & 2 & 7353 & O56502020 & 1999-10-02 & 2775 & 0.28 & 4818-5104 & 0.051 & 38.05 & -0.4\\
NGC 1068 & HST & STIS & G430M & 3 & 7353 & O56502030 & 1999-10-02 & 2775 & 0.28 & 4818-5104 & 0.051 & 38.05 & -0.2\\
NGC 1068 & HST & STIS & G430M & 4 & 7353 & O56502040 & 1999-10-02 & 2775 & 0.28 & 4818-5104 & 0.051 & 38.05 & 0.0\\
NGC 1068 & HST & STIS & G430M & 5 & 7353 & O56502050 & 1999-10-02 & 2775 & 0.28 & 4818-5104 & 0.051 & 38.05 & 0.2\\
NGC 1068 & HST & STIS & G430M & 6 & 7353 & O56503010 & 2000-09-22 & 2294 & 0.28 & 4818-5104 & 0.051 & 38.05 & 0.2\\
NGC 1068 & HST & STIS & G430M & 7 & 7353 & O56503020 & 2000-09-22 & 2853 & 0.28 & 4818-5104 & 0.051 & 38.05 & 0.4\\
NGC 1068 & HST & STIS & G430M & 8 & 8185 & O5LJ01070 & 2000-01-14 & 960 & 0.28 & 4818-5104 & 0.051 & 10 & 0.51\\
NGC 1068 & HST & WFPC2/PC & F502N & ... & 5754 & U2M30103T & 1995-01-17 & 300 & ... & 4969-5044 & 0.045 & ... & ...\\
NGC 1068 & HST & WFPC2/PC & F502N & ... & 5754 & U2M30104T & 1995-01-17 & 600 & ... & 4969-5044 & 0.045 & ... & ...\\
NGC 1068 & HST & WFPC2/PC & F547M & ... & 5754 & U2M30101T & 1995-01-17 & 140 & ... & 5060-5885 & 0.045 & ... & ...\\
NGC 1068 & HST & WFPC2/PC & F547M & ... & 5754 & U2M30102T & 1995-01-17 & 300 & ... & 5060-5885 & 0.045 & ... & ...\\
\hline
NGC 4151 & HST & STIS & G430M & 1 & 8473 & O5KT01010 & 2000-07-02 & 2379 & 0.28 & 4818-5104 & 0.051 & 58 & 0.0 \\
NGC 4151 & HST & STIS & G430M & 2 & 8473 & O5KT01020 & 2000-07-02 & 2865 & 0.28 & 4818-5104 & 0.051 & 58 & 0.2\\
NGC 4151 & HST & STIS & G430M & 3 & 8473 & O5KT01030 & 2000-07-02 & 2865 & 0.28 & 4818-5104 & 0.051 & 58 & 0.4\\
NGC 4151 & HST & STIS & G430M & 4 & 8473 & O5KT01040 & 2000-07-02 & 2865 & 0.28 & 4818-5104 & 0.051 & 58 & -0.2\\
NGC 4151 & HST & STIS & G430M & 5 & 8473 & O5KT01050 & 2000-07-02 & 2865 & 0.28 & 4818-5104 & 0.051 & 58 & -0.2\\
NGC 4151 & HST & WFPC2/WF & F502N & ... & 7569 & U423A103M & 1997-07-15 & 900 & ... & 4998-5025 & 0.05 & ... & ...\\
NGC 4151 & HST & WFPC2/WF & F502N & ... & 7569 & U423A104M & 1997-07-15 & 900 & ... & 4998-5025 & 0.05 & ... & ...\\
NGC 4151 & HST & WFPC2/WF & F555W & ... & 7569 & U423A102M & 1997-07-15 & 3 & ... & 4825-6053 & 0.05 & ... & ...\\
NGC 4151 & HST & WFPC2/WF & F555W & ... & 7569 & U423A101M & 1997-07-15 & 3 & ... & 4825-6053 & 0.05 & ... & ...\\
\enddata
\tablecomments{A summary of the HST observations used in this study. The columns list the (1) Galaxy name, (2) HST Instrument, (3) filter (imaging) or grating (spectra), (4) slit number (spectra), (5) Program ID, (6) Observation ID, (7) observation date, (8) total exposure time for each data set, (9) spectral dispersion of the long-slit grating, (10) wavelength range (for spectra) or bandpass (for imaging), (11) spatial scale of the image/spectra, (12) position angle (PA) of the STIS slits, and (13) their spatial offsets from nucleus. The values in column (9)-(11) were obtained from their respective instrument handbooks (STIS: \citealp{stisihb}; FOC: \citealp{Nota1996}; WFPC2: \citealp{McMaster2008}; ACS: \citealp{Ryon2021}) and the file headers. The exact spatial scale of STIS is 0\farcs05078~pixel$^{-1}$, and the combined images of NGC~4151 (F502, F555W) are sampled at 0\farcs1~pixel$^{-1}$. This data can be obtained from MAST using the DOI:\dataset[10.17909/zvk2-f030]{\doi{10.17909/zvk2-f030}}.
}
\vspace{-2em}
\end{deluxetable*}

\begin{figure*}[ht!]
\begin{center}
\subfigure{\includegraphics[width=0.49\textwidth]{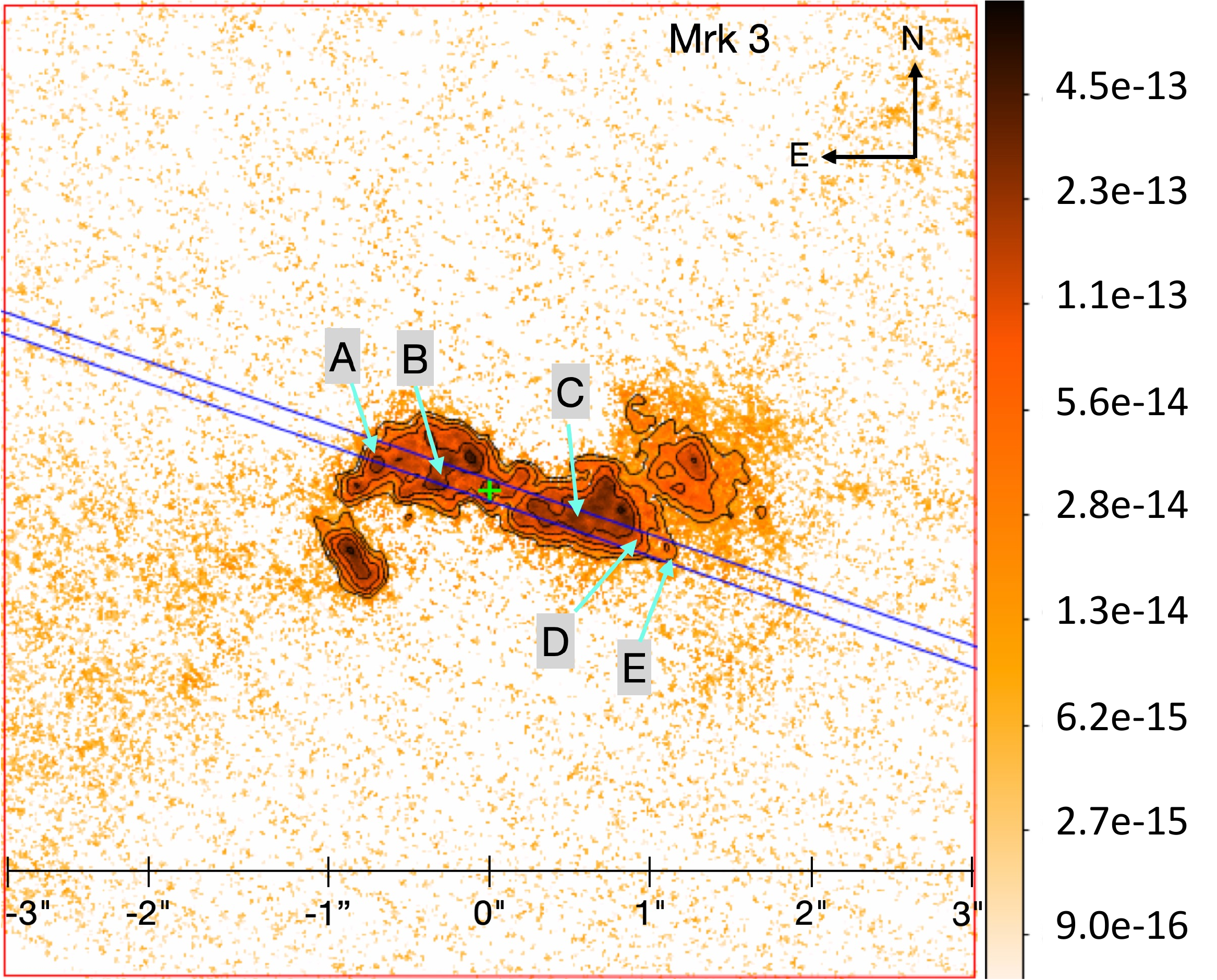}}
\subfigure{\includegraphics[width=0.49\textwidth]{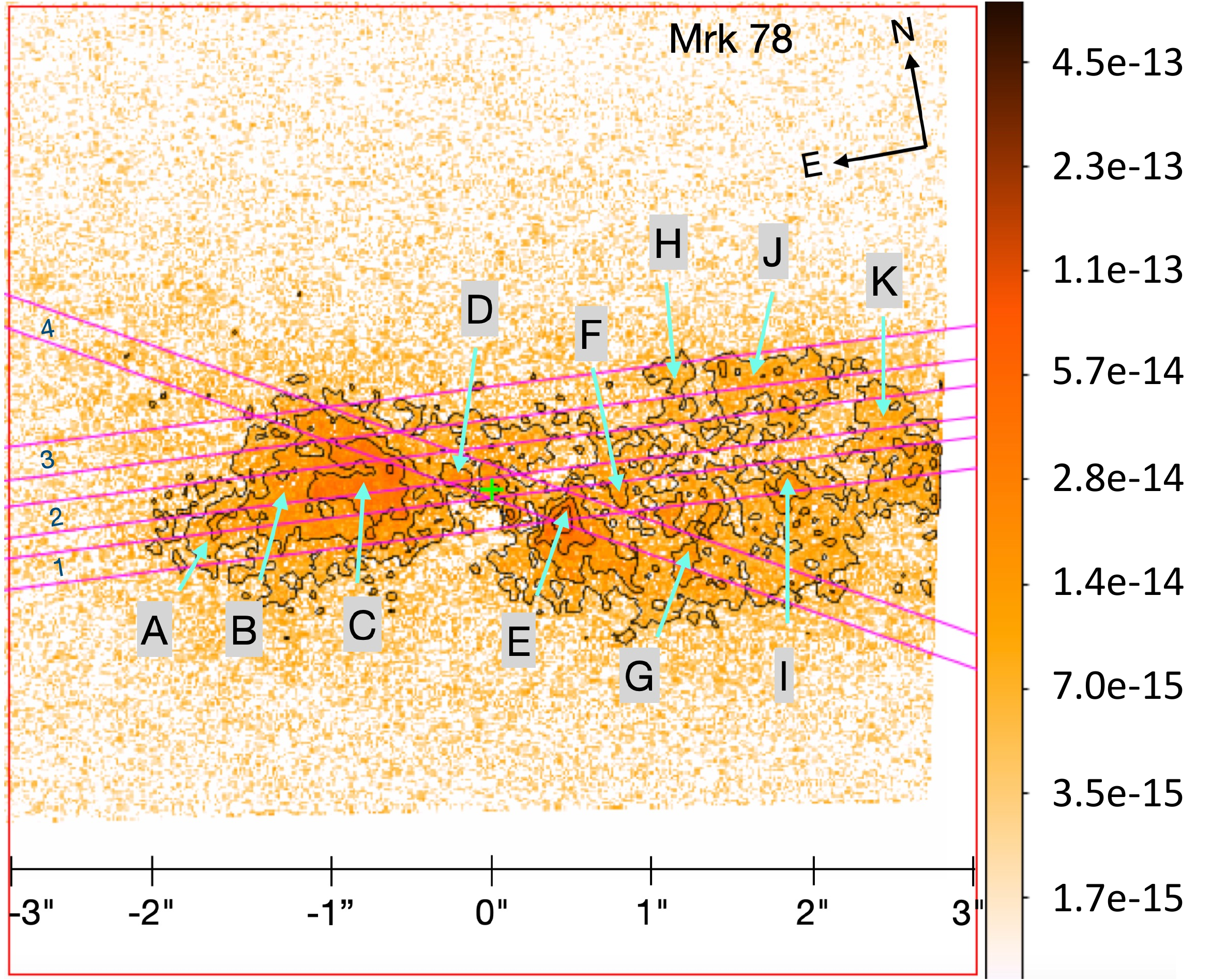}}
\subfigure{\includegraphics[width=0.49\textwidth]{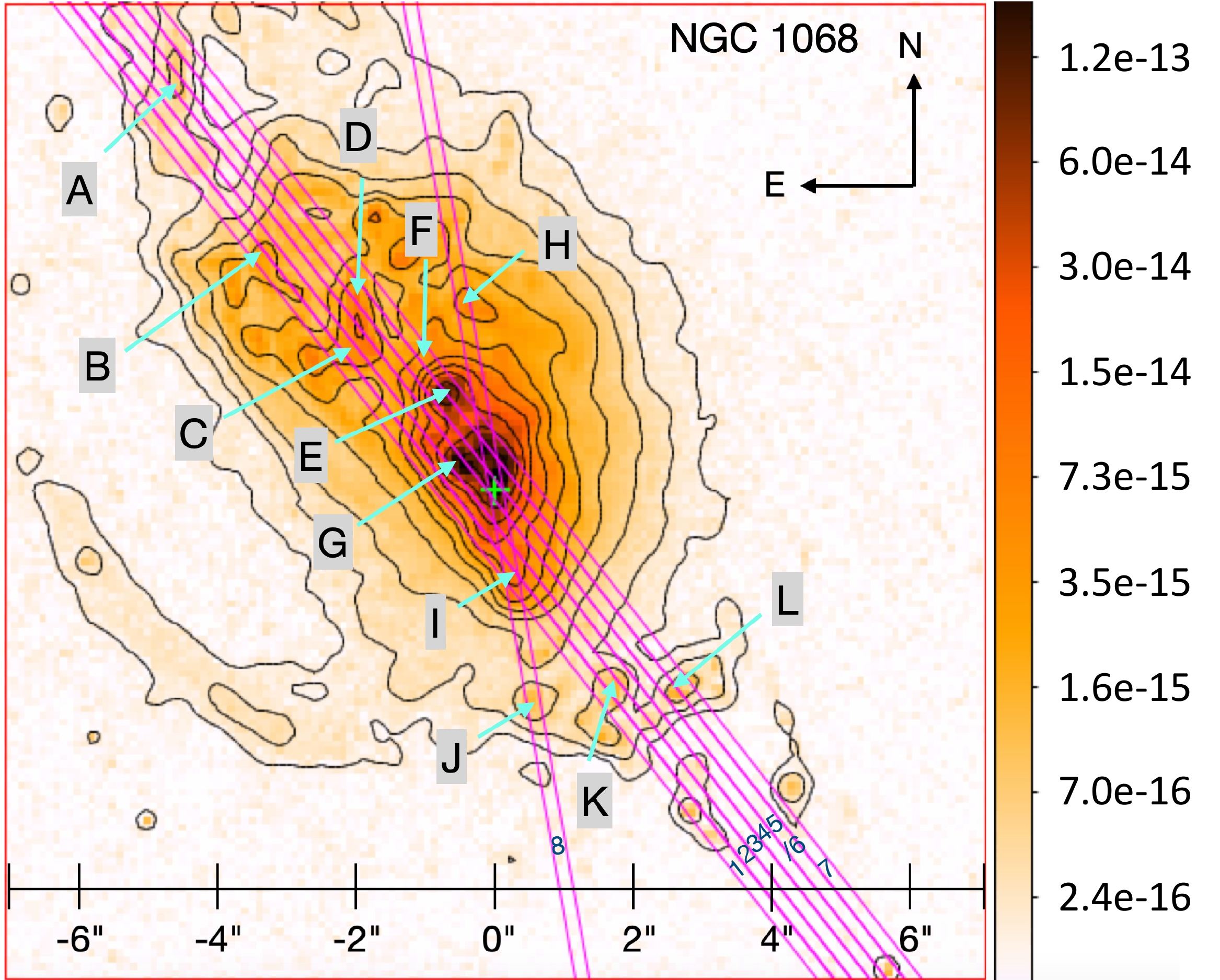}}
\subfigure{\includegraphics[width=0.49\textwidth]{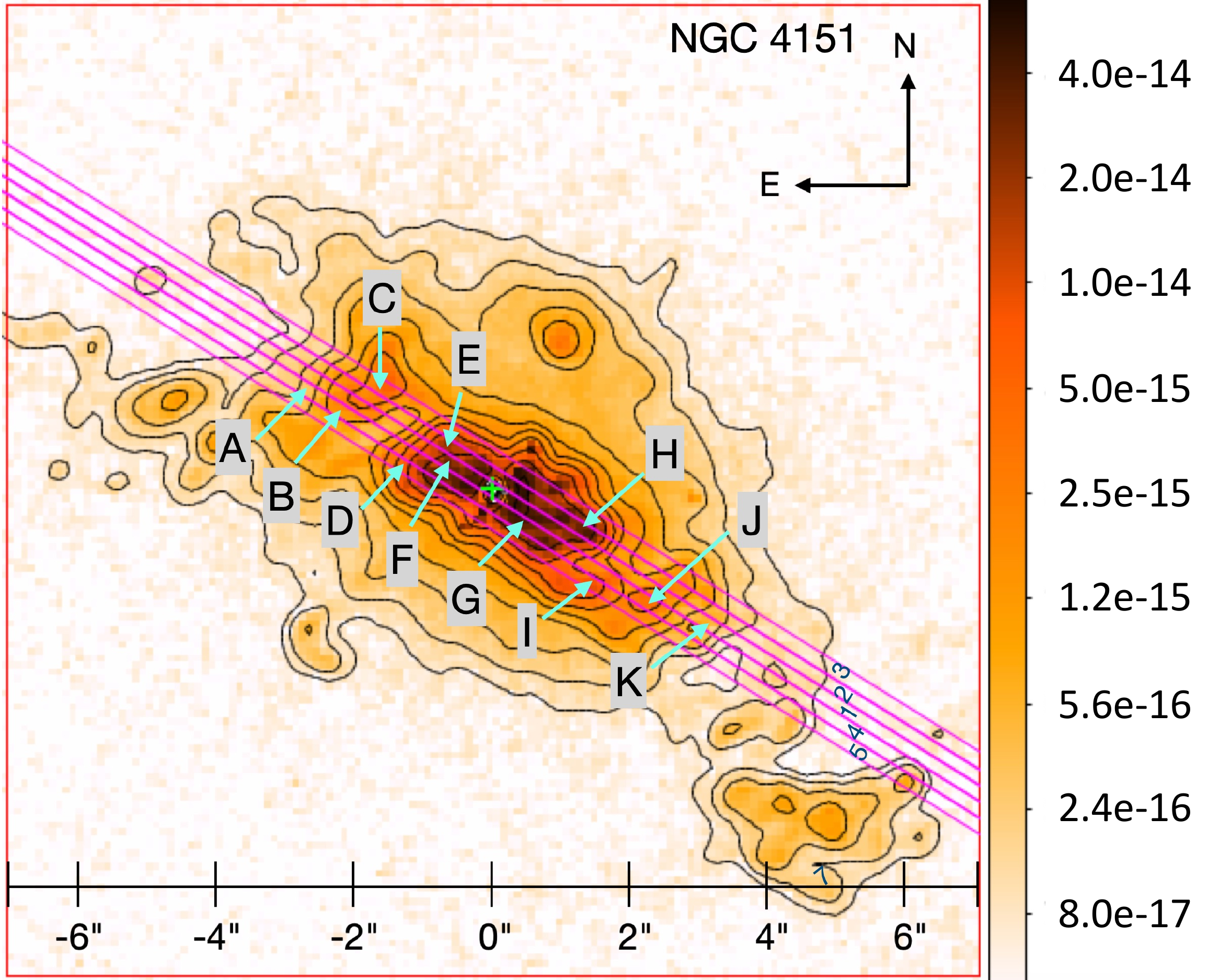}}
\end{center}
\vspace{-0.5em}
\caption{The continuum-subtracted \othree images of Mrk~3 (top-left), Mrk~78 (top-right), NGC~1068 (bottom-left) and NGC~4151 (bottom-right). The image dimensions are $6\arcsec\times6\arcsec$ for Mrk~3 and Mrk~78 and $14\arcsec\times14\arcsec$ for NGC~1068 and NGC~4151. The continuum centroid peaks are shown by green cross symbols, and the STIS long-slits are represented in blue for G430L (only Mrk~3) and magenta for G430M gratings. The slit PAs and offsets from the centroids are given in Table \ref{tab:hst-obs}. The parallel slits are numbered as 1-3 from bottom-top in Mrk~78 and 1-7 from left-right in NGC~1068. Slit 4 in Mrk~78 and slit 8 in NGC~1068 are at different PAs than the parallel slits, but pass through the nuclei. In NGC~4151, slit 1 is aligned on the continuum centroid, while slits 4 and 5 are placed to the south, and slits 2 and 3 are placed on the north.
The flux contours are generated starting at 3$\sigma$ above the background, and increase in powers of 2 $\times$ 3$\sigma$ for each inner contour. The color bars on the right indicate the fluxes in units of erg s$^{-1}$ cm$^{-2}$ pix$^{-1}$. The standard deviations ($\sigma$) were calculated from a portion of the background image outside the respective galaxies. The individual knots of emission are labeled with alphabetical letters and were identified using the flux contours and \othree kinematics. North is up for Mrk~3, NGC~1068 and NGC~4151, while the image of Mrk~78 is rotated 10$\degr$ to the East because the extended emission falls off the edge of the chip towards the SW.}
\label{fig:O3}
\end{figure*}

To study the gas kinematics, we utilized HST STIS long-slit spectroscopy encompassing the [O~III] \llothree~emission line doublet, which is the strongest tracer of AGN ionized gas and outflows in the optical.
Mrk~78, NGC~1068, and NGC~4151 have multiple parallel slit (52\arcsec $\times$ 0\farcs2) observations with the medium dispersion G430M grating, which provides a velocity resolution of $\sim$30~\kms~that is useful for distinguishing multiple kinematic gas components. In the absence of G430M observations, we employed the lower dispersion G430L grating (velocity resolution of 300~$-$~400~\kms) with a 52\arcsec $\times$ 0\farcs1~slit for Mrk~3, sufficient for distinguishing components with large kinematic differences. We retrieved multiple exposures from MAST and further processed and combined them into the final data files using the Interactive Data Language (IDL). More details of the data reduction and calibration are given in \cite{Fischer2013}.

To map the structure of the emission line gas, we complemented our spectroscopic observations with emission line images using narrow/medium band-pass filters centered on the \othree emission. Furthermore, we employed medium/wide band filter images of the same targets centered on the continuum-part of the spectrum.
These images contain minimal or no emission/absorption features and we used them to subtract the continuum emission from each \othree image. For Mrk~3 and Mrk~78, we employed the FOC images using the F502M ([O III]) and F550M (continuum) filters. For NGC~1068 and NGC~4151, we used images taken with the WFPC2 F502N filter for [O III], and the F547M and F555W filters to measure the continuum emission. 

The continuum-subtracted \othree images are shown in Figure~\ref{fig:O3}, along with the  orientations of the STIS slits.
The slit PA and offsets from the nuclei in the spatial direction are given in Table~\ref{tab:hst-obs}.
In most cases, the slit positions are aligned close to the major axis of the NLR emission.
Whereas Mrk~3 has a single slit position, the remaining AGN have multiple slit locations to cover significant fractions of their NLRs.
For NGC~1068, slits 5 and 6 overlap and cover the same portion of the NLR with an offset along the slit in the spatial direction.

Finally, to obtain the surface brightness profiles and radial mass distributions of Mrk~78 and NGC~1068, we used the wide band F814W images. We acquired the relevant parameters for Mrk~3 and NGC~4151 from the literature (see \S \ref{subsec:galfit} for more details).
We retrieved all of the combined and drizzled images from the Hubble Legacy Archive (HLA), and then analysed them using SAOImage DS9 and the Astropy library in Python \citep{ds92000, astropy:2013}.

\subsection{Apache Point Observatory (APO)}\label{subsec:apoobs}

\setlength{\tabcolsep}{0.06in}
\renewcommand{\arraystretch}{1.05}
\tabletypesize{\small}
\begin{deluxetable*}{ccccccccccc}[ht!]
\tablecaption{ARC 3.5m Telescope Observations of the sample \label{tab:apo-obs}}
\tablehead{
\colhead{Target} & \colhead{Instrument} & \colhead{Filter} & \colhead{Date} & \colhead{Exposure} & \colhead{Spectral} & \colhead{Wavelength} & \colhead{Spatial} & \colhead{Position} & \colhead{Mean} & \colhead {Mean} \vspace{-2ex}\\
\colhead{Name} & \colhead{Name} & \colhead{} & \colhead{(UT)} & \colhead{Time} & \colhead{Dispersion} & \colhead{Range} & \colhead{Scale} & \colhead{Angle} & \colhead{Air Mass} &\colhead {Seeing} \vspace{-2ex}\\
\colhead{} & \colhead{} & \colhead{} & \colhead{} & \colhead{(s)} & \colhead{(\AA~pix$^{-1}$)} & \colhead{(\AA)} & \colhead{($\arcsec$~pix$^{-1}$)} & \colhead{(deg)} & \colhead{} &\colhead {($\arcsec$)}
}
\startdata
Mrk 3 & ARCTIC & J-C B & 2021-09-08 & 400 & ... & 3400-6000 & 0.228 & ... & 1.6 & 1.1\\
Mrk 3 & ARCTIC & J-C V & 2021-09-08 & 45 & ... & 4500-7000 & 0.228 & ... & ... &  1.5\\
\hline
Mrk 78 & ARCTIC & J-C B & 2021-09-08 & 900 & ... & 3400-6000 & 0.228 & ... & 1.6 & 1.1\\
Mrk 78 & ARCTIC & J-C V & 2021-09-10 & 250 & ... & 4500-7000 & 0.228 & ... & 1.7 & 1.2\\
Mrk 78 & DIS & B1200 & 2016-01-02 & 2137 & 0.615 & 4257-5517 & 0.42 & 24 & 1.4 & 1.5\\
Mrk 78 & DIS & B1200 & 2014-10-25 & 2400 & 0.615 & 4760-6000 & 0.42 & 76 & 1.2 & 1.5\\
Mrk 78 & DIS & B1200 & 2015-02-19 & 2400 & 0.615 & 4481-5721 & 0.42 & 100 & 1.2 & 1.5\\
Mrk 78 & DIS & B1200 & 2015-12-03 & 2700 & 0.615 & 4278-5518 & 0.42 & 152 & 1.5 & 1.5\\
\hline
NGC 1068 & ARCTIC & J-C B & 2021-09-08 & 25 & ... & 3400-6000 & 0.228 & ... & 1.4 & 1.1\\
NGC 1068 & ARCTIC & J-C V & 2021-09-08 & 9 & ... & 4500-7000 & 0.228 & ... & 1.3 &  1.1\\
NGC 1068 & ARCTIC & J-C I & 2021-09-08 & 7 & ... & 7000-1200 & 0.228 & ... & 1.3 & 1.1\\
NGC 1068 & DIS & B1200\tablenotemark{$\star$} & 2014-08-20 & 2700 & 0.615 & 4734-5533 & 0.42 & 40 & 2.4 & 1.3\\
NGC 1068 & DIS & B1200\tablenotemark{$\star$} & 2015-12-03 & 2700 & 0.615 & 4264-5536 & 0.42 & 64 & 1.9 & 1.5\\
NGC 1068 & DIS & B1200\tablenotemark{$\star$} & 2015-12-03 & 2700 & 0.615 & 4264-5536 & 0.42 & 106 & 1.9 & 1.2\\
NGC 1068 & DIS & B1200\tablenotemark{$\star$} & 2014-08-20 & 2700 & 0.615 & 4262-5534 & 0.42 & 154 & 1.9 & 1.3\\
\hline
NGC 4151 & DIS & B1200\tablenotemark{$\star$} & 2016-03-05 & 2700 & 0.615 & 4261-5533 & 0.42 & 221 & 1.4 & 1.4\\
NGC 4151 & DIS & B1200\tablenotemark{$\star$} & 2016-03-05 & 2700 & 0.615 & 4261-5533 & 0.42 & 250 & 1.4 & 1.2\\
NGC 4151 & DIS & B1200\tablenotemark{$\star$} & 2016-03-05 & 2700 & 0.615 & 4261-5533 & 0.42 & 288 & 1.4 & 1.2
\enddata
\tablecomments{A summary of the ground-based imaging and spectroscopy used in this study. The columns list (1) Galaxy name (2) the APO instruments, (3) filters (for imaging)/ gratings (for spectra) used, (4) the observations dates, (5) total exposure times for each data set, (6) spectral dispersion (for spectra) (7) wavelength range (for spectra) or bandpass (for imaging), (8) spatial scales of the images/spectra, (9) PAs for the long-slits, (10) mean air mass and (11) mean seeing for the nights of observations. \tablenotemark{$\star$}Affected by the instrument scattered light.}
\vspace{-2em}
\end{deluxetable*}

We obtained ground-based long-slit spectra and broad-band imaging  of our targets using the Apache Point Observatory (APO) 3.5m telescope. 
We have previously utilized observations using APO's Dual Imaging Spectrograph (DIS) to study the kinematics of ionized gas in nearby Seyfert galaxies such as Mrk 573 \citep{Fischer2017, Revalski2018a}, Mrk~34 \citep{Revalski2018b}, Mrk~3 \citep{Gnilka2020}, Mrk~78 \citep{Revalski2021} and NGC 4051 \citepalias{Meena2021}.
The 6\arcmin$\times$2\arcsec wide DIS long-slit enables the detection of fainter emission from the extended ionized gas that is missed by the narrow HST slits, and traces the host galaxy rotation with an angular resolution of $\sim$1\arcsec. 
We use multiple slit PAs to fully sample the velocity fields. The DIS splits the light into a blue and a red channel to provide \hbeta and \halpha portions of the spectra at the same time. We use the medium dispersion B1200 and R1200 gratings with velocity resolutions of $\sim$75 \kms. The two-dimensional spectral images were reduced using an IRAF \citep{Tody1986, Tody1993} routine consisting of standard reduction steps such as bias subtraction, flat-field adjustments, cosmic ray removal, and correction for atmospheric extinction. The wavelength and flux calibration were performed using same night exposures of arc lamps and a standard star. A detailed description of the reduction and calibration processes can be found in our previous publications, including \citetalias{Meena2021}.

For this study, we adopted the DIS kinematics for Mrk~3 and Mrk~78 from \cite{Gnilka2020} and \cite{Revalski2021}, respectively. For NGC~1068 and NGC~4151, we identified a broad wing to the emission lines caused by contamination from a dewar leak that was previously noted in some Mrk~3 observations \citep{Gnilka2020} and the red channel spectra of NGC 4051 \citepalias{Meena2021}. The DIS spectra for NGC~1068 and NGC~4151 were considered inadequate to precisely measure the outflow kinematics or to determine the transition from outflow to rotation. However, after a careful examination of the spectra, we were able to distinguish the contaminating emission due to scattered light from the strong central \othree lines from the rotating gas outside the inner 20\arcsec (in NGC~1068) and 9\arcsec (in NGC~4151) around the nucleus.

Furthermore, we observed Mrk~3, Mrk~78 and NGC~1068 with the Astrophysical Research Consortium Telescope Imaging Camera (ARCTIC) using Johnson-Cousins (JC) B,V,R,I filters on the APO 3.5m telescope. We used these broad-band images to supplement the HST wide-band F814W (comparable to the JC-I filter) images and constrain the $M/L$ of different morphological components of the host galaxies determined using two-dimensional surface brightness decomposition as discussed in \S \ref{subsec:galfit}. 
For this work, we only used B and V band images to determine the color (B-V), which provides a suitable indicator of $M/L$ for I band images \citep{McGaugh2014}.

ARCTIC provides a field of view of $7.85\arcmin \times $7.85\arcmin using a $4096 \times 4096$ pixel CCD with a plate scale of 0\farcs228 pixel$^{-1}$ for $2 \times 2$ binning to avoid excessive over-sampling of the seeing.
We used an open source automatic reduction pipeline named `Acronym' \citep{Weisenburger2017} that was developed at APO to reduce the raw exposures including the dark, bias, and flat-field corrections. We then combined the exposures for each target and bandpass into final images using the IMCOMBINE task in IRAF. Detailed information of the APO DIS and ARCTIC and observations can be found in Table~\ref{tab:apo-obs}.

\section{ANALYSIS} \label{sec:analysis}

\subsection{Ionized Gas Kinematics}\label{subsec:gas_kinematics}

In order to derive the launch distances of AGN-driven outflows using radiative-gravity formalism similar to \cite{Meena2021}, we require the observed positions and velocities of emitting ionized gas. The ionized gas kinematics are also needed for determining the galaxy rotation curves (see \S \ref{subsec:rotation}) that are used to separate outflows from rotating gas components and to compare with the host galaxy mass distribution (see \S \ref{subsec:mass_vel}). As discussed in the \S \ref{sec:obs}, we utilized the strong \othree emission lines from archival HST (imaging and spectroscopy) observations to measure the spatially-resolved  velocities of the NLR gas and from APO spectroscopy to map the large-scale galaxy rotation.
 
\begin{figure*}[ht!]
\centering
\subfigure{
\includegraphics[width=0.425\textwidth]{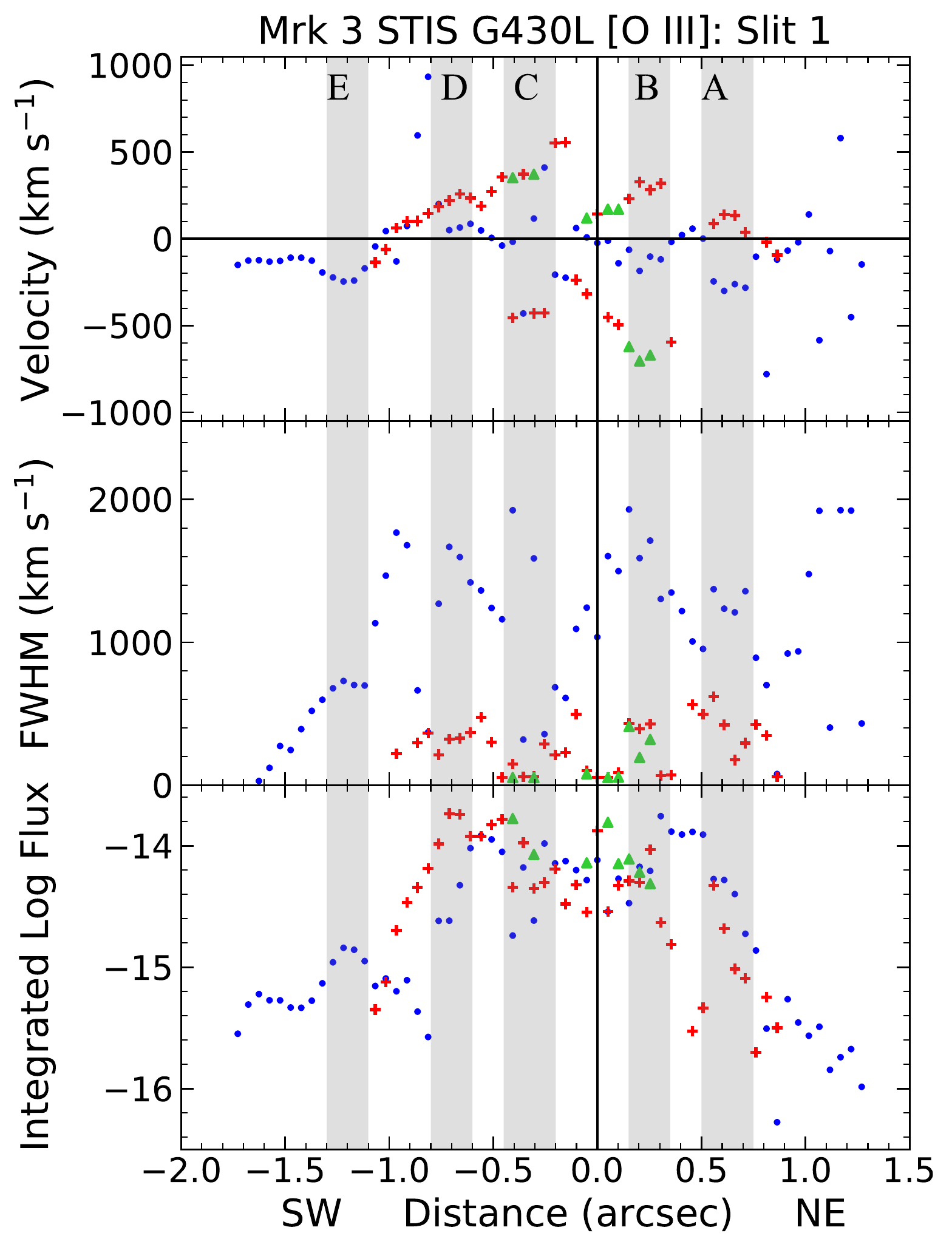}}
\subfigure{
\includegraphics[width=0.415\textwidth]{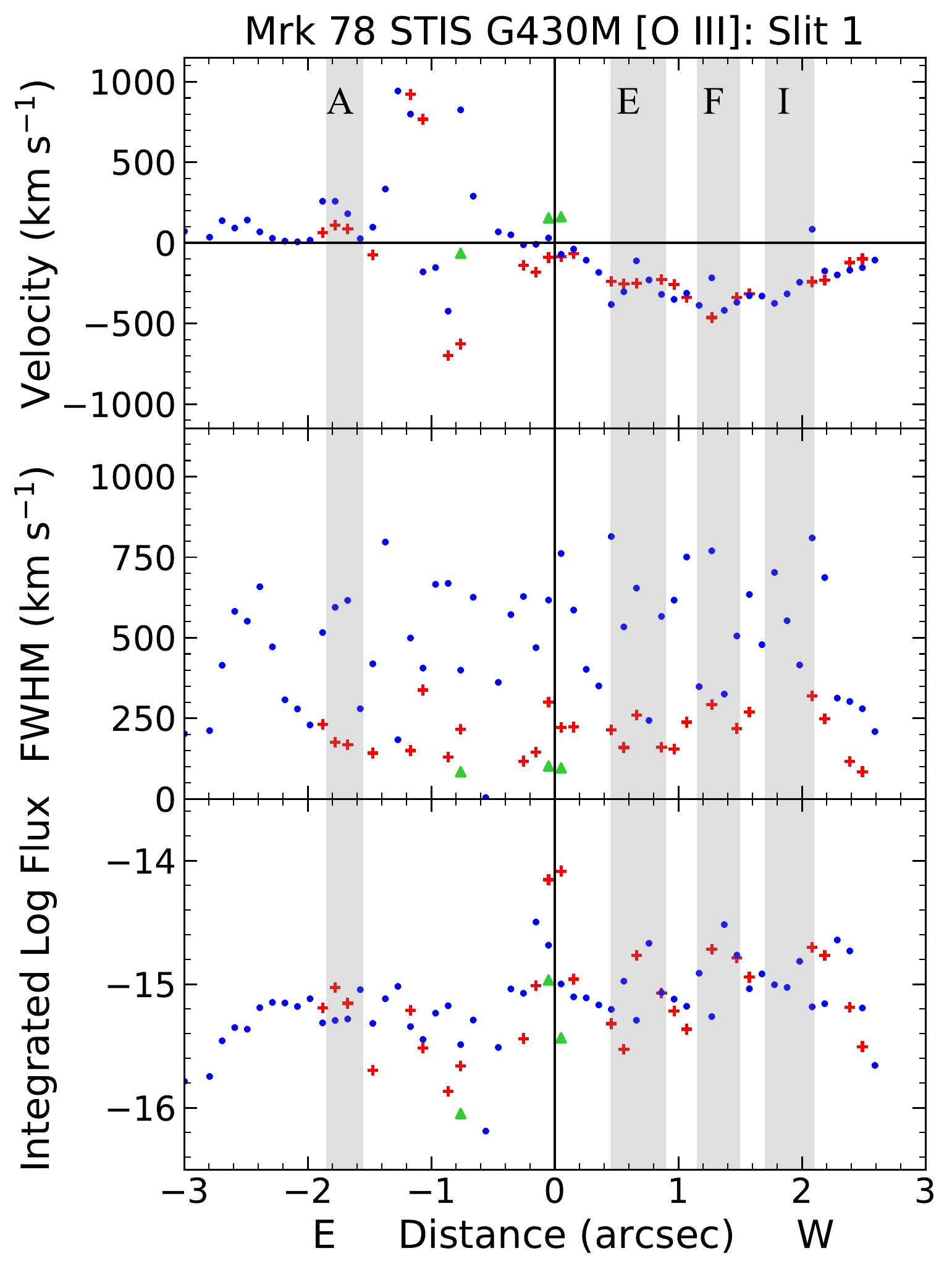}}
\vspace{1ex}
\subfigure{
\includegraphics[width=0.425\textwidth]{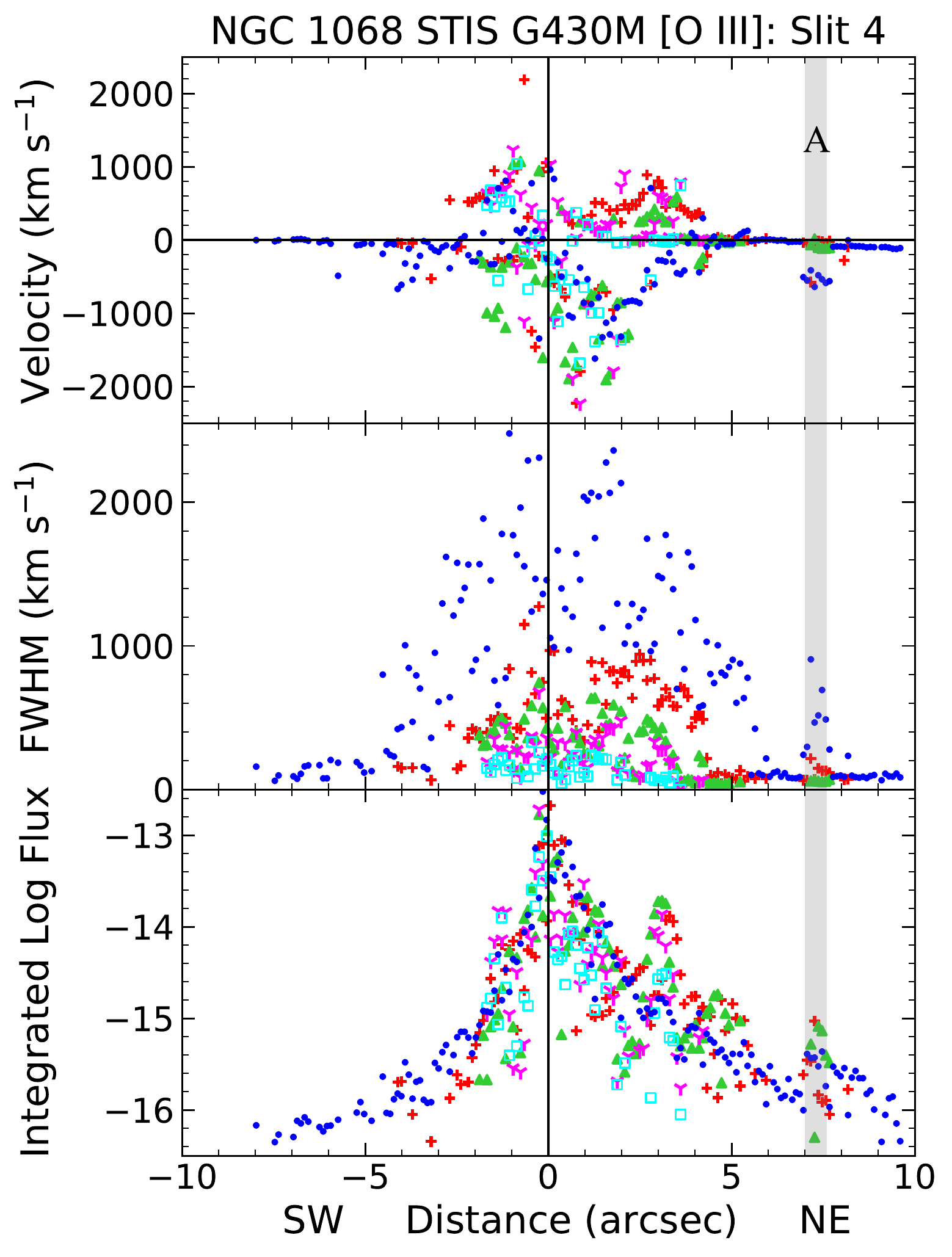}}
\subfigure{
\includegraphics[width=0.415\textwidth]{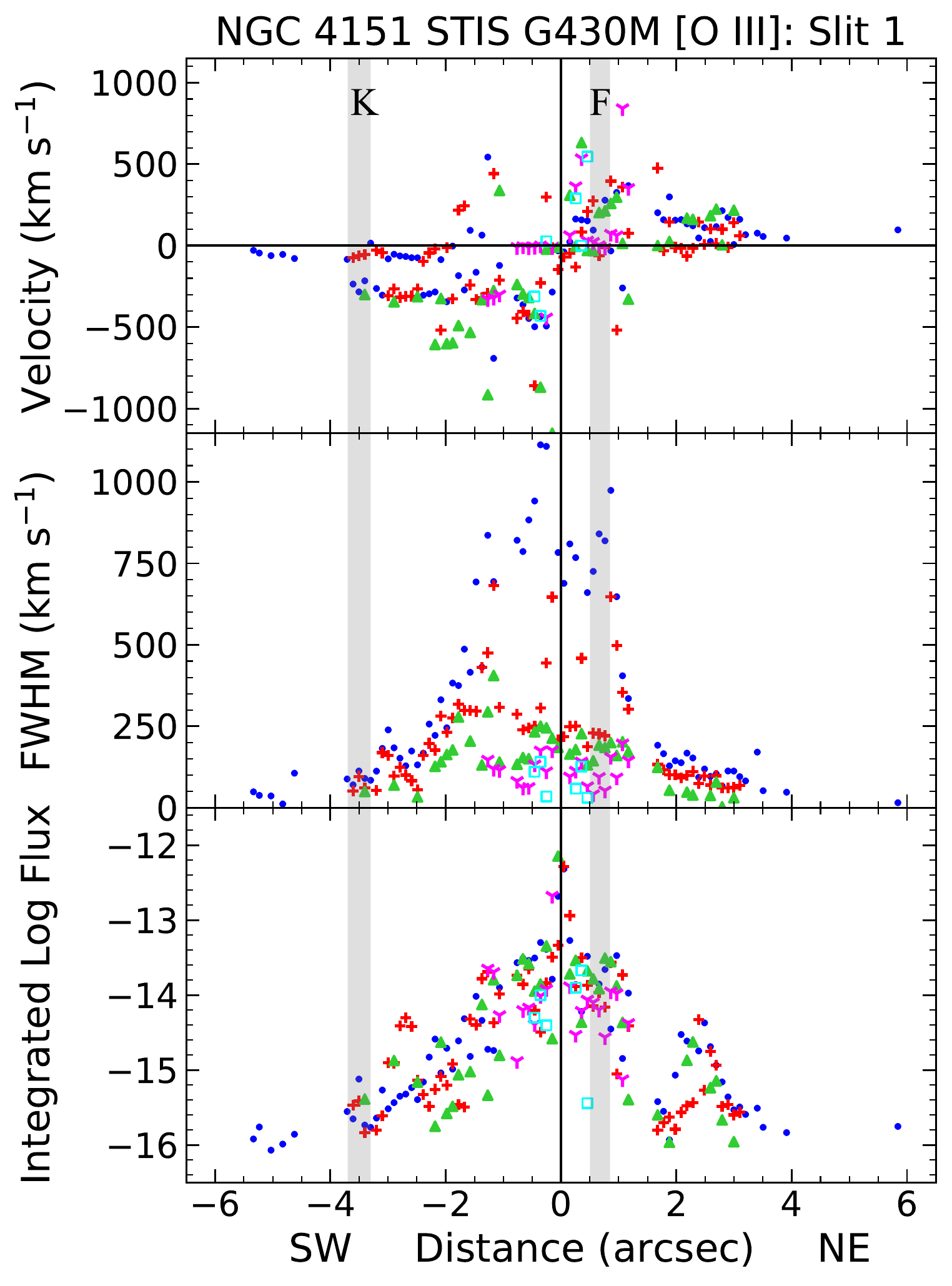}}
\caption{The observed velocity centroids, FWHM, and integrated flux (erg s$^{-1}$ cm$^{-2}$) of the \othree ionized gas measured for STIS long-slit spectra centered on or close to the nucleus in each target. The velocities are presented for galaxy rest frame where velocity of 0 \kms~corresponds to the redshift of the galaxy and the position of 0\arcsec~represents the continuum centroid. The FWHM is corrected for the instrument's LSF. The points in each extracted bin of 0.1\arcsec~(and $\sim$0.05\arcsec~for Mrk~3) are sorted for highest to lowest FWHM in blue (dots), red (pluses), limegreen ($\bigtriangleup$), magenta (Y) and cyan (squares). The grey shaded regions show some of the \othree emission knots identified in the Figure~\ref{fig:O3}. The full sets of kinematics and rest of the identified emitting knots for Mrk~78, NGC~1068 and NGC~4151 are shown in Appendix~\ref{app_kinematics}.}
\label{fig:kinematics}
\end{figure*}

The kinematics of the \othree ionized gas, using the same HST datasets as in this work, have been previously presented for Mrk~3 \citep{Ruiz2001, Collins2005}, Mrk~78 \citep{Fischer2011}, NGC~1068 \citep{Das2006} and NGC~4151 \citep{Das2005}. 
In order to acquire the values for the velocities and distances of individual emitting gas, we again fit the \othree\llothree\ lines using the same observations. However, in this work, we utilize an automated fitting routine based on a Bayesian technique, first introduced in \cite{Fischer2017} to fit the spectral lines using multiple Gaussian profiles. The previous studies used visual identification of bumps in the profile to determine the number of Gaussian components. The spectral fitting procedure is discussed in detail in Appendix \S \ref{subsec:app_spec_fit}.
Our multi-Gaussian fitting routine provides a significant improvement over the previous methods by analytically determining the most significant number of kinematic components \citep{Fischer2017}. This routine is highly efficient at identifying kinematic components as either rotation, outflow or turbulent gas, and also allow us to perform spectral fitting for large datasets simultaneously using pre-defined parameters. 
The detailed descriptions of the kinematic observations for individual galaxies are provided in Appendix~\ref{app_kinematics} including the similarities and differences with the previous measurements. An example of the observed velocity maps for one of the HST slits (centered on the brightest continuum peak) for Mrk~3, Mrk~78, NGC~1068 and NGC~4151 is shown in Figures~\ref{fig:kinematics}. 
We matched the knots identified in the \othree images (Figure~\ref{fig:O3}) in our observed velocity maps and labeled them as shown in Figures~\ref{fig:kinematics}. Since there is only one slit observation for Mrk~3, all of the identified knots are presented in  Figures~\ref{fig:kinematics}. For Mrk~78, NGC~1068 and NGC~4151, the kinematic maps along with the identified \othree knots for the rest of the HST slit observations are provided in Appendix~\ref{app_kinematics}. 
The knots as labeled in alphabetical letters on the kinematics maps are chosen from similar adjacent kinematics (velocities, FWHM, and fluxes) in velocity distribution for adjacent pixels. If more than one component at a position qualified, we selected the one with higher velocity at that location. We use an average of the distances and velocities to define each knot. These values were further used to determine their velocity profiles using radiative-gravity models (see \S \ref{subsec:radiative_driving}). The uncertainties in the positions and velocities are also discussed in \S \ref{subsec:radiative_driving}.

\subsection{Galaxy Rotation Curves}  \label{subsec:rotation}

\begin{figure*}[ht!]
\centering
\includegraphics[width=0.325\textwidth]{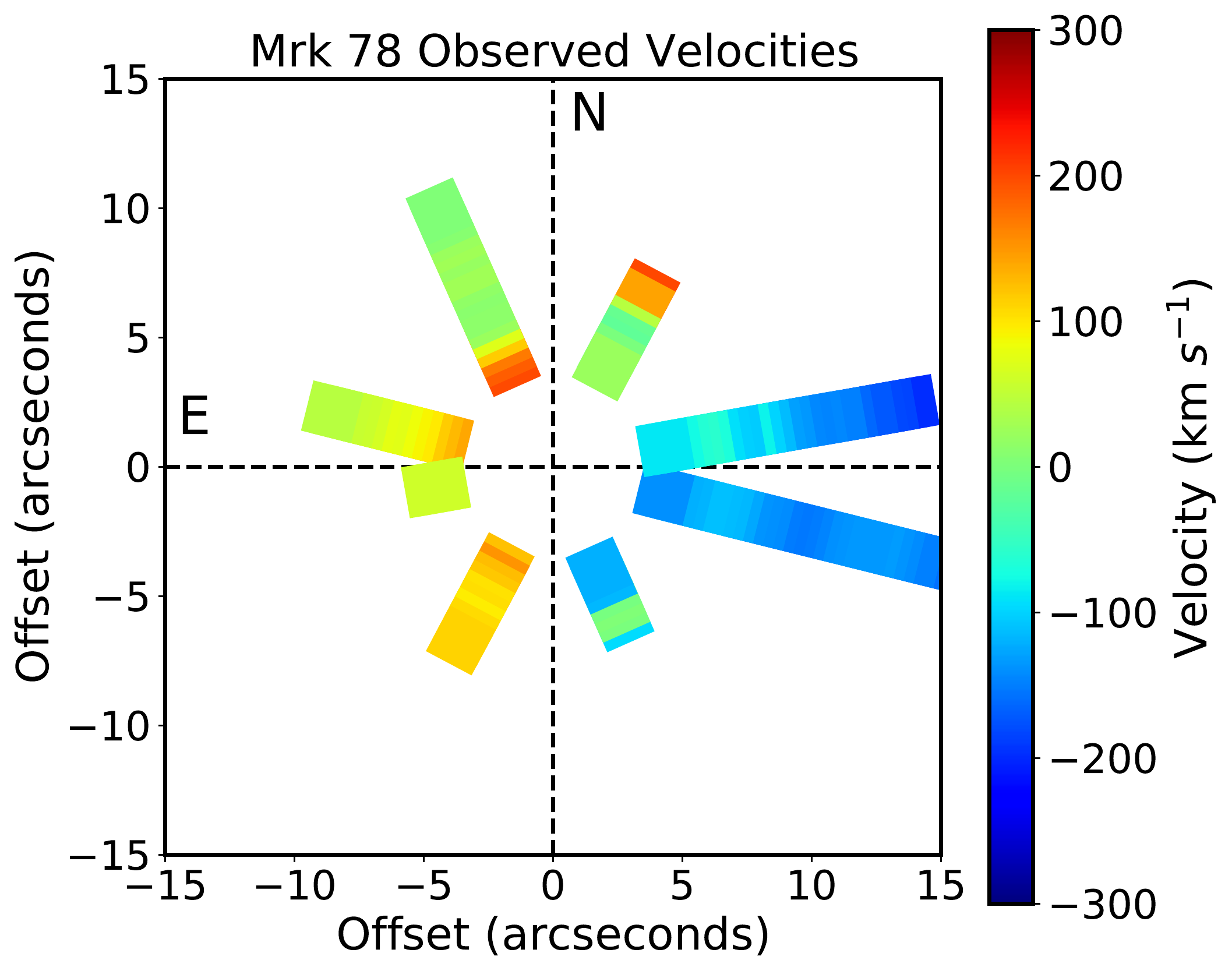}
\includegraphics[width=0.325\textwidth]{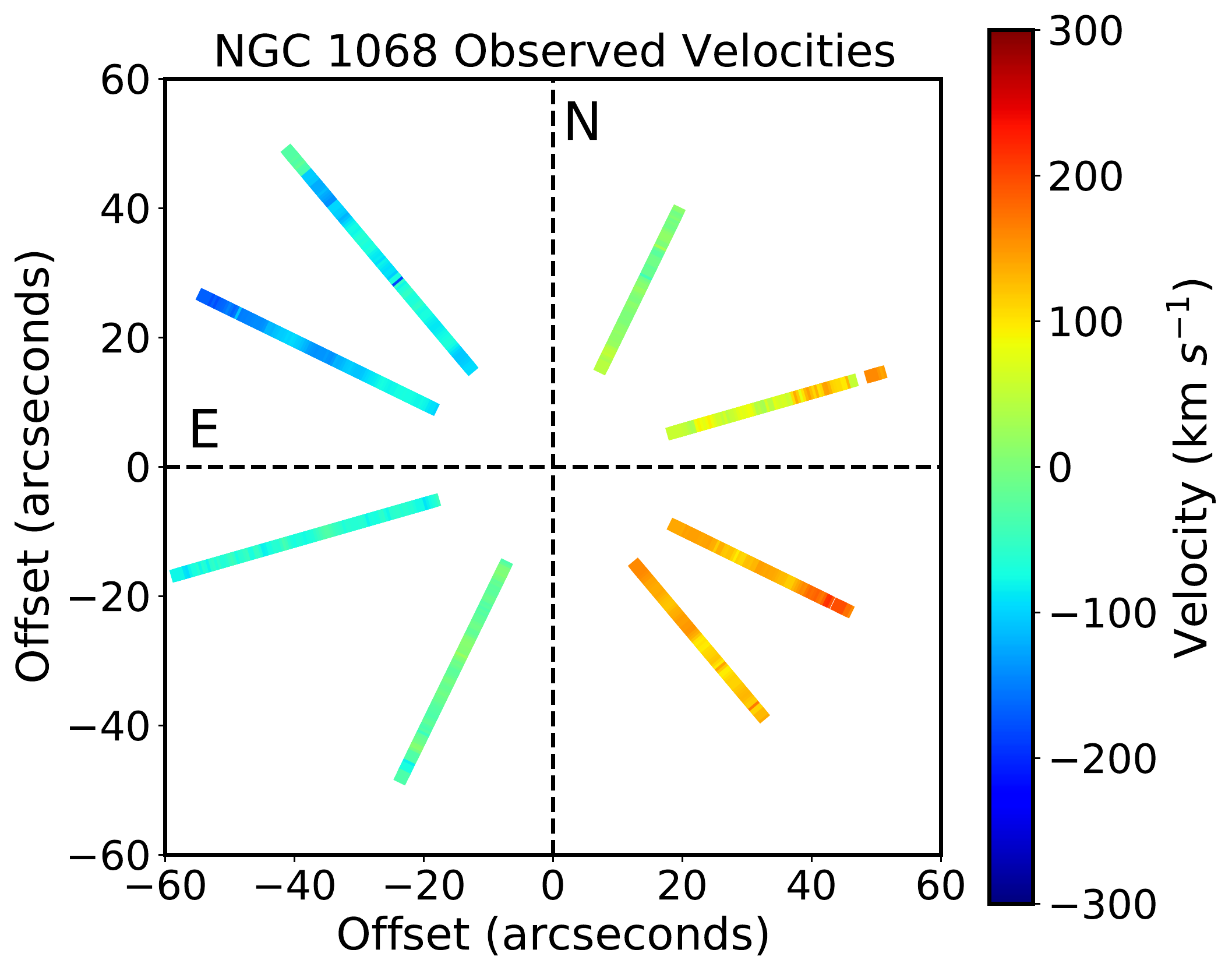}
\includegraphics[width=0.325\textwidth]{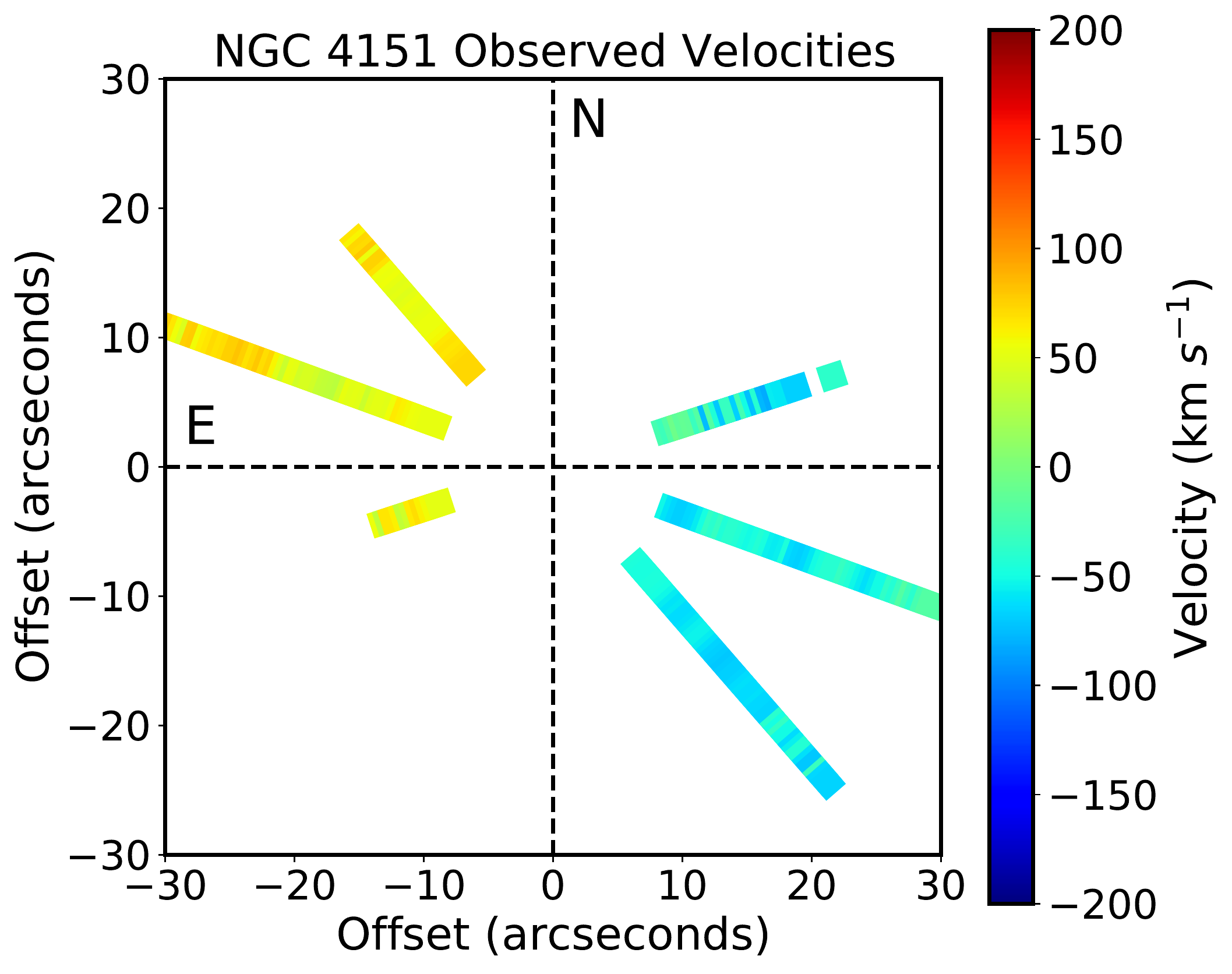}
\includegraphics[width=0.325\textwidth]{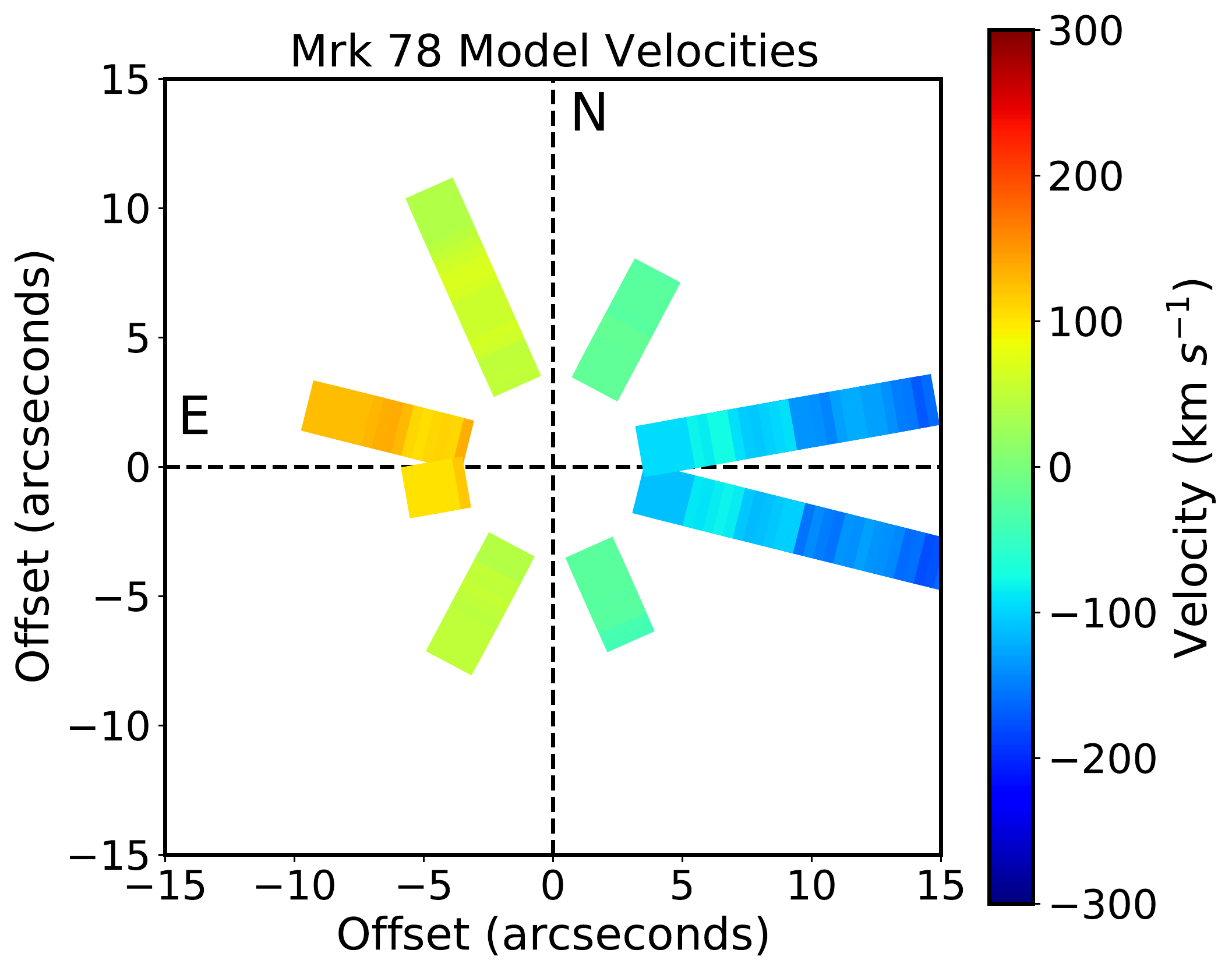}
\includegraphics[width=0.325\textwidth]{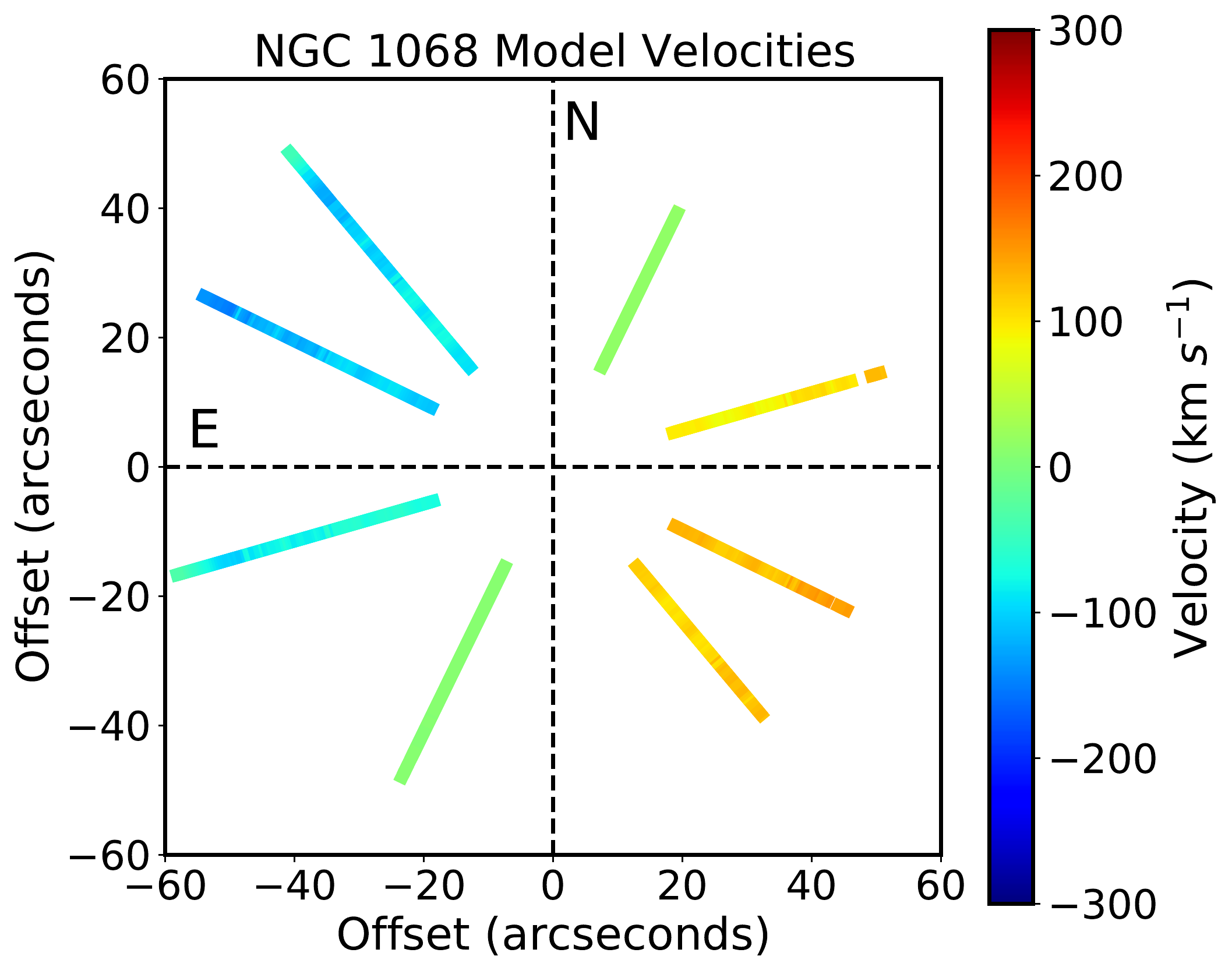}
\includegraphics[width=0.325\textwidth]{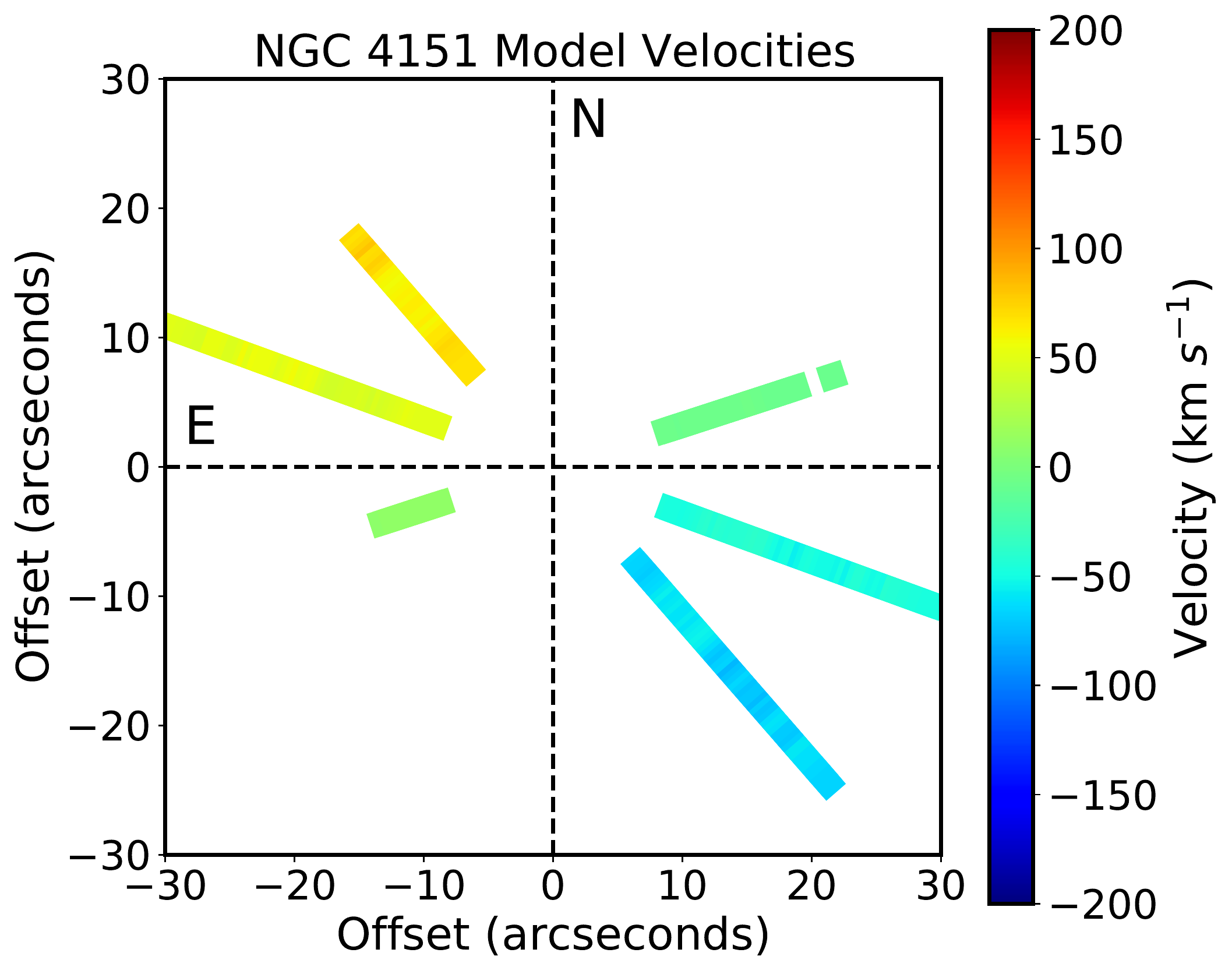}
\includegraphics[width=0.325\textwidth]{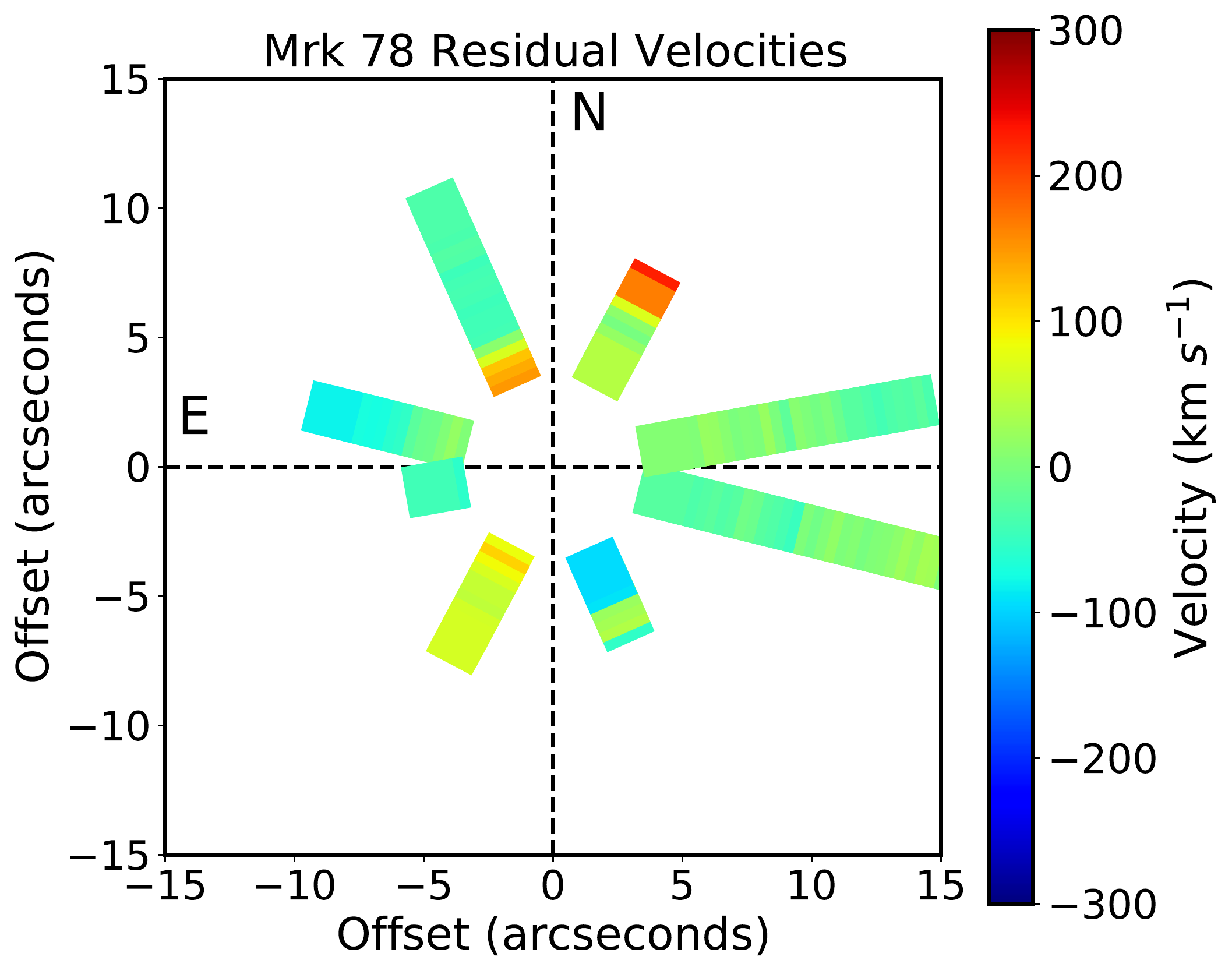}
\includegraphics[width=0.325\textwidth]{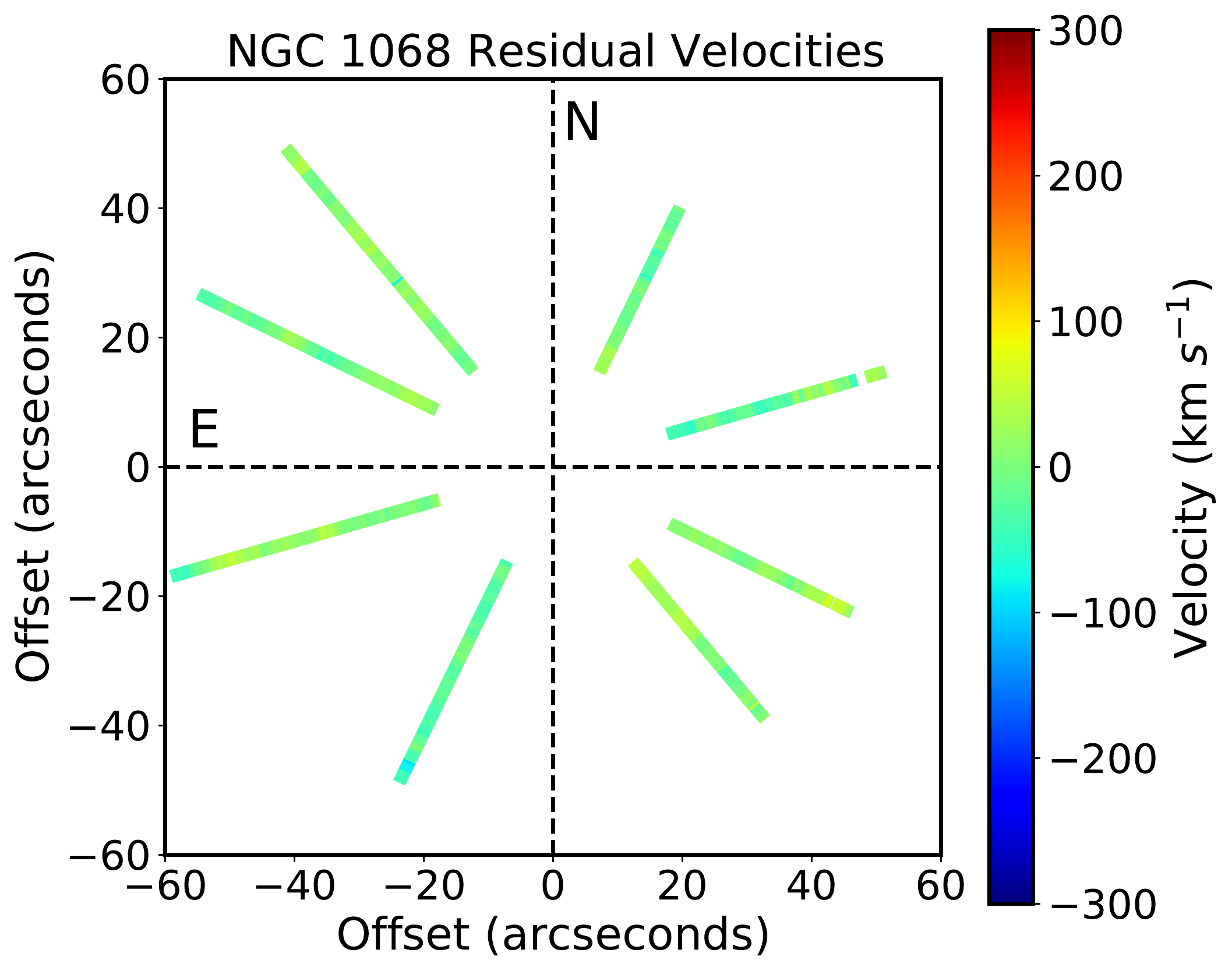}
\includegraphics[width=0.325\textwidth]{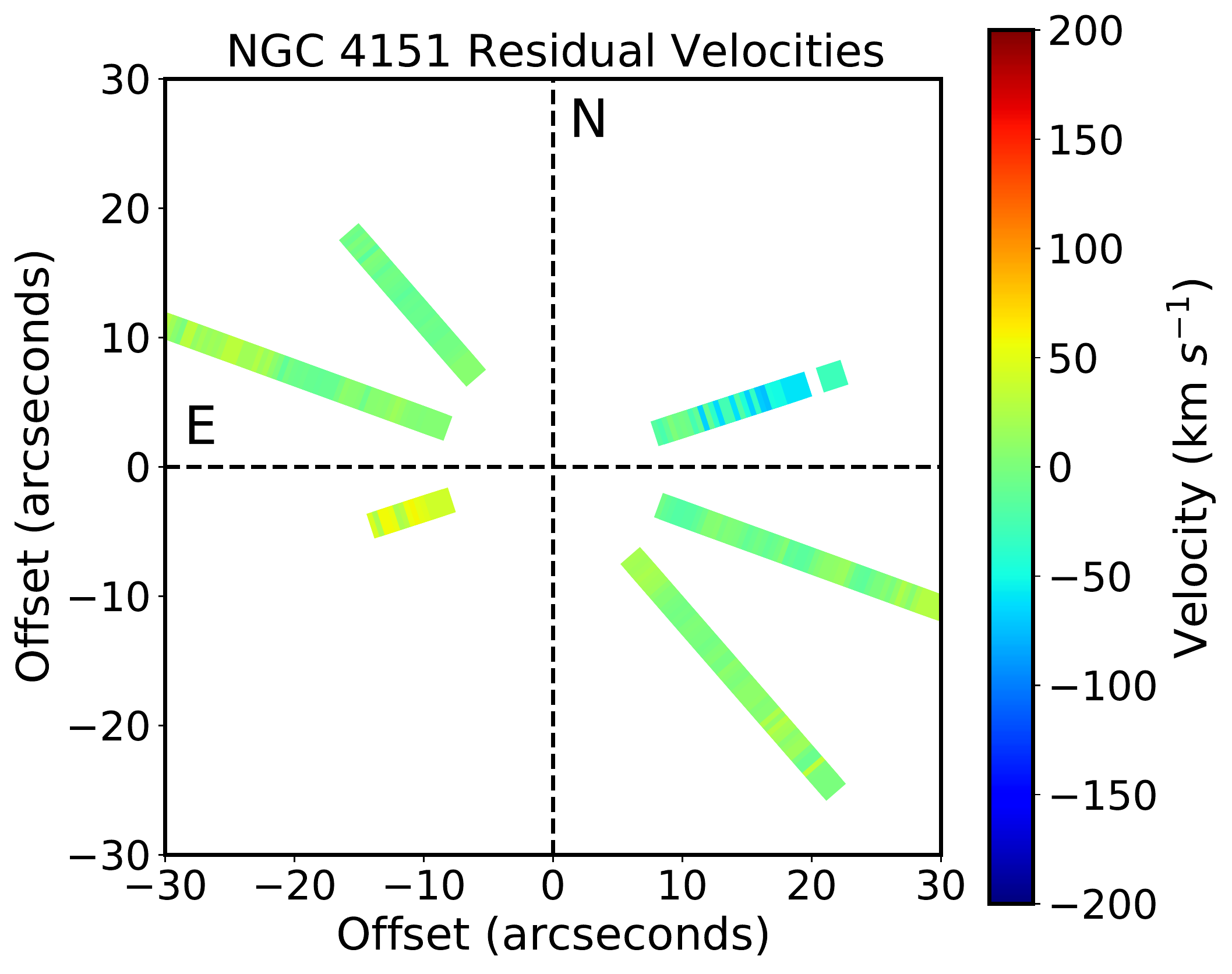}
\caption{Top row: Pseudo-IFU velocity maps of the large-scale ionized gas, determined using APO DIS \othree spectroscopic fits. The velocity map for Mrk~78 is adopted from Figure~4 of \cite{Revalski2021} for the 1st component (highest flux), which demonstrates rotational structure in the kinematics. For NGC 4051 and NGC~1068, the velocity maps constitute the narrow component of the two Gaussian fit to extended emission. The width represents the relative width of the slits on the sky. The middle row shows the velocities derived by fitting rotational models to the galaxies using DiskFit \citep{Sellwood2010,Sellwood..Spekkens2015}. The bottom row shows the residual (data-model) velocities. The data from the inner 4\arcsec~in Mrk~78, 20\arcsec~in NGC~1068 and 9\arcsec~in NGC~4151 were masked to avoid contamination from outflows and scattered light. For Mrk~3, we previously presented the \othree velocity maps and \textsc{GALFIT} models in \cite{Gnilka2020}, and adopt those results.}
\label{fig:velmaps}
\end{figure*}

We determined the large-scale rotation of the host galaxies using the APO DIS observations. These rotation curves will be compared with our enclosed mass-profiles and associated Keplerian rotation in \S \ref{subsec:mass_vel}. 
The stellar and ionized gas rotation of Mrk~3 is provided in our previous work by \cite{Gnilka2020} using APO DIS Ca II triplet absorption lines and \othree emission, respectively.

In \cite{Revalski2021}, we presented the kinematics of the \othree ionized gas in Mrk~78 using four APO DIS slits, which demonstrate large-scale rotation in the highest flux component as shown in the pseudo-Integral Field Unit (IFU) velocity maps in Figure \ref{fig:velmaps}. We removed the inner $\sim$4\arcsec~data-points to avoid the high-velocity outflow signature that is present in all four slits. We then used DiskFit \citep{Sellwood2010,Sellwood..Spekkens2015,Kuzio2012soft,Peters2017} to model the observed velocities and determine the de-projected values. We fit a simple rotational model to the host galaxy, assuming circular motion of the ionized gas outside $\pm$4\arcsec~from the nucleus. Due to a low number and unequal distribution of velocity points in different slits, we were not able to achieve a reasonable model by fitting all the parameters such as PA, disk ellipticity/inclination,  disk center and systemic velocity as variables. Therefore, we fixed the inclination of the host galaxy to 55.6\arcdeg~based on its I-band photometry \citep{Schmitt2000}. The disk center was also fixed to the continuum centroid for each slit and all other parameters were left as free. The model outputs a disk major axis PA of 85$\pm$3\arcdeg, which is in an excellent agreement with the photometric PA of 84\arcdeg~given in \cite{Schmitt2000}.

We derive the rotation curves of the \othree ionized gas in NGC~1068 and NGC~4151 using the narrow-component fits to the APO DIS spectra as discussed in \S \ref{subsec:app_spec_fit}. The pseudo-IFU velocity fields are shown in Figure \ref{fig:velmaps}. While four different APO DIS slit are used to map the ionized gas velocity distribution in the host galaxy of NGC~1068, only three long-slit observations are available for NGC~4151. We removed the data points from the inner $\sim$ $\pm$ 20\arcsec~in NGC~1068 and $\sim$ $\pm$ 9\arcsec~in NGC~4151 to avoid the high-velocities associated with outflows and/or scattered light (\S\ref{subsec:app_spec_fit}).

Similar to Mrk~78, we model these large-scale velocities to fit simple rotation to the host galaxies using DiskFit. The four APO DIS slits for NGC~1068 sufficiently cover the velocity field, so all the model parameters were allowed to vary except the disk centers that were fixed to the continuum centroids for each slit. The best fit model provides PA = 66$\pm$1\arcdeg, inclination = 35$\pm$1\arcdeg~and a systemic velocity offset of 12$\pm$1 \kms. 
The major axis PA from our model based on the ionized gas kinematics is almost 90\arcdeg offset from I-band photometric PA (-67\arcdeg) given in \cite{Schmitt2000}. However, our modeled PA more closely matches with the inner oval structure than the outer galactic disk \citep{Schinnerer2000}. Our modeled inclination is closer to B-band photometric value  (i$_{B}$ = 36\arcdeg) in \cite{Schmitt2000} and comparable to the 40\arcdeg adopted in \cite{Brinks1997} and \cite{Schinnerer2000}.
For NGC~4151, an optimal fit was not attainable with only three slit positions. Therefore, we fixed the major axis PA and inclination of the host galaxy to 26\arcdeg~and 22\arcdeg, respectively \citep{Davies1973, Mundell1999}. The disk center was fixed to the continuum centroid of each slit and an offset of $\sim$2$\pm$2 \kms~was output from the model.

\begin{figure*}[ht!]
\centering
\includegraphics[width=0.325\textwidth]{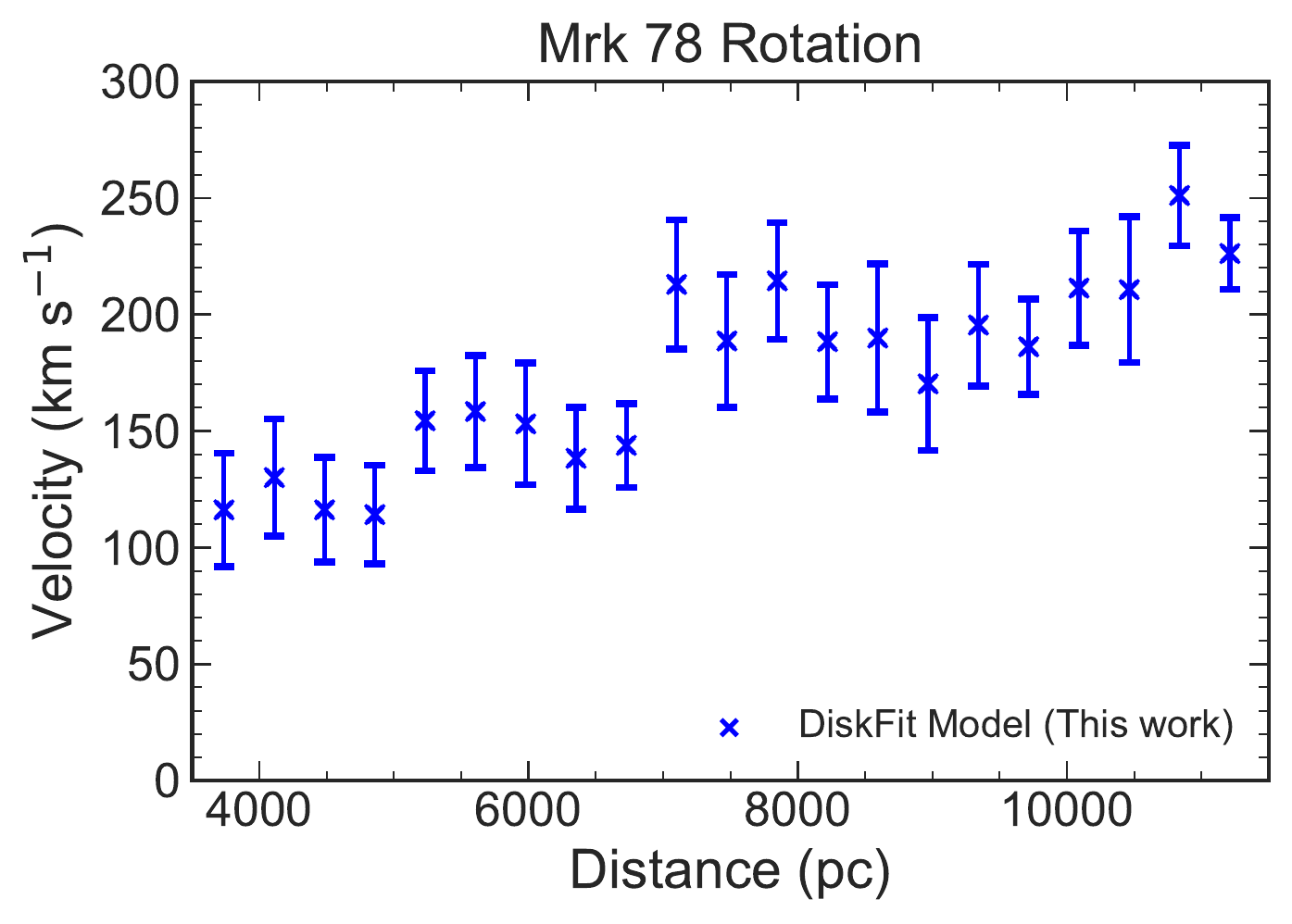}
\includegraphics[width=0.325\textwidth]{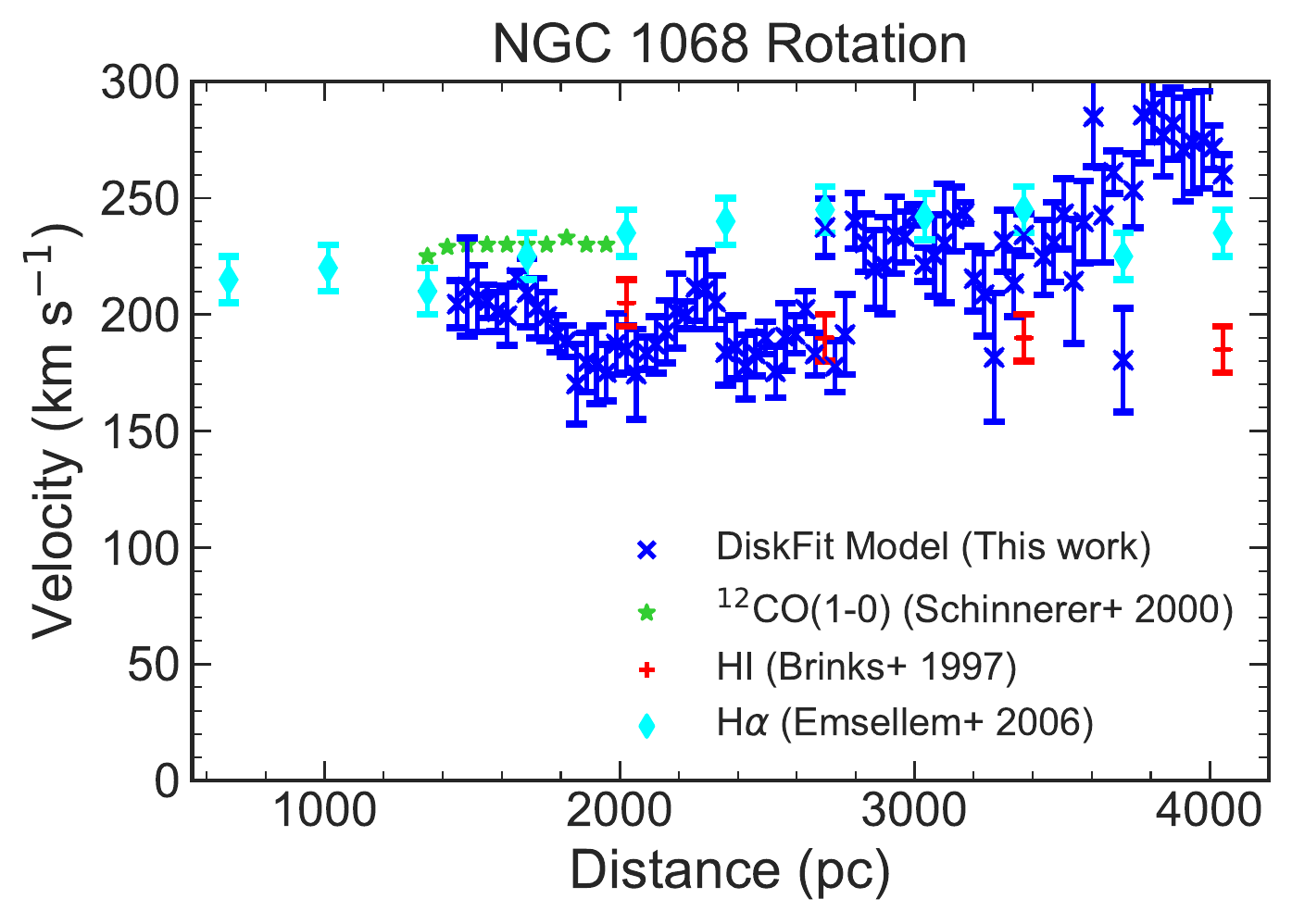}
\includegraphics[width=0.325\textwidth]{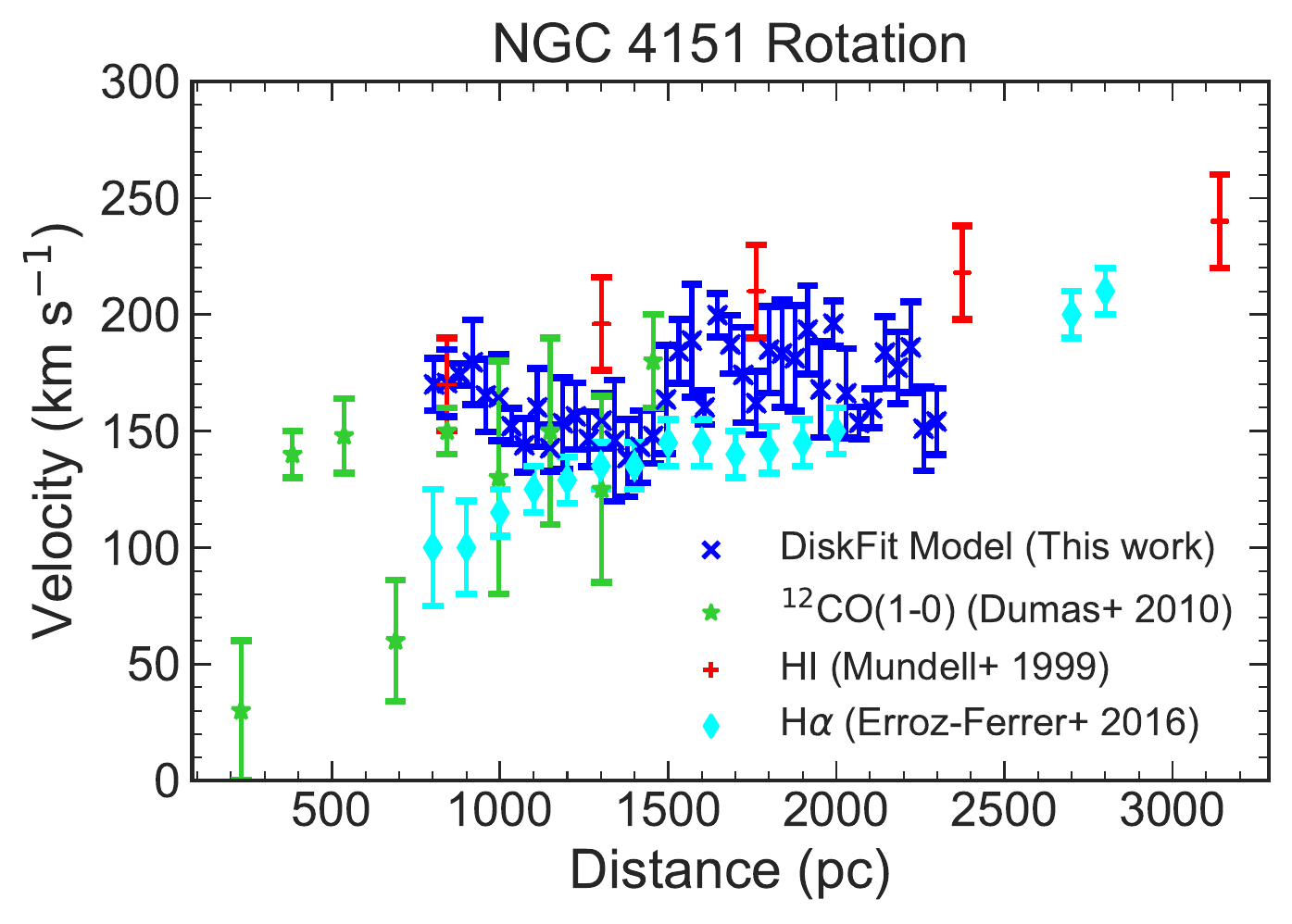}
\caption{The average rotation curves obtained from DiskFit models of the \othree kinematics (blue crosses), compared with literature observations of the CO (green stars), HI (red pluses) and \halpha (cyan diamonds) emission. A rotation curve for Mrk~3 is provided in \cite{Gnilka2020}. The adopted or modeled (using DiskFit) values of the PA and inclinations are: Mrk~78 (PA = 84\arcdeg, \textit{i} = 55.6\arcdeg), NGC~1068 (PA = 66\arcdeg, \textit{i} = 35\arcdeg), NGC~4151 (PA = 26\arcdeg, \textit{i} = 22\arcdeg).}
\label{fig:rotation}
\end{figure*}

The observed DiskFit model and residual velocities for Mrk~78, NGC~1068, and NGC~4151 are shown in Figure \ref{fig:velmaps}, and we present their rotation curves from the DiskFit models in Figure~\ref{fig:rotation}. We compare them with the velocity fields from various sources in the literature for neutral hydrogen (HI) 21-cm, cold molecular gas using CO emission, and ionized gas using \halpha emission. There are no prior large-scale velocities available for Mrk~78, and so we adopt our DiskFit model velocities as the basis for Mrk~78 rotation.

Our modeled circular rotation for NGC~1068 is in agreement with the \halpha rotation derived using SAURON integral field spectrograph \citep{Emsellem2006}, the $^{12}$CO(1-0) velocity fields using the IRAM interferometer on Plateau de Bure, France \citep{Schinnerer2000}, and the rotation curve based on VLA HI data \citep{Brinks1997} to within $\sim$30 \kms. Our model velocities show more structure than the literature values, which may not be real given the uncertainties, agreeing with some previous values more than others at different scales. We also observe an increase in velocities around 4 kpc unlike HI data, which shows a gradual decrease. This difference is likely attributed to noise in the ends of the APO spectra.

For NGC~4151, our model rotational velocities based on \othree kinematics are slightly higher than those obtained using \halpha kinematics with the GH$\alpha$FaS (Galaxy \halpha Fabry–Perot System) instrument on the William Herschel Telescope (WHT) \citep{Erroz-Ferrer2016}. This difference is due to their adopted host galaxy inclination ($\sim$48\arcdeg) used to derive the rotation curve, which is higher than ours (22\arcdeg, adopted from \citealp{Mundell1999}). On the other hand, for a similar inclination angle, our rotation curve is mostly within the uncertainties of the velocity fields of the VLA-HI \citep{Mundell1999} and CO observations \citep{Dumas2010}. The dip in our velocities toward large radii are likely due to low S/N data. Overall, our rotation velocities agree well with previous values, indicating their reliability for evaluating our mass models of the host galaxies (\S\ref{subsec:mass_vel}).

\subsection{Surface Brightness Decomposition} \label{subsec:galfit}

We determined the radial stellar mass distribution of each galaxy from 2D surface brightness profiles using the two-dimensional image decomposition software \textsc{GALFIT} \citep{Peng2002}. \textsc{GALFIT} employs multiple analytical functions to fit the different brightness components of the galaxy that may correspond to morphological features such as bars, bulges, disks and rings. We fit the radial brightness of each component using variations of the Sérsic \citep{Sersic1968} profile, which is  defined as 

\begin{equation}
\Sigma(r) = \Sigma_{e} \exp\left[-\kappa\left(\left(\frac{r}{r_{e}}\right)^{1/n} -1\right)\right]
\label{eq:sersic}
\end{equation}
\noindent
where $\Sigma_{e}$ is the surface brightness of a given component at its effective radius $r_{e}$, $n$ is the Sérsic index that defines a power law distribution and the value of constant $\kappa$ is calculated based on $n$ using incomplete gamma function $\Gamma(2n) = 2\gamma(2n, k)$. The Sérsic function for $n$ = 1 represents an exponential disk profile. We modeled each galaxy using a combination of HST wide-band images and ARCTIC broadband images. Mrk~3 was modeled in our previous work by \cite{Gnilka2020} using a HST WFPC2 F814W image. We retrieved the model parameters and used them to guide the fits for ARCTIC B- and V-band images.

\begin{figure*}[ht!]
\begin{center}
\includegraphics[width=0.33\textwidth]{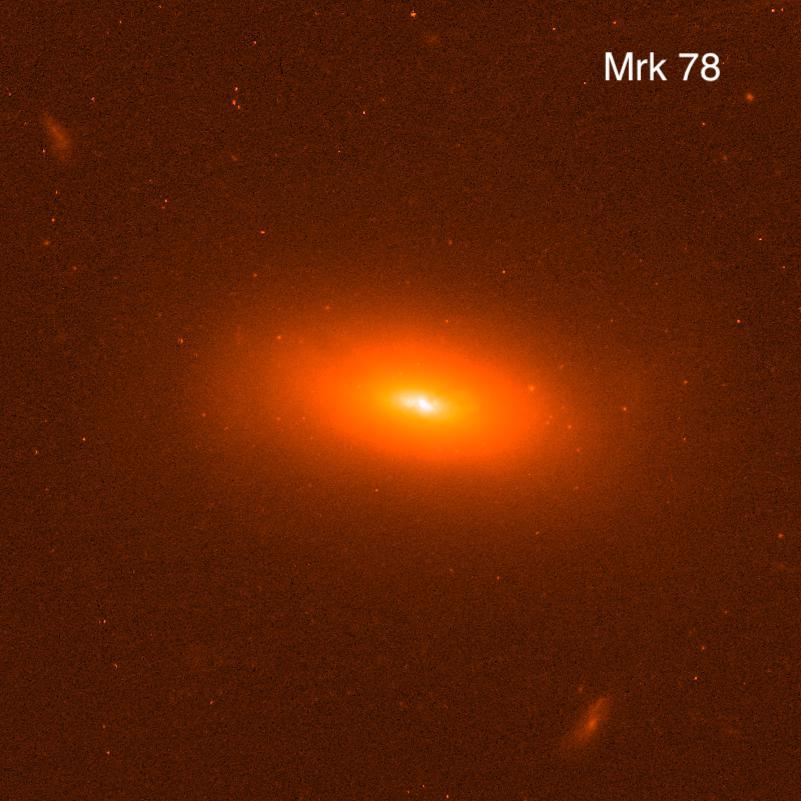}
\includegraphics[width=0.33\textwidth]{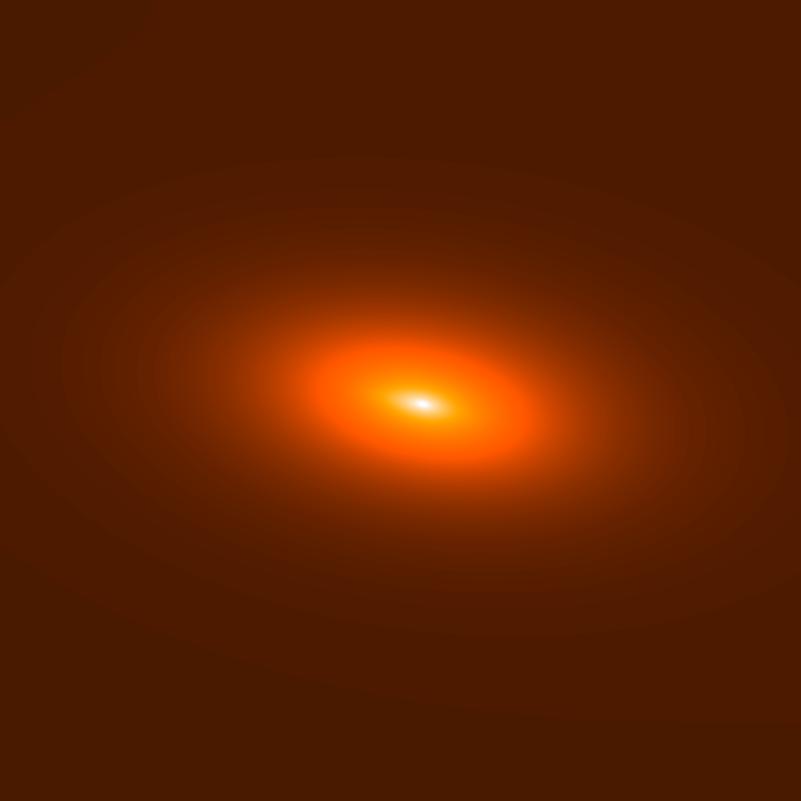}
\includegraphics[width=0.33\textwidth]{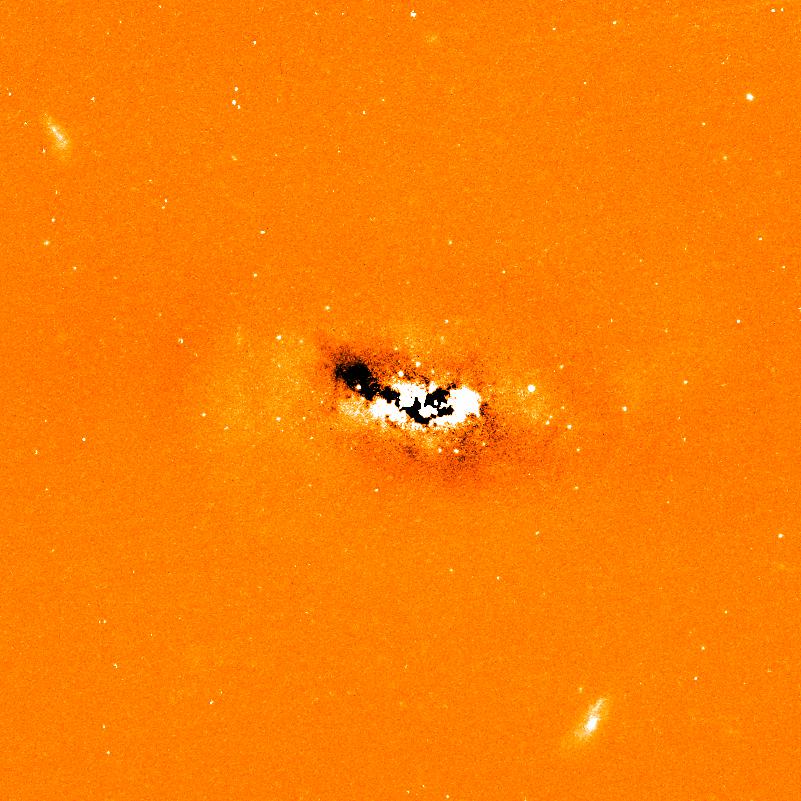}
\vspace{3ex}
\includegraphics[width=0.33\textwidth]{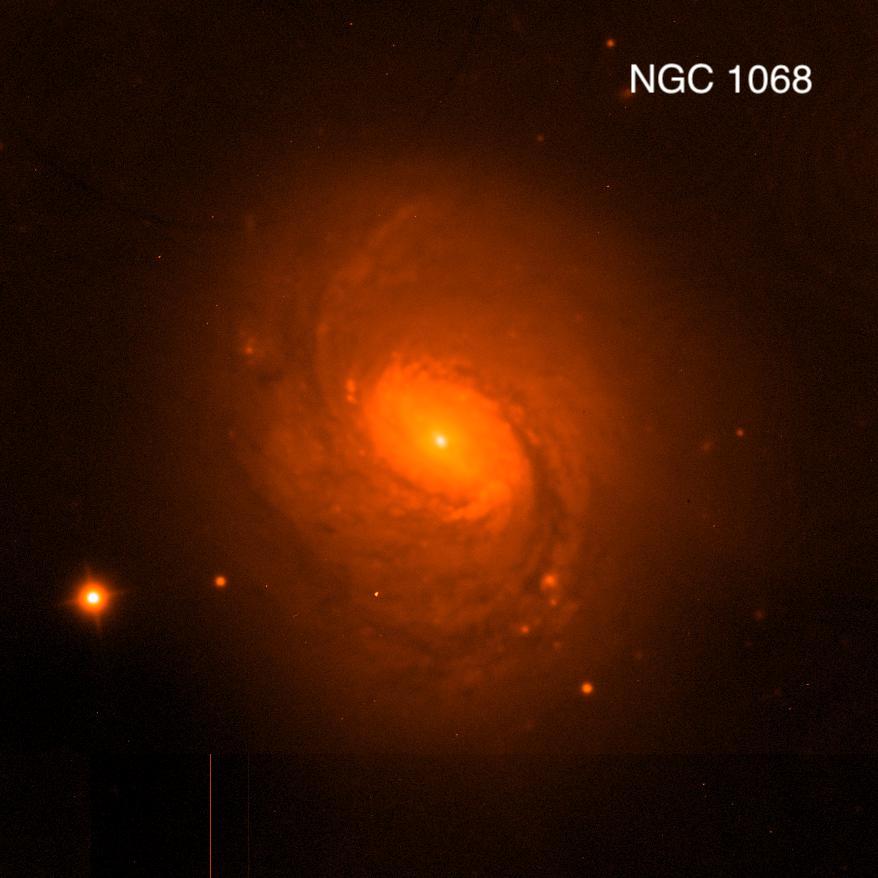}
\includegraphics[width=0.33\textwidth]{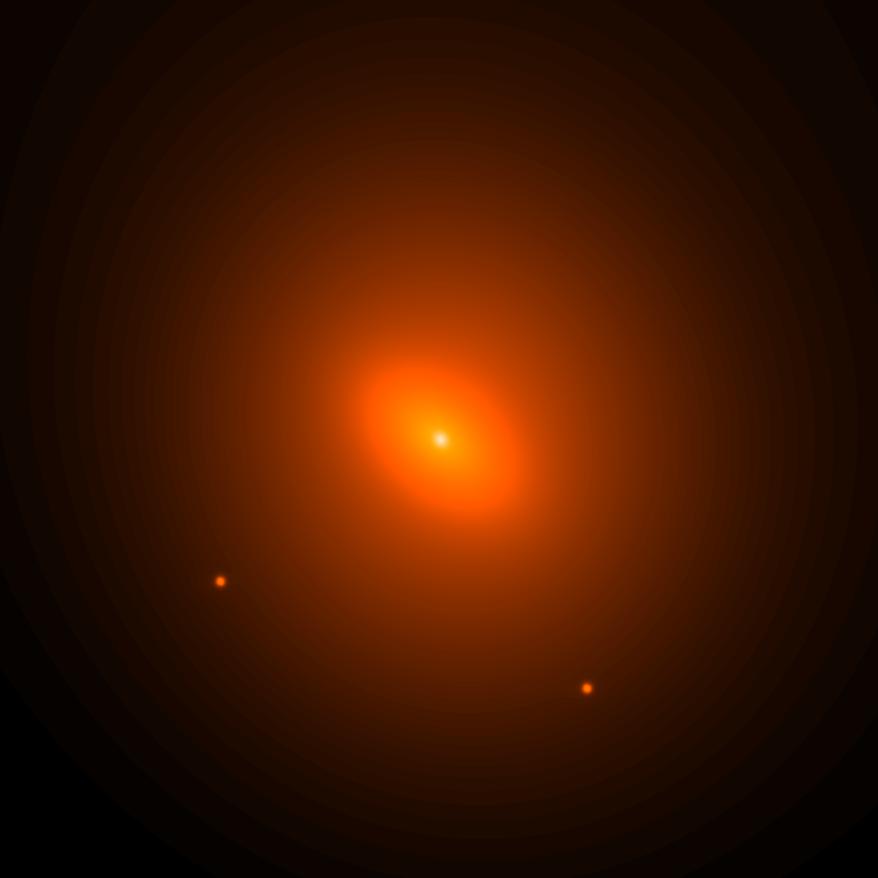}
\includegraphics[width=0.33\textwidth]{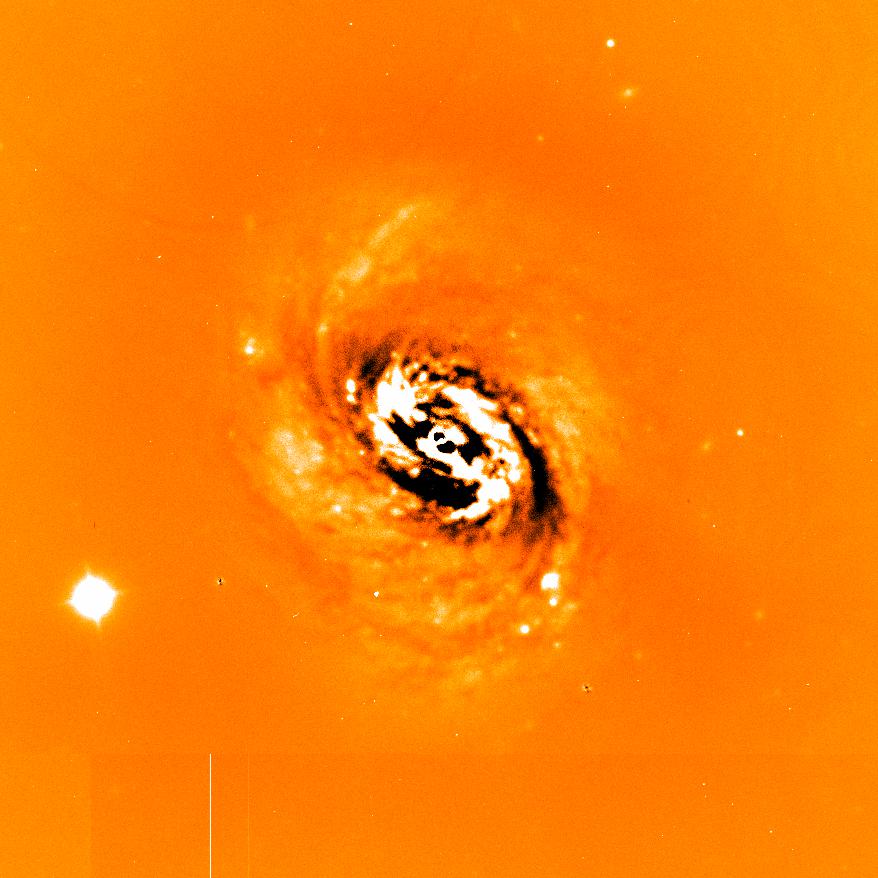}
\end{center}
\vspace{-0.5em}
\caption{Top: Mrk~78 $40 \times 40$\arcsec~HST ACS F814W image (left), \textsc{GALFIT} model (middle), and residual (right). Bottom: NGC~1068 $200 \times 200$\arcsec~ APO-ARCTIC I-band image (left), \textsc{GALFIT} model (middle), and residual (right). The brighter regions in the residual are due to excess NLR emission and/or star-forming regions, whereas the dark regions correspond to the dust lanes. } 
\label{fig:galfit}
\end{figure*}

\begin{figure*}[ht!]
\begin{center}
\includegraphics[width=0.49\textwidth]{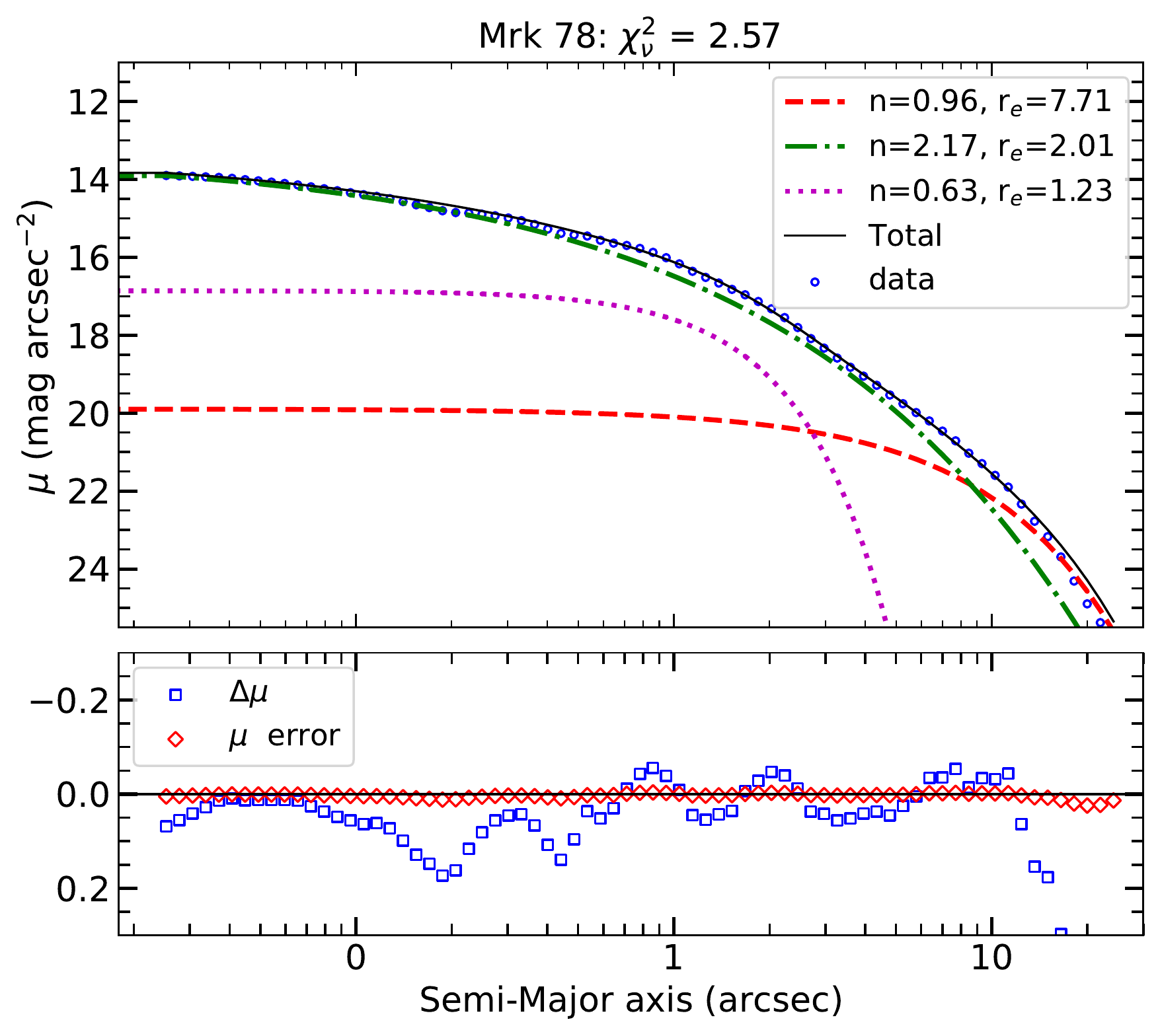}
\includegraphics[width=0.5\textwidth]{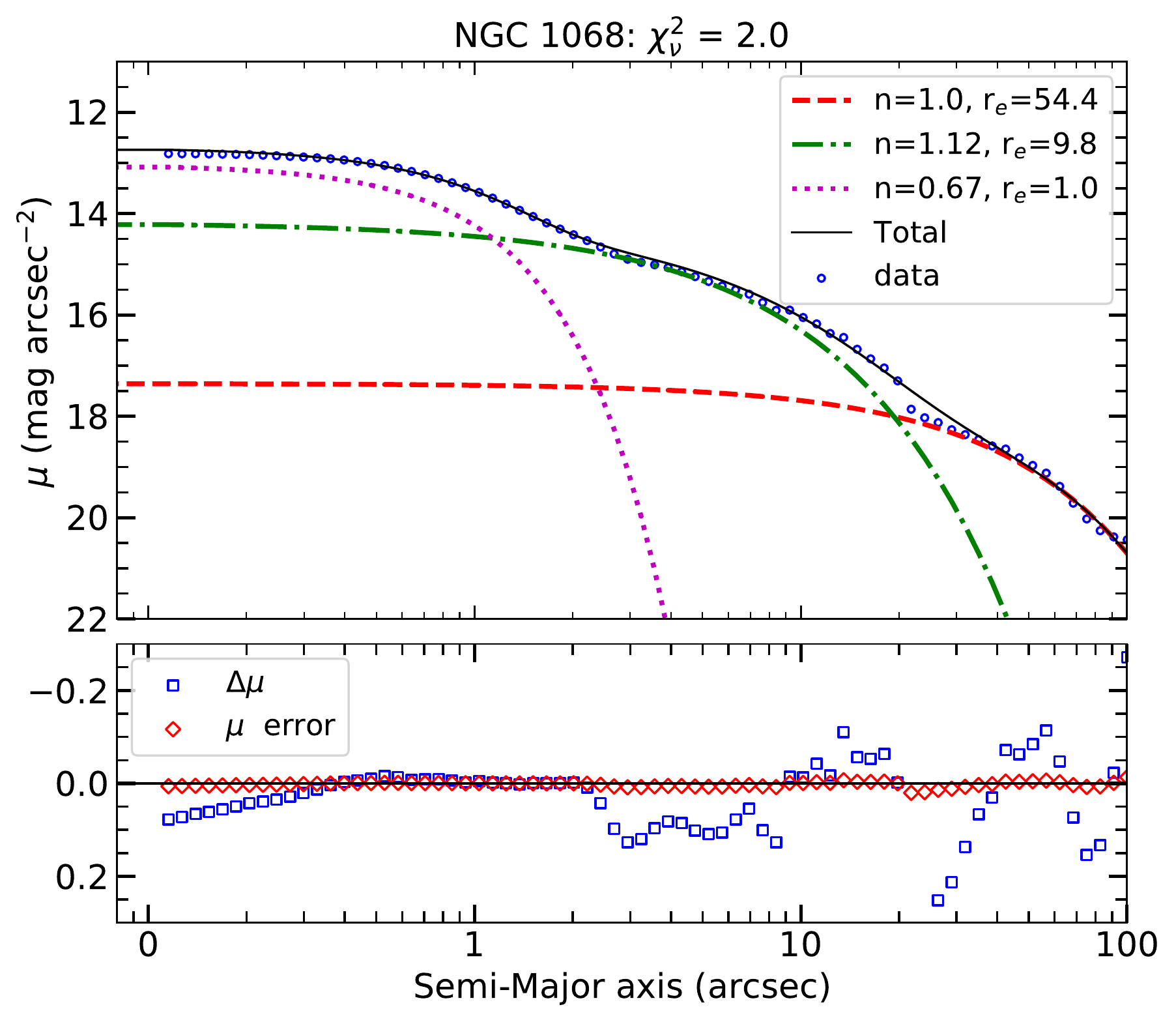}
\end{center}
\caption{The one-dimensional surface brightness profiles for Mrk~78 F814W image (left) and NGC~1068 ARCTIC-I band image (right). The data points obtained using the isophotal analysis of the host galaxy image are shown in blue circles and the best-fit model is shown as the solid black curve.  The text boxes on the upper left of each plot show the Sérsic indices ($n$) and effective radii (r$_{e}$, in arcsec) of individual components (red, green and magenta dashed lines). The panel below shows the magnitude difference ($\Delta\mu$ = data $-$ model) and fractional error = ($\mu$ error = (data$-$model)/data) in blue squares and red diamonds respectively.}
\label{fig:surf_bright}
\end{figure*}

To model Mrk~78, we used a HST ACS F814W image, which is equivalent to a JC-I filter and is largely free of any strong emission lines from the AGN while only sampling light from the host galaxy. The 
integrated magnitude of each component was calculated using $m =  -$2.5log$(counts/sec) + zpt$, where the $zpt$ is the photometric zeropoint of the instrument. We retrieved the zeropoint using ACS Zeropoints Calculator (acszpt) from python package ACStools \citep{acstools2020} for the observation date. The best-fit model was constructed using a pseudo inner bulge/bar, an outer bulge, and a less luminous disk component. The Sérsic indices for each component were determined using an iterative process where some of the parameters, such as the sky level, were initially fixed to a mean value obtained from a background portion of the image lacking light from stars/galaxies. Finally, all of the parameters were set free to minimize the $\chi^{2}$ and to achieve best fit to a 1D brightness profile as well as a reasonable residual. The top panel in Figure \ref{fig:galfit} shows the Mrk~78 input image, the \textsc{GALFIT} model, and the residual.
The residuals show strong dust-lane features close to the nucleus as well as areas of excess luminosity due to the presence of the large-scale NLR. Figure~\ref{fig:surf_bright} (left) shows the 1D surface brightness for Mrk~78 obtained by fitting elliptical isophotes to a galaxy image using the Astropy package, Photutils \citep{larry_bradley_2020_4044744}. The radial brightness profile for individual components from \textsc{GALFIT} model as well as the total model brightness are also over-plotted. Our model fits are in excellent agreement with the data, yielding a difference of less than 0.1 mag in the radial surface brightness distribution. As evidenced by the 1D surface brightness profile and the Sérsic components, the majority of the galaxy light belongs to the large galaxy bulge ($n$ = 2.17, $r_{e}$ = 1.5 kpc). Similar to Mrk~3 \citep{Gnilka2020}, we do not find any spiral arms; 
however, the \othree kinematics \citep{Revalski2021} show a distinct rotational signature at large radii, suggesting a potential disk - which agrees with our \textsc{GALFIT} model and confirms the SB classification of the host galaxy.

Due to NGC~1068's large size on the sky, and its off-center placement on the detector, a significant portion of the galaxy falls outside of the HST ACS F814W image. Therefore, we instead used an ARCTIC I band image to model the surface brightness of the host galaxy, which, similar to HST wide-band filters, offers the host galaxy's stellar light distribution without strong emission or absorption features. The best fit \textsc{GALFIT} model consists of an inner bulge or fat bar, an outer bulge, and a disk component. The bright field star on the left of the galaxy was masked and a few foreground stars were modeled using a point-spread function (PSF). The detailed procedure to build the PSF using ground-based images is provided in \cite{Robinson2021}. The ARCTIC I-band image, \textsc{GALFIT} model and residual are shown in Figure \ref{fig:galfit}. The spiral arms are clearly visible in the residuals, including knots of star formation. The dark black regions show the dust lane features in the nuclear regions. Figure \ref{fig:surf_bright} (right) shows the one-dimensional (1D) distribution of the surface brightness for NGC~1068, using elliptical isophotes to the galaxy image as well as the distribution for the individual sérsic components and the total model profile. The difference between the data and the modeled surface brightness in the inner 20\arcsec~is $<$ 0.1 mag, indicating an acceptable fit using a simple combination of three Sérsic profiles.

We used ARCTIC B and V band images to constrain the $M/L$ of different components in our targets. Due to the higher angular resolution and sensitivity provided by HST, we used the \textsc{GALFIT} parameters determined from the HST F814W image as a guide to fit the ARCTIC images of Mrk~3 and Mrk~78. We fixed the bulge Sérsic indices and effective radii to their HST values, while the magnitude of each component was left as a free parameter. For NGC~1068, the modeling was done independently for B, V, and I band images, which yielded similar results for the Sérsic indices, effective radii, PA, and axis ratio.

The magnitude of each component varied based on the different filters.  
To constrain the unique zeropoints for each ground-based image, we first modeled stars in each field, which had reported magnitudes in optical and near-IR filters. The catalogue values for the B and V bands were obtained from the VizieR database \citep{Ochsenbein2000} using AAVSO Photometric All-Sky Survey \citep{Henden2014, Henden2016}.
For the I band observations, we gathered the SDSS-\textit{i} band magnitudes of each star from the Sloan Digital Sky Survey (SDSS) data release 16 \citep{Ahumada2020}. We then determined the magnitude difference between SDSS-\textit{i} and JC-I of each star using their g-i colors and spectral type \citep{Covey2007}, utilizing the stellar template from the Kurucz 1993 Atlas of Model Atmospheres \citep{Kurucz1993} in IRAF task SYNPHOT. Finally, the \textsc{GALFIT} zeropoints in each filter were adjusted to minimize the difference between the modeled and catalogued magnitudes of the field stars. More details of the process to determine the zero-point for ground-based observations can be found in \cite{Bentz2016} and \cite{Robinson2021}.

The \textsc{GALFIT} model parameters and integrated magnitudes of individual components in the F814W filter (for Mrk~3 and Mrk~78) and JC-I band (for NGC~1068), as well as the corresponding B and V band magnitudes, are given in Table~\ref{tab:galfit}. We emphasize that the parameters for the innermost components obtained using ground-based images should be accepted cautiously, because the effective radii of these components are equivalent to the mean seeing of the observations. These regions also largely encompass the NLR emission (particularly in the B and V bands), which may skew the 
magnitudes of these components associated with stellar light.

Finally, for NGC~4151, we adopted the galaxy decomposition from  \cite{Bentz2018}, which used HST WFC3 F547M imaging for a medium V band image, and WIYN High-Resolution Infrared Camera (WHIRC) H-band imaging. We retrieved the model parameters and the photometric corrected V-H colors from Tables~2 and 4 of \cite{Bentz2018} to calculate the radial mass distribution (see \S \ref{subsec:mass_vel}). 

\renewcommand{\arraystretch}{1.05}
\begin{deluxetable*}{ccccccccc}[ht!]
\tablecaption{\textsc{GALFIT} Parameters \label{tab:galfit}}
\tablehead{
\colhead{Component} & \colhead{r$_{e}$} & \colhead{n} & \colhead{b/a} & \colhead{PA} & \colhead{$m_{_{F814W}}$} & \colhead{$m_{_{B}}$}  & \colhead{$m_{_{V}}$} & \colhead{$M/L$} \vspace{-2ex}\\
\colhead{Number} & \colhead{(pc)} & \colhead{} & \colhead{} & \colhead{($\degr$)} & \colhead{ or $m_{_{I}}$} & \colhead{} & \colhead{} & \colhead{} \vspace{-4ex}\\
}
\startdata
 &  &  & & Mrk 3 &  &  &  & \\
1 & 330 & 1.7 & 0.85 & 23 & 13.08 & 16.19 & 15.29 & 2.2\\
2 & 2130 & 1.1 & 0.42 & 24 & 12.93 & 15.63 & 14.59 & 3.1\\
3 & 5260 & 1.7 & 0.86 & 27 & 11.79 & 14.08 & 13.31 & 1.6\\
\hline
&  &  &  & Mrk 78 & &  &  & \\
1 & 920 & 0.63 & 0.3 & 77 & 16.29 & 16.23 & 15.24 & 2.7\\
2 & 1500 & 2.17 & 0.49 & 78 & 13.87 & 16.05 & 15.46 & 1.0\\
3 & 5760 & 0.96 & 0.52 & 84 & 15.23 & 17.22 & 16.01 & 4.7\\
\hline
&  &  &  & NGC 1068 & &  &  & \\
1 & 80 & 0.67 & 0.79 & 15 & 11.65 & 13.34 & 12.5 & 1.9\\
2 & 685 & 1.12 & 0.64 & 50 & 9.23 & 11.18 & 10.49 & 1.3\\
3 & 3800 & 1 & 0.82 & 20 & 8.13 & 10.05  & 9.29 & 1.5\\
\enddata
\tablecomments{The host galaxy surface brightness fits obtained using \textsc{GALFIT}. The columns list the (1) Component, (2) Effective Radius of the component, (3) Sérsic index, (4) PA, (5) magnitude in HST F814W (Mrk~3, Mrk~78) or I-band (NGC~1068) filter, (6) B-band magnitude, (7) V-band magnitude and (8) Mass-to-Light ratio for I-band and B-V color using Table~1 in \cite{Bell2001}. The parameters for the F814W image decomposition for Mrk~3 have been adopted from \cite{Gnilka2020} and the rest are derived in this work. The magnitude of each component is calculated as $m =$ -2.5log$(counts/sec) + zpt$. The Sérsic parameters and $M/L$ for different surface brightness components of NGC~4151 were collected from \cite{Bentz2018}.}
\vspace{-2em}
\end{deluxetable*}

\section{RESULTS}\label{sec:results}

\subsection{Host Galaxy Mass Profiles} \label{subsec:mass_vel}

Following the technique used in \cite{Fischer2017} and thereafter in \citetalias{Meena2021}, the radial density distribution for individual Sérsic components can be determined using Equation 4 in \cite{Terzi2005}, given as below

\begin{equation}
\rho(r) = \rho_{0} \left(\frac{r}{r_{e}}\right)^{-p} ~e^{-\kappa(\frac{r}{r_{e}})^{1/n}},
\label{eq:density1}
\end{equation}

where

\begin{equation}
 \rho_{0} = \frac{M}{L} \Sigma_{e}\kappa^{n(1-p)} \frac{\Gamma(2n)}{2r_{e}\Gamma(n(3-p))},
\label{eq:density2}
\end{equation}

and

\begin{equation}
p = 1 - \frac{0.6097}{n} + \frac{0.05563}{n^{2}}.
\label{eq:density3}
\end{equation}

The parameters $r$, $r_{e}$, $\kappa$ and $n$ have the same definitions as in Equation~\ref{eq:sersic} for a spherical distribution. For Mrk~3, Mrk~78, and NGC~1068, we calculated the $M/L$ values in the I band using B-V colors and the relationship provided in Table~1 of \cite{Bell2001}. The corresponding values for individual surface brightness components of each galaxy are provided in Table \ref{tab:galfit}. We were unable to locate $M/L$ values for Mrk~3 and Mrk~78 in the literature, and our values for the bulge and disk of NGC~1068 are close to those presented in \cite{Yoshino2008} as  $M/L$$_{(bulge)}$ = 1.46$\pm$0.46 and $M/L$$_{(disk)}$  = 1.29$\pm$0.43. For NGC~4151, we calculated the V-band $M/L$ values for individual components using associated V-H colors retrieved from Table~4 of \cite{Bentz2018}, and the relationship provided in Table~1 of \cite{Bell2001}.

\begin{figure*}[ht!]
\begin{center}
\includegraphics[width=0.475\textwidth]{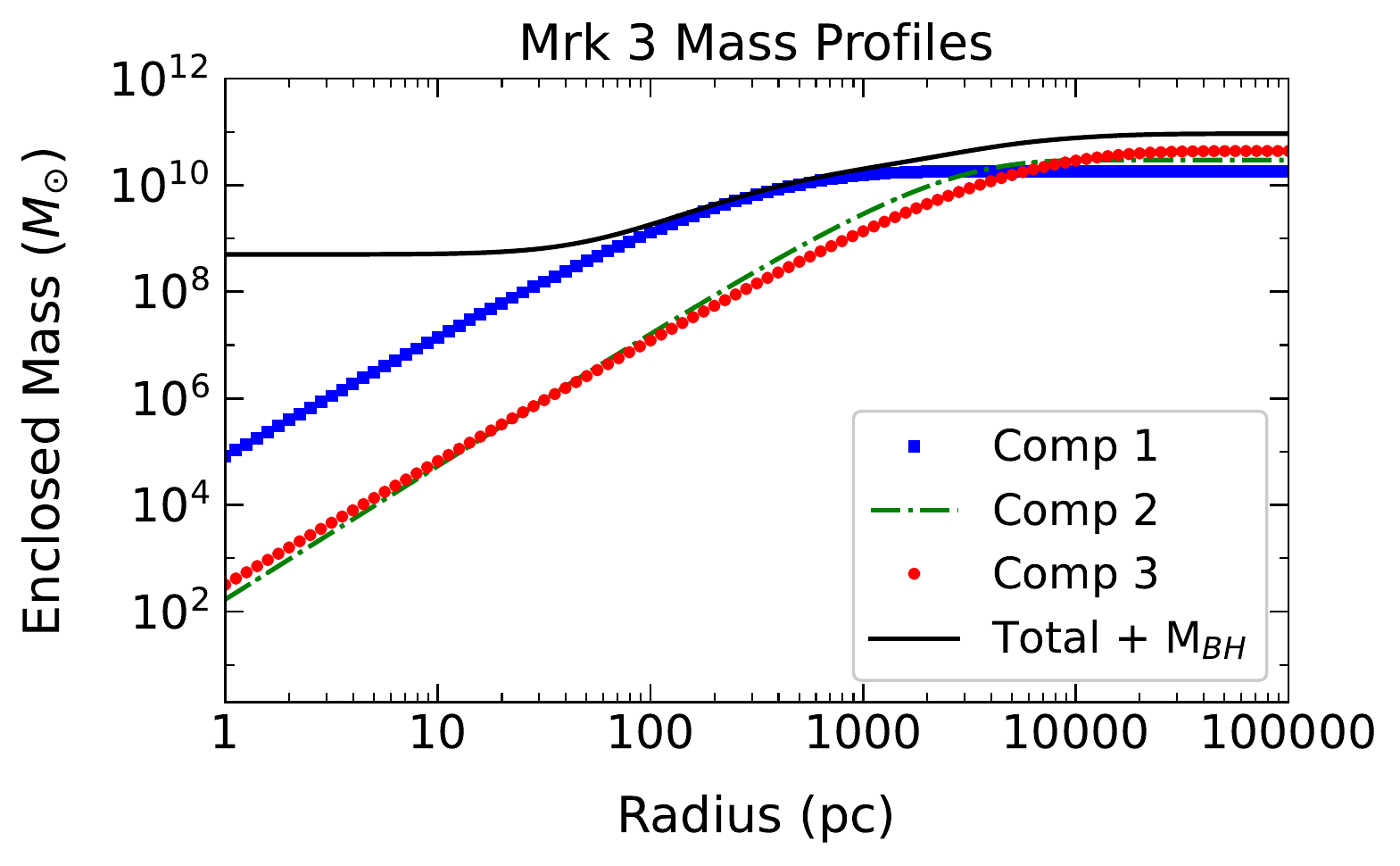}\hspace{2ex}
\includegraphics[width=0.465\textwidth]{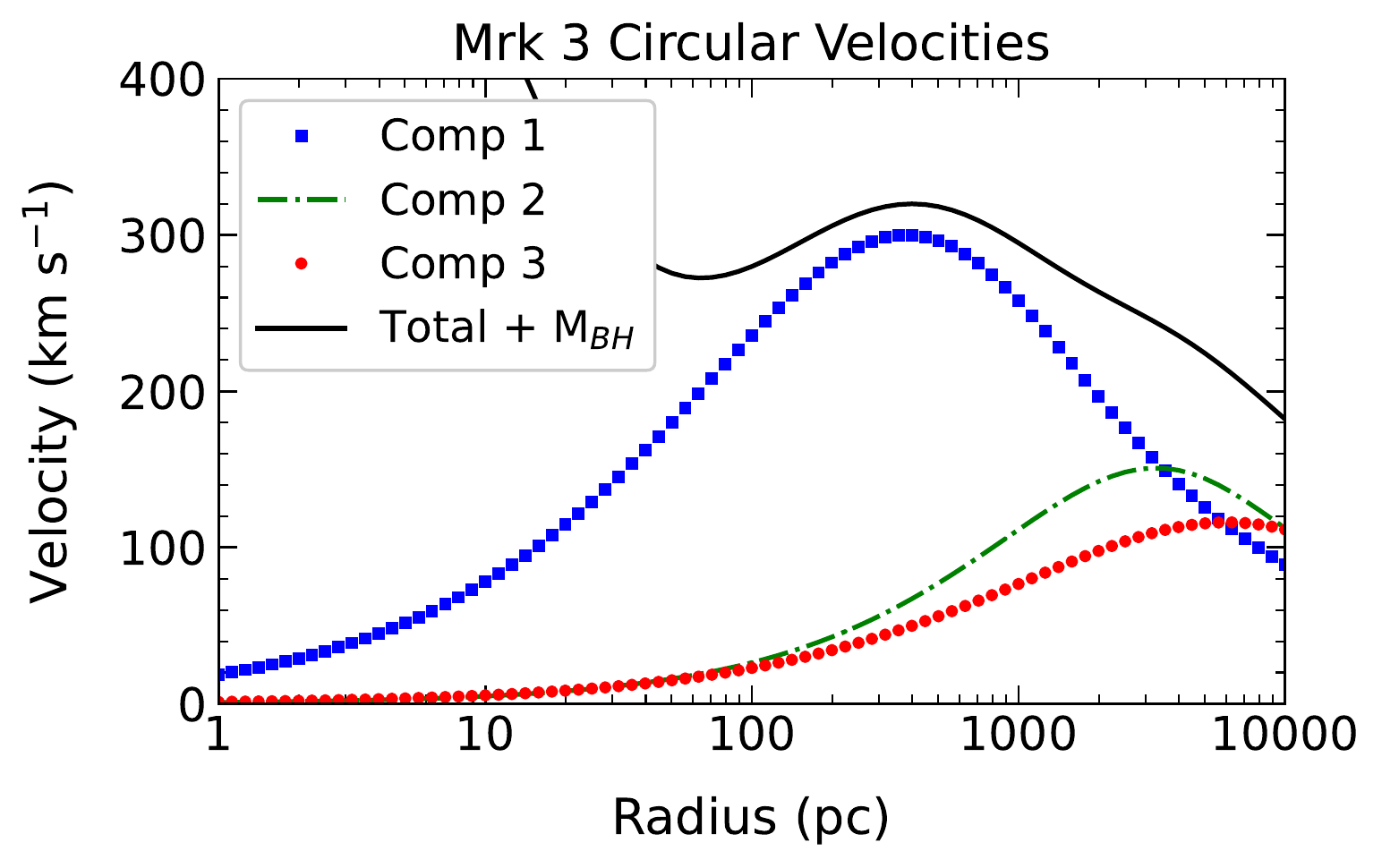}
\includegraphics[width=0.475\textwidth]{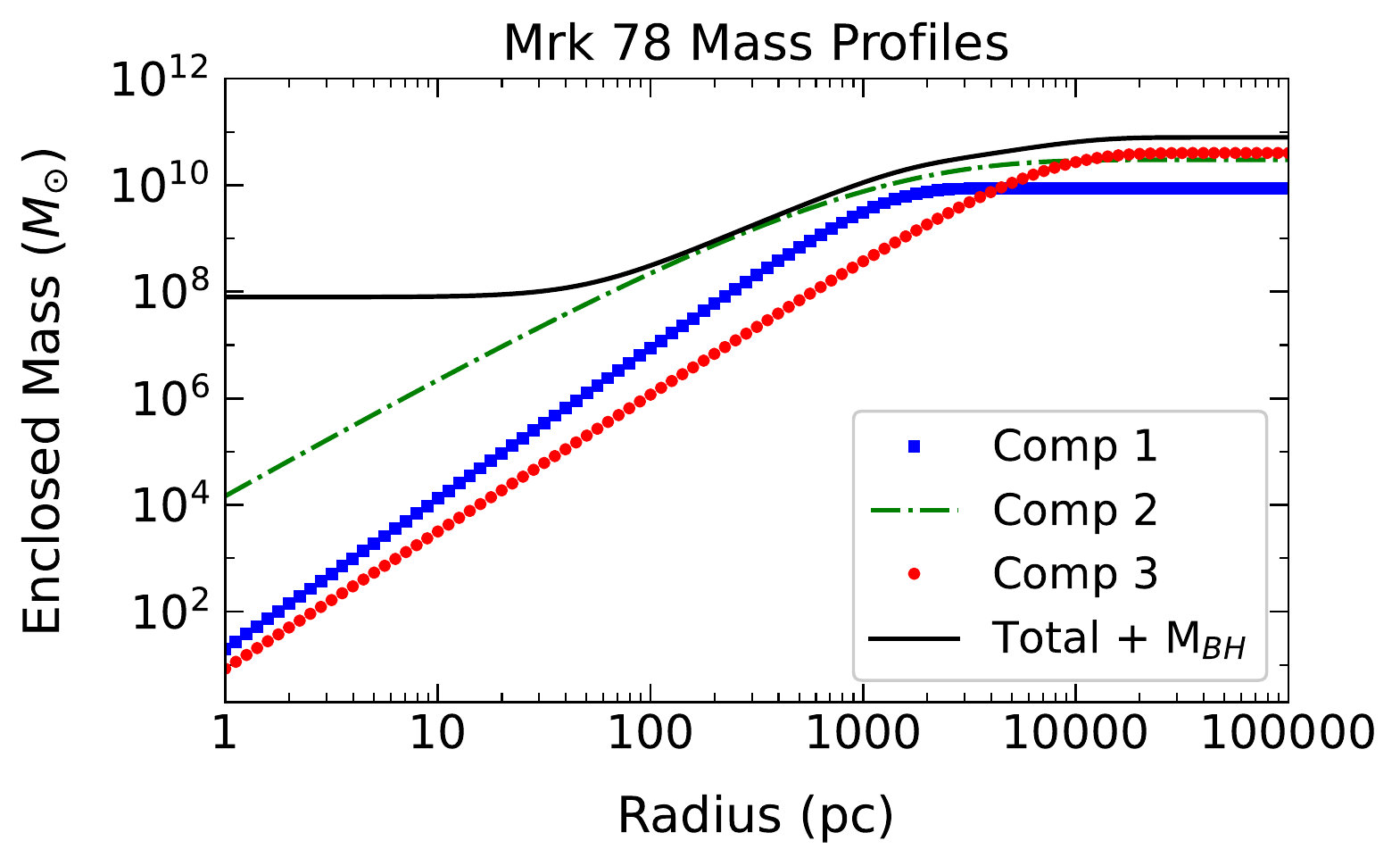}\hspace{2ex}
\includegraphics[width=0.465\textwidth]{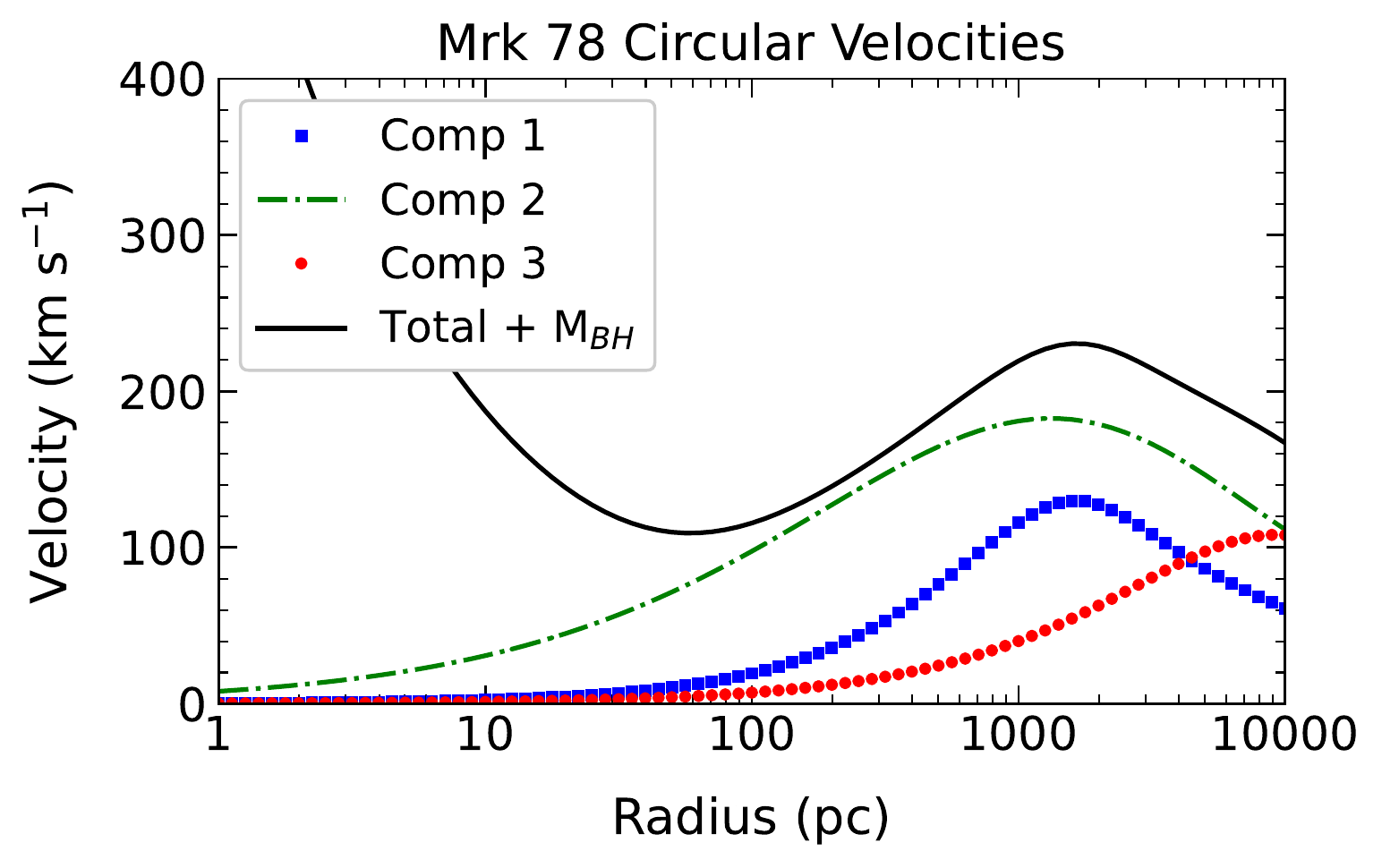}
\includegraphics[width=0.475\textwidth]{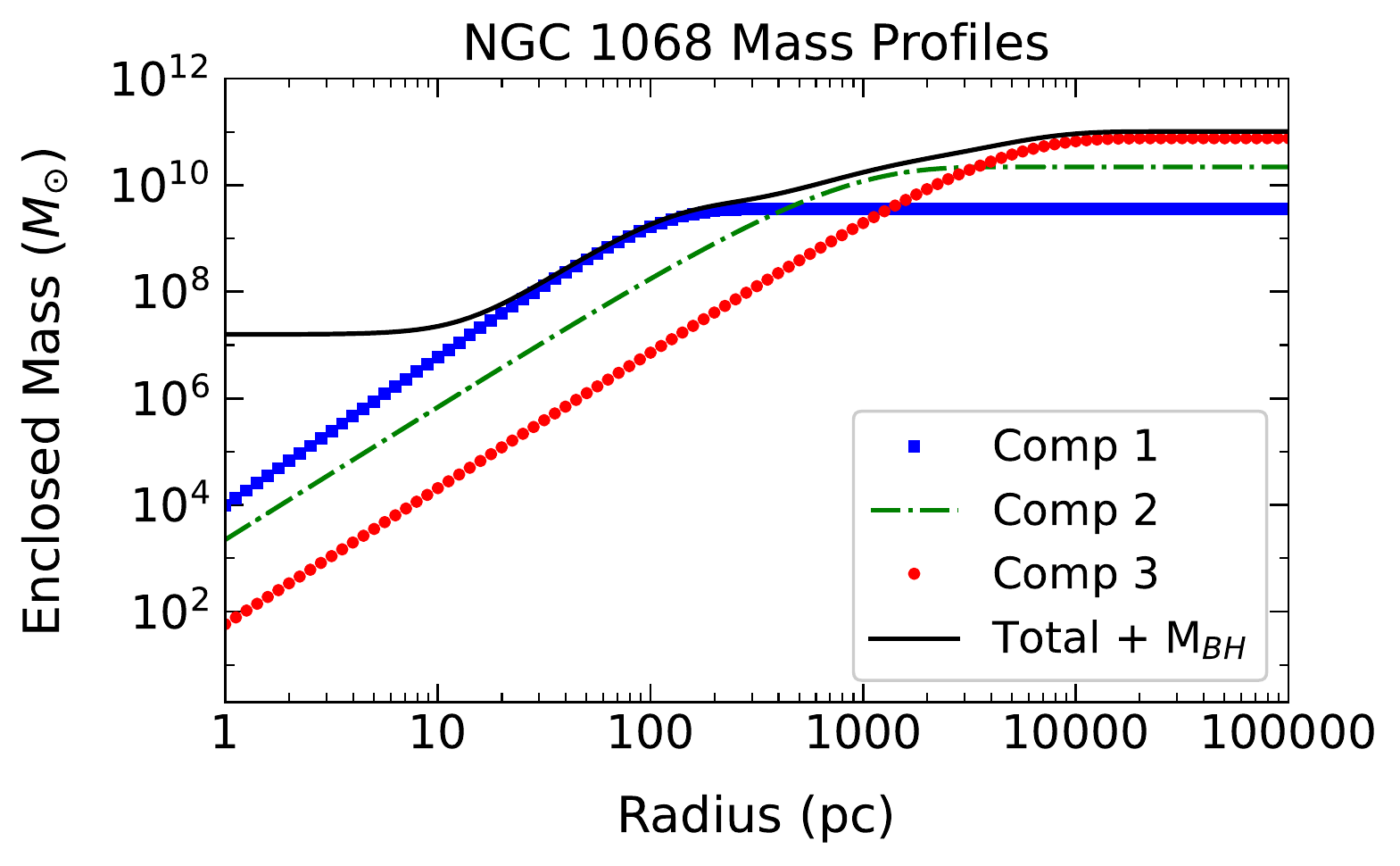}\hspace{2ex}
\includegraphics[width=0.465\textwidth]{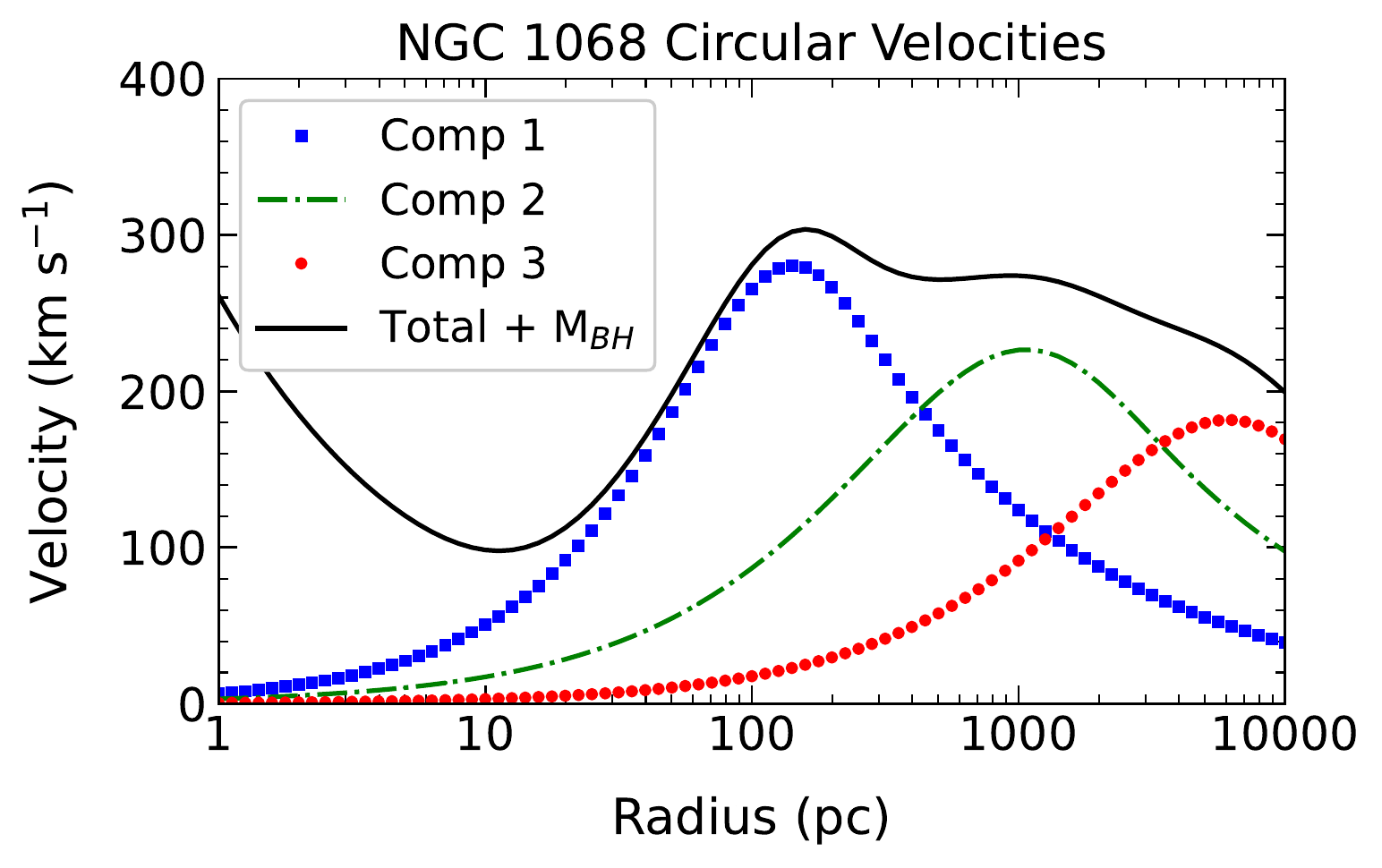}
\includegraphics[width=0.475\textwidth]{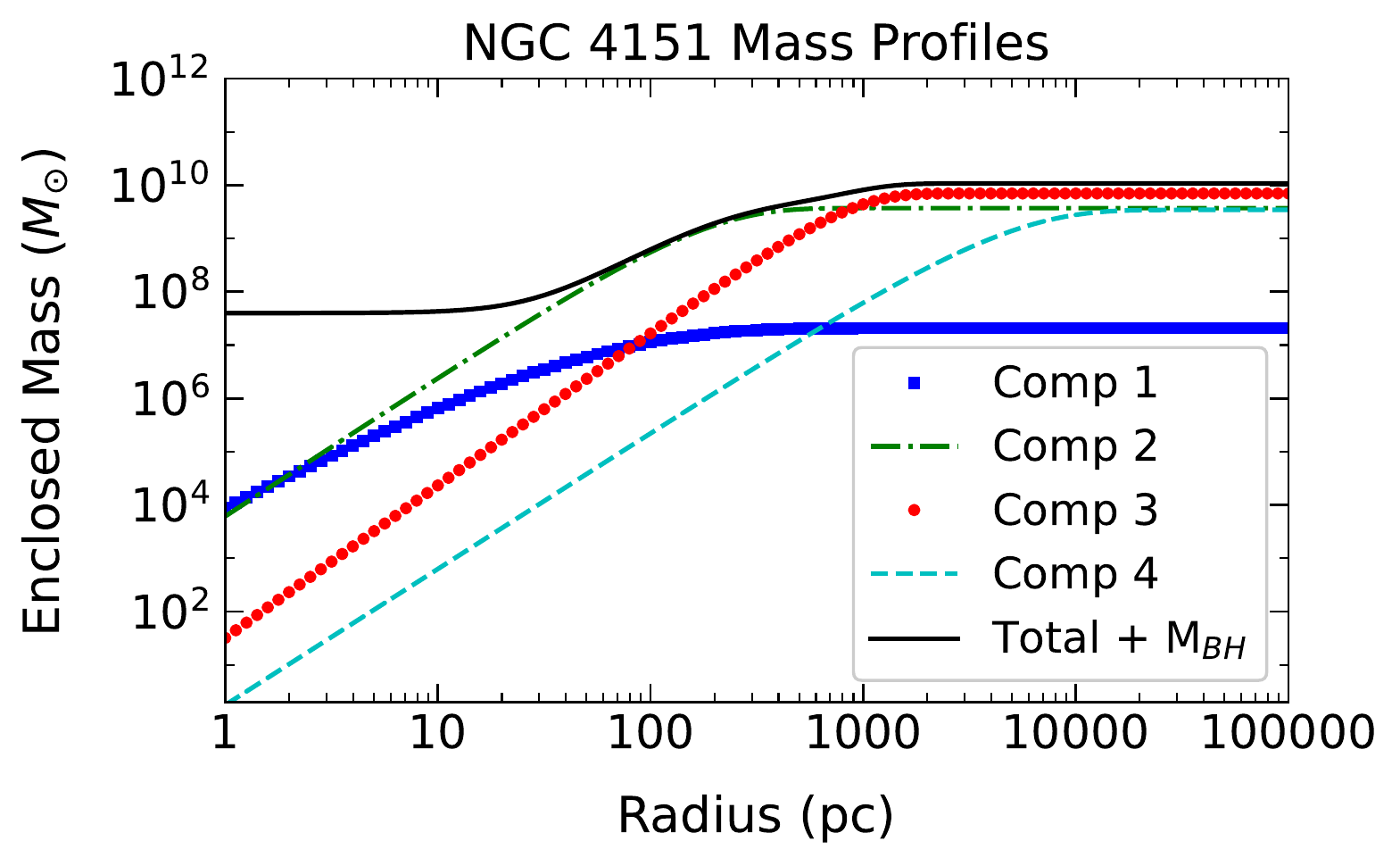}\hspace{2ex}
\includegraphics[width=0.465\textwidth]{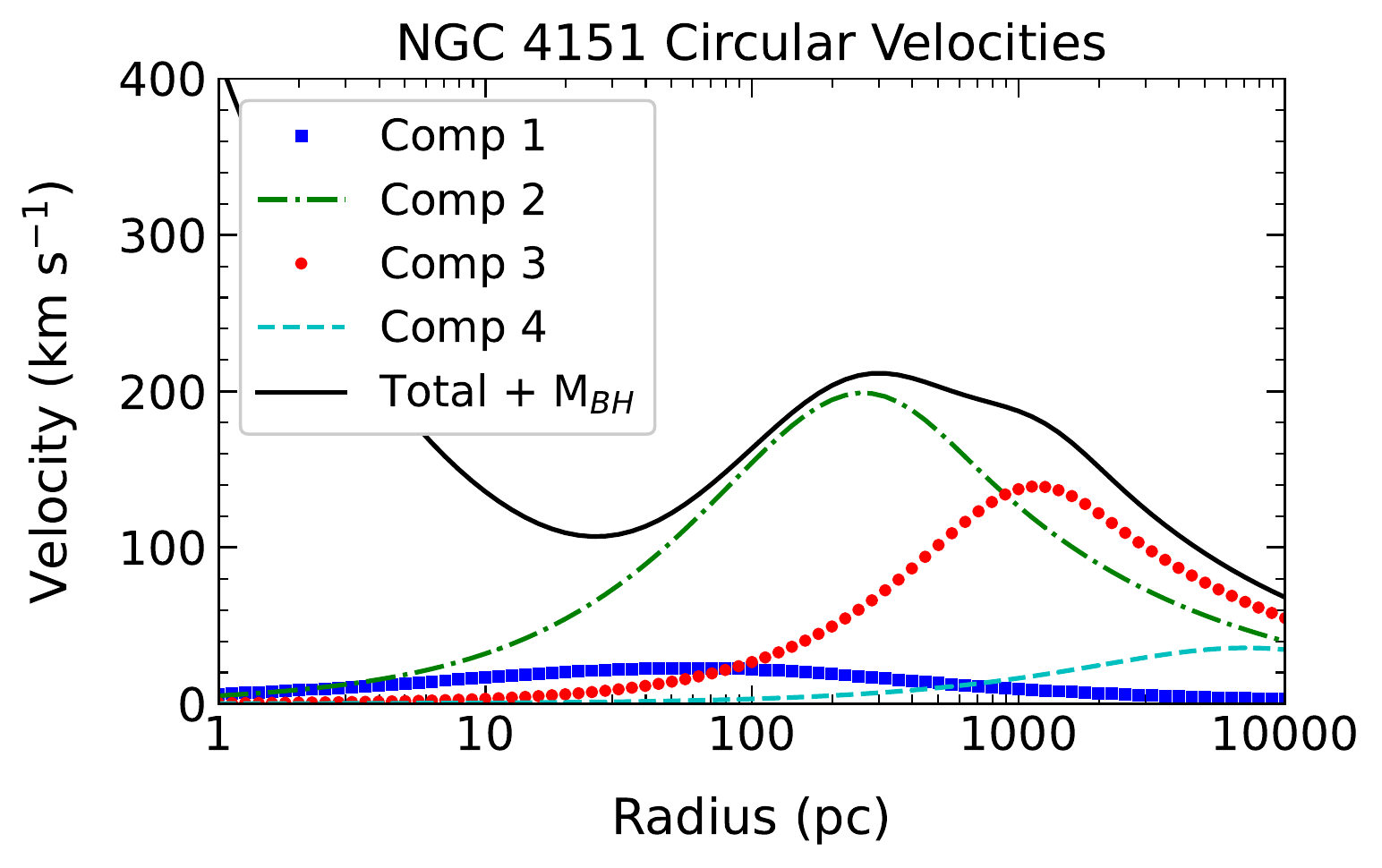}
\end{center}
\vspace{-3ex}
\caption{The left panel shows the radial mass profiles of individual surface brightness components for each galaxy, calculated using \textsc{GALFIT} model parameters given in Table \ref{tab:galfit} and Equations \ref{eq:density1} - \ref{eq:mass}. The smallest-to-largest radii components are plotted in blue-green-red curves. The interpretation of each component is discussed in \S \ref{subsec:galfit}. The integrated masses (including the masses of SMBH) are plotted in solid black lines. The right panel shows the corresponding Keplerian rotation curves, calculated for individual components and for the integrated mass using Equation \ref{eq:rot}.}
\label{fig:masses}
\end{figure*}

Assuming spherical symmetry, we calculated the radial mass profiles of individual components using Equation~A2 in \cite{Terzi2005}:

\begin{equation}
M(r) = 4\pi \rho_{0}r_{e}^2 n\kappa^{n(3-p)} \Gamma(n(3-p),Z).
\label{eq:mass}
\end{equation}

Here, $\Gamma(n(3-p),Z)$ is an incomplete gamma function that can be solved analytically. The radial mass distribution for individual Sérsic components, as well as the total enclosed mass profiles of the host galaxies (including the contribution from the SMBH) are shown as functions of radius in the left panel of Figure~\ref{fig:masses}. We note that the gravitational influence of the SMBHs (with masses given in Table~\ref{tab:sample}) dominate only the inner tens of pc in these galaxies and in general, the inner kpc is dominated by roughly spherical components. For Mrk~3, the majority of the enclosed mass is contributed by a pseudo-bulge (1st component) in the inner $\sim$1~kpc, and the outer bulge (3rd component) at larger radii. The total enclosed mass in Mrk~78 is dominated by the large bulge (2nd component) over the full extent of the galaxy. The faint exponential disk (3rd component) contributes to the mass beyond 10~kpc, whereas the innermost component (bar/lense-like structure) contributes little to the total mass profile. NGC~1068 contains a nuclear stellar cluster \citep{Thatte1997,Jaffe2004,Rouan2019}, which is responsible for the slight bump seen in the total mass profile (or the 1st component) at $\sim$150 pc. 
Beyond this radius, the mass of the bulge (2nd component) builds up to $\sim$1~kpc before being overtaken by the galaxy disk at larger distances. Similarly, in NGC~4151, the inner bulge (2nd component) carries the majority of the mass up to a few hundred pc, immediately followed by the 3rd component - defined as a `Barlens' in \cite{Bentz2018} or as an outer bulge in \cite{Bentz2009}. The exponential disk (4th component) adds to the total mass at $>$100~kpc, but there is not a significant contribution to the total mass from the innermost bar component.

In general, all four of these galaxies show an enclosed stellar mass of 10$^{10} - 10^{11}$ $M_{\odot}$ within 100~kpc, which is slightly higher than in NGC 4051 \citepalias{Meena2021}. Our enclosed mass profiles for NGC~1068 are also sightly higher than the ones presented in \cite{Das2006}, who analytically calculated the radial mass distribution using the total bulge mass provided in \cite{Marconi2003}. 
We determined the total bulge mass for NGC~1068 to be $\sim$2.2$\times$ 10$^{10}$ $M_{\odot}$ within 100 kpc for an effective radius of $r_{e}$ = 9\farcs8 (685 pc). While the overall bulge mass is almost half of the value ($M_{bulge}$ = 5$\times$10$^{10}$ $M_{\odot}$) provided in \cite{Marconi2003} based on surface brightness deconvolution of Two Micron All Sky Survey (2MASS) atlas images, their calculated effective bulge radius is $\sim$4 times larger ($r_{e}$ = 3.1 kpc), possibly due to the lower angular resolution of the images. Our values are in excellent agreement with the recent results by \cite{Davis2019}, who calculated a bulge mass $M_{bulge}$ $\approx$ 2$\times$10$^{10}$ $M_{\odot}$ with an effective bulge radius $r_{e}$ = 10\farcs5 ($\sim$814 pc). For NGC~4151, we measure a total stellar mass of $\sim$2.5$\times$ 10$^{10}$ $M_{\odot}$, which is close to the total stellar mass ($\sim$2.2$\times$ 10$^{10}$ $M_{\odot}$) provided in \cite{Robinson2021}. We do not have prior information for the stellar mass distributions in Mrk~3 and Mrk~78.

The right panel in Figure \ref{fig:masses} shows the circular velocities determined for the individual surface brightness components and the total enclosed mass distribution of each galaxy using

\begin{equation}
v_{rot}(r) = \sqrt{\mathrm{G} \frac{M(r)}{r}}
\label{eq:rot}
\end{equation}
\noindent

We compare these circular velocities with the rotation curves discussed in \S \ref{subsec:rotation} to further verify our mass measurements, particularly at the bulge radii. For Mrk~3, \cite{Gnilka2020} reports an average rotation of $\sim$160 \kms~at a projected distance of 5\arcsec~($\sim$1375 pc) using APO DIS Ca II triplet absorption lines along PA = 33\arcdeg, which corresponds to the major axis of the stellar disk. Correcting for the galaxy inclination (33\arcdeg), the maximum rotation at this distance is $\sim$290 \kms, which is close to the value determined using the total enclosed mass in this work, within a maximum uncertainty of $\pm$20 \kms(see Figure~\ref{fig:masses}). They also present the large-scale rotation of the ionized gas using APO DIS \othree line measurements $\sim$180 \kms~at $\sim$10\arcsec~(2745 pc) for the counter-rotating disk, with major axis PA of 129\arcdeg~and an inclination of 64\arcdeg. This corresponds to a true rotational velocity of $\sim$200 \kms, which closely agrees with our circular velocity at that radius. 
There is no prior rotation curve available for Mrk~78, although we obtained a large-scale rotational velocity map using APO DIS \othree kinematics outside $\pm$4\arcsec~of the nucleus, as discussed in \S \ref{subsec:rotation} (see Figure~\ref{fig:velmaps}). The de-projected (corrected for inclination) rotational velocity curve, as modeled using DiskFit based on these kinematics, is shown in Figure~\ref{fig:rotation}. At $\sim$5000 pc, the rotational velocities are $\sim$140 \kms, whereas the circular velocity calculated based on the enclosed mass at this distance is close to $\sim$190 \kms. This indicates a slight overestimation in our mass models, possibly due to higher $M/L$ calculation of the galaxy bulge.
The circular velocity for NGC~1068 at the bulge radius is $\sim$290 \kms, which is again $\sim$50 \kms~higher than the observed rotational velocities in Figure~\ref{fig:rotation}. On the other hand, for NGC~4151, our circular velocities based on enclosed mass are in close agreement with the rotation curves ($\sim$190 \kms) presented in Figure~\ref{fig:rotation}, giving confidence to our mass measurements for NGC~4151.

It must be noted that we do not take into account dark matter in our mass profiles, only stellar masses. Consequently, we see a decline in the rotational/circular velocities at radii $>$ 1 kpc in Figure~\ref{eq:mass}, as opposed to the constant or slightly increasing velocities observed in various kinematic measurements, such as those shown in Figure~\ref{fig:rotation}. Nonetheless, these distances are larger than the NLR scales of their galaxies, and therefore the host galaxy masses at these radii do not affect our radiative driving models discussed in the next section.

\subsection{Radiation - Gravity Coupling}
\label{subsec:radiative_driving}

\begin{figure*}[ht!]
\begin{center}
\subfigure{
\includegraphics[width=0.485\textwidth]{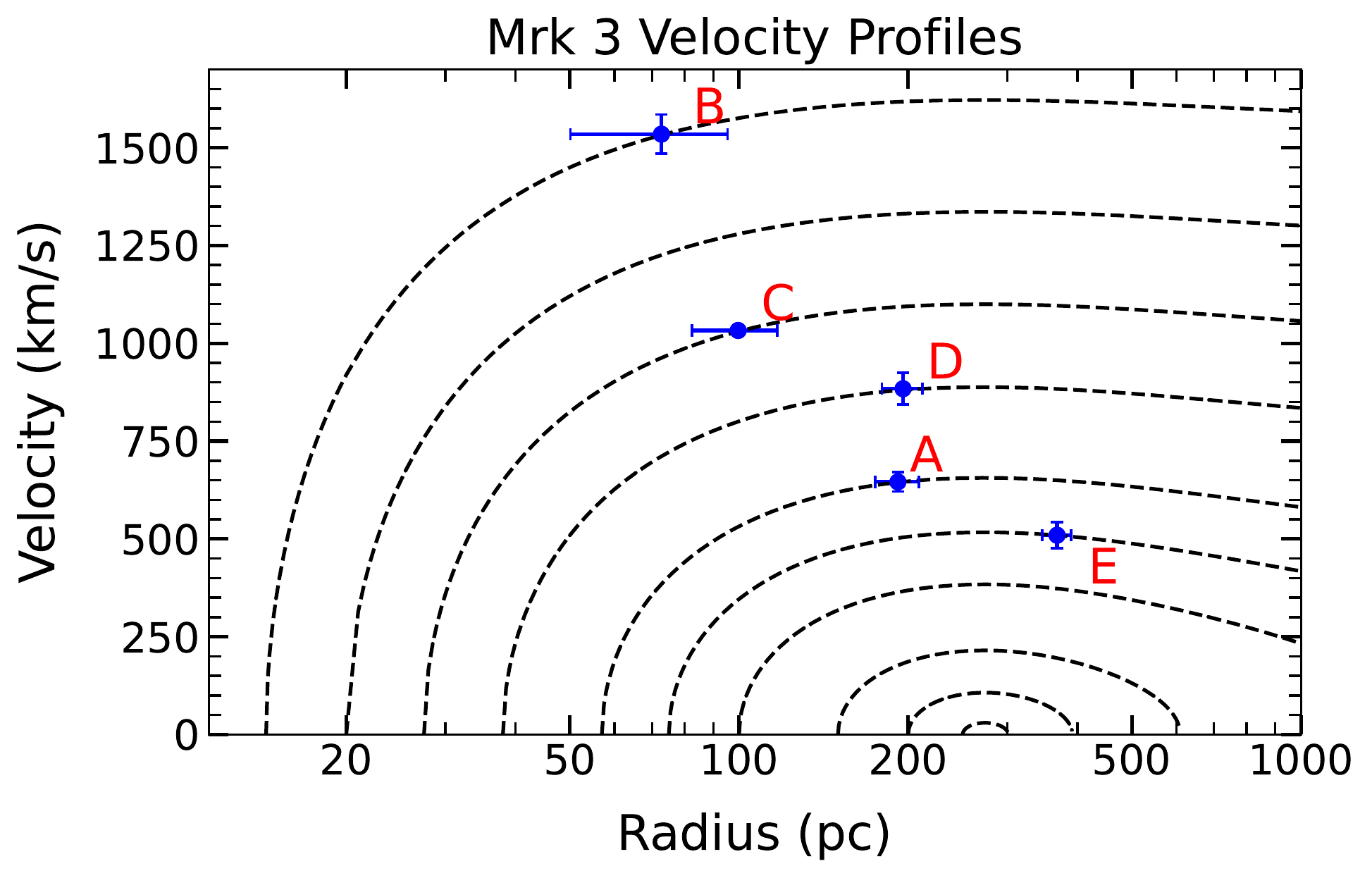}}\hspace{2ex}
\subfigure{
\includegraphics[width=0.485\textwidth]{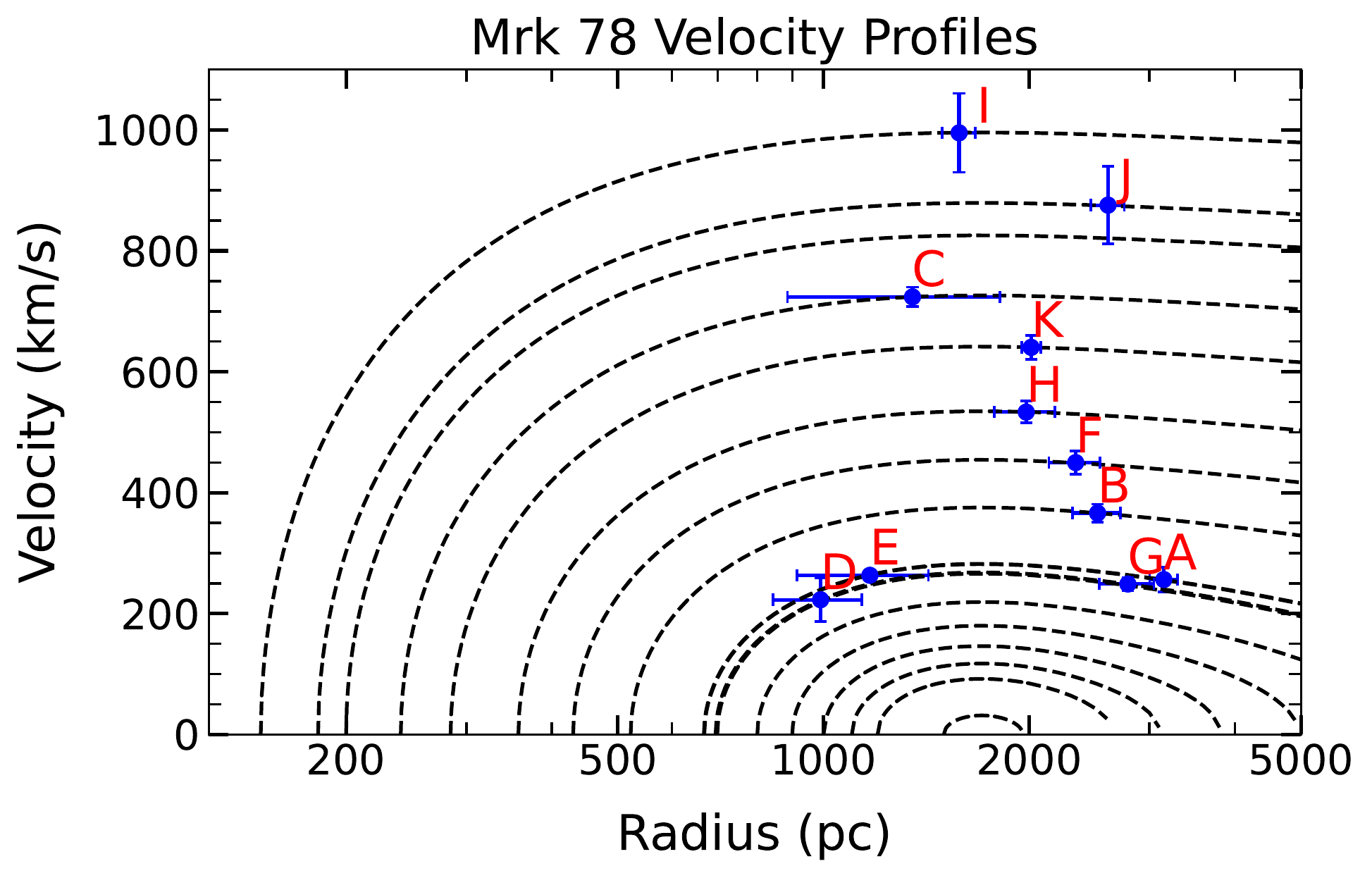}}
\subfigure{
\includegraphics[width=0.485\textwidth]{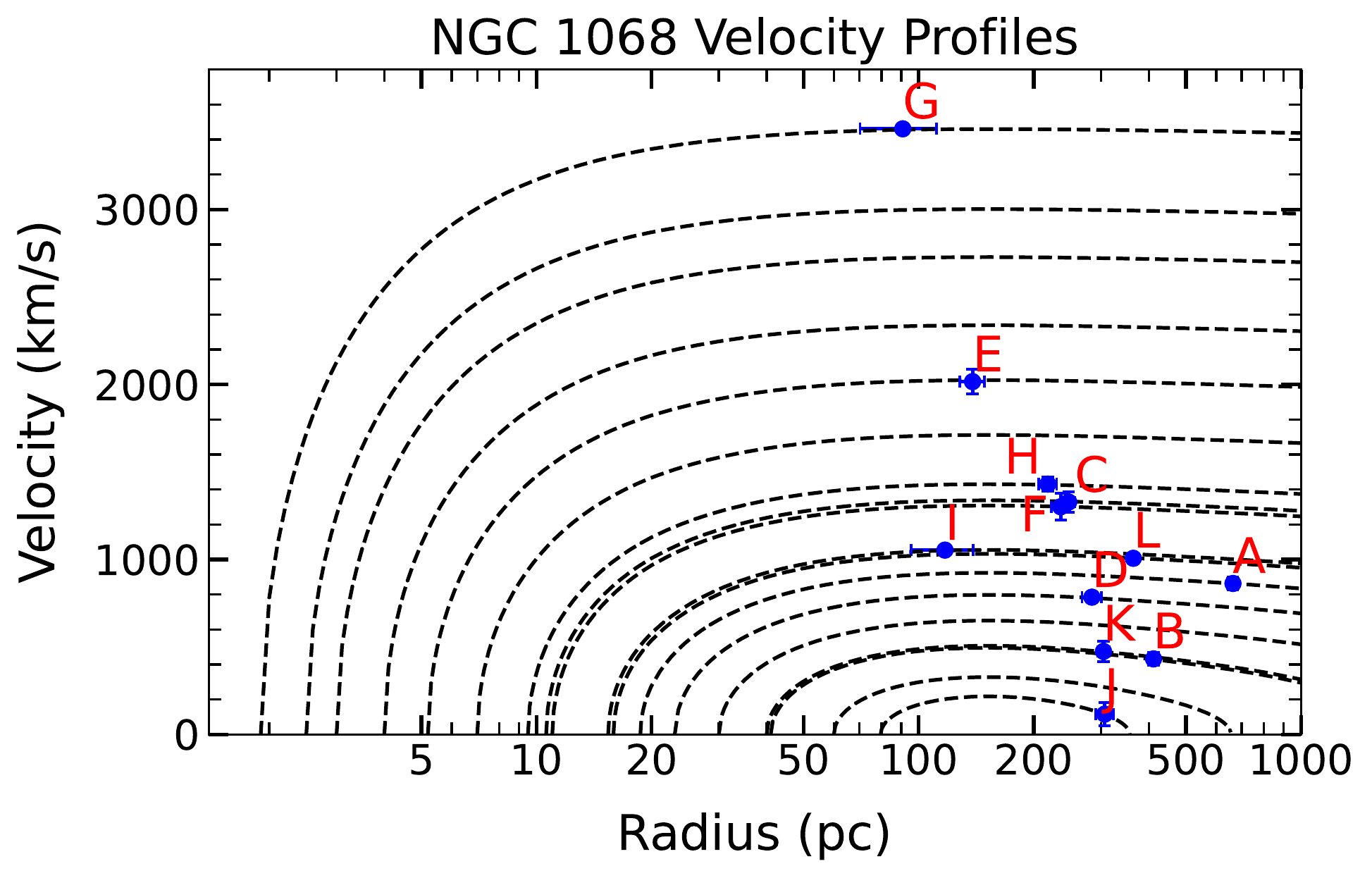}}\hspace{2ex}
\subfigure{
\includegraphics[width=0.485\textwidth]{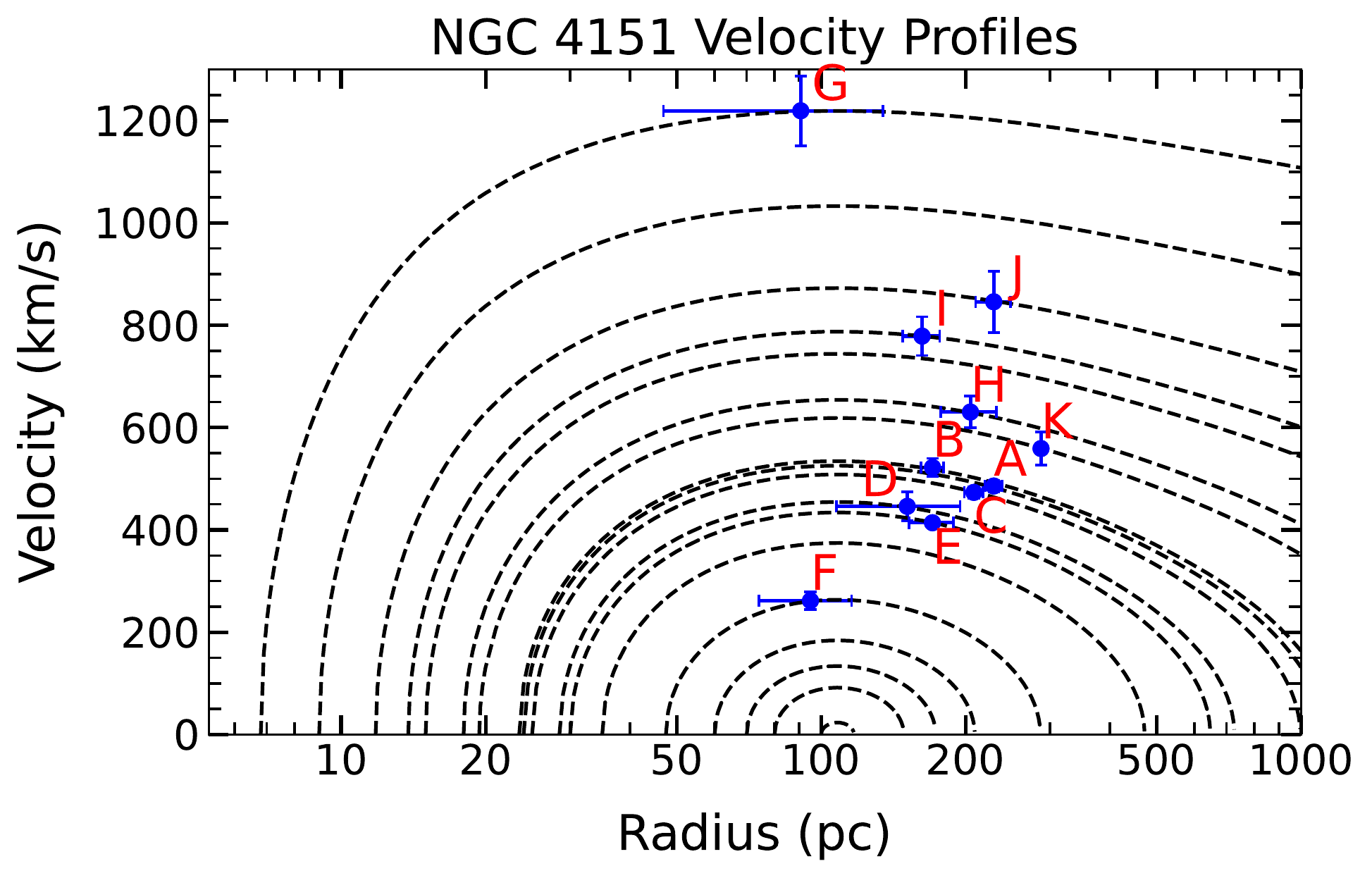}}
\end{center}
\vspace{-1em}
\caption{The de-projected velocities of the outflowing gas knots (blue points), with the radial velocity profiles predicted from our radiative acceleration plus gravitational deceleration model (Equation~\ref{eq:vel_pro}). The alphabetical letters represent the true velocities and distances of each knot from the central AGN, with each knot identified as in Figure~\ref{fig:O3}, Figure~\ref{fig:kinematics} and Appendix~\ref{app_kinematics}. The uncertainties in the distances and velocities are calculated from the data points assigned to the given knots.}
\label{fig:rad_driv_model}
\end{figure*}

Assuming that the NLR gas is radiatively accelerated by the AGN \citep{Proga2000,Chelouche2001,Crenshaw2003} and then slows down under the gravitational influence of the galaxy, we can derive the radial velocity profiles of the outflowing gas clouds using a numerical expression given by \cite{Das2007}:

\begin{equation}
v(r) = \sqrt{ \int_{r_{1}}^{r} \Big[ 4885 \frac{L_{44} \mathcal{M}}{r^{2}} - 8.6 \times 10^{-3}\frac{M(r)}{r^{2}} \Big] dr },
\label{eq:vel_pro}
\end{equation}

where the outflow velocity $v(r)$ of the gas is in \kms~at a distance $r$ (in pc), $L_{44}$ is the bolometric luminosity of the AGN in units of $10^{44}$ erg s$^{-1}$, and $r_{1}$ is the launch distance from the central SMBH where the outflow originates. $M(r)$ is again the total stellar mass in solar masses ($M_{\odot}$) as obtained from Equation~\ref{eq:mass}. The force multiplier $\mathcal{M}$ depends on the ionization parameter ($U$) for a given spectral energy distribution (SED) and relates the effectiveness of the bound-bound, bound-free and continuum opacities in enhancing the radiative driving efficiency beyond that of pure Thomson scattering ($\mathcal{M}$ = 1). A detailed derivation of Equation~\ref{eq:vel_pro} is provided in \cite{Das2007} and \citetalias{Meena2021}.
Given the velocity and distance of an ionized gas knot (or cloud) from an AGN's SMBH, corrected for projection effects, the only unknown in the above equation is the launch distance $r_{1}$, which can be solved for numerically.

While the original model by \cite{Das2007} also includes deceleration of the gas due to drag, it would significantly reduce the outflow velocities and subsequently the launch distances, and there is no independent evidence for this effect. Therefore, many of the subsequent studies \citep{Fischer2017, Fischer2019, Garcia2021, Meena2021} ignored this term and only considered the forces due to AGN radiation pressure and gravity. We adopt the same model for this work.

Similar to \citetalias{Meena2021}, we adopt a force multiplier value of $\mathcal{M}$ = 500 (Kraemer et. al, private communication), which accounts for the decrease in ionizing radiation with the increase in column density and optical depth as radiation propagates deeper into the gas clouds (see also \citealp{Trindade2021}). We used the $L_{bol}$ provided in Table~\ref{tab:sample} in our model. The implication of the adopted force multipliers and luminosities will also be discussed in \S\ref{subsec:assum}.

Next, we determined the true distances and velocities of the observed outflowing gas using the de-projection factors obtained from the biconical models of outflows, developed by \cite{Ruiz2001, Crenshaw2010} for Mrk~3, \cite{Fischer2011} for Mrk~78, \cite{Das2006} for NGC~1068, and \cite{Das2005} for NGC~4151. The velocity and distance de-projection factors for Mrk~3, Mrk~78 and NGC~1068 can be retrieved from Table~3 of \cite{Revalski2021}. For NGC~4151, the distance de-projection factors are 1.88 $\&$ 1.07 and velocity de-projection factors are 1.07 $\&$ 2.28 for the near (r $<$1\farcs5) and far (r $>$1\farcs5) sides of the bicone, respectively.  

Finally, we determined the radial velocity profiles of the outflows, as shown in Figure~\ref{fig:rad_driv_model}, for each AGN, using Equation~\ref{eq:vel_pro}. The individual curves on each plot represent the trajectory of an outflowing cloud for the given AGN luminosity and enclosed mass of the host galaxy at each radius from the center (zero~pc). Some of the selected outflow knots as identified in the \othree emissions (Figure~\ref{fig:O3}) and kinematics (Figure~\ref{fig:kinematics} and Appendix~\ref{app_kinematics}) are labeled with alphabetical letters. Their positions on the velocity curves represent their true velocity and distances. The zero-velocity starting positions of the curves represent the launch distances of the outflows.

As seen from the velocity profiles, the outflows are launched at multiple radii, which is consistent with in-situ acceleration of the NLR gas in these targets, as discussed in \cite{Revalski2021}. Our model indicates that the outflowing clouds launched closer to the nucleus have higher velocities, and their initial velocities decrease with increasing launch radius due to the gravitational potential of the host galaxy. At a critical turnover distance, the gravitational pull of the enclosed mass of the galaxy exceeds the force of the radiation pressure, and clouds can no longer be launched radially outward at larger distances. We define this distance as the maximum theoretical launch radius (max. $R_{Launch}$), or equivalently a ``model turnover" distance for clouds launched from inside max. $R_{Launch}$. For a given AGN, this point of initial deceleration is independent of the launch distance, which can be seen in the velocity curves in Figure~\ref{fig:rad_driv_model}. For the highest velocity knots, this decrease in velocity is subtle, but a significant deceleration can be seen in the lower velocity curves at the model turnover distance. The turnover distance can be calculated numerically from Equation~\ref{eq:vel_pro} by determining the point where gravitational deceleration from the host mass overtakes the AGN radiation pressure. We assume that no outflows originate beyond this location. 

We calculated the launch distances of the outflowing knots by numerically solving Equation~\ref{eq:vel_pro} for $r_{1}$ based on their measured velocities and positions. Using the distance and velocity at each point on the curve, we also calculated the total travel time of each knot from their launch position to their current distance. For each galaxy, the assigned name of the knots, their observed and de-projected velocities and distances, model velocities, launch distances from the center, and integrated travel times, are given in Table~\ref{tab:rad_drive}.

\setlength{\tabcolsep}{0.11in}
\renewcommand{\arraystretch}{1.2}
\tabletypesize{\footnotesize}
\begin{deluxetable*}{cccccchcc}[ht!]
\tablecaption{Parameters for the Radiative Driving Models\label{tab:rad_drive}}
\tablehead{
\colhead{Galaxy} & \colhead{Knot} & \colhead{Observed} & \colhead{Observed} & \colhead{True} & \colhead{True} & \nocolhead{Model} & \colhead{Launch} & \colhead{Travel} \vspace{-2ex}\\
\colhead{Name} & \colhead{Name} & \colhead{Distance} & \colhead{Velocity} & \colhead{Distance} & \colhead{Velocity} & \nocolhead{Velocity} & \colhead{Distance} & \colhead{Time} \vspace{-2ex}\\
 \colhead{} &  \colhead{} & \colhead{(\arcsec)} & \colhead{(km s$^{-1}$)}  & \colhead{(pc)}  & \colhead{(km s$^{-1}$)} & \nocolhead{(km s$^{-1}$)} & \colhead{(pc)} & \colhead{(10$^6$ years)}
}
\startdata
Mrk 3 &    A &  0.64 &  -272 & 191 &    645 &     644 &        57 &  0.27 \\
Mrk 3 &    B &  0.24 &  -647 &  72 &   1534 &    1531 &        14 &  0.05 \\
Mrk 3 &    C & -0.33 &  -435 &  99 &   1032 &    1030 &        27 &  0.09 \\
Mrk 3 &    D & -0.69 &   228 & 195 &    883 &     880 &        38 &  0.22 \\
Mrk 3 &    E & -1.22 &  -214 & 368 &    509 &     509 &        75 &  0.65 \\
\hline
Mrk 78 &    A & -1.78 &   233 & 3149 &    256 &     256 &       669 & 10.16 \\
Mrk 78 &    B & -1.42 &   333 & 2518 &    366 &     366 &       522 &  6.29 \\
Mrk78 &    C & -0.76 &   658 & 1349 &    723 &     723 &       240 &  1.84 \\
Mrk 78 &    D & -0.56 &   202 &  990 &    222 &     222 &       695 &  2.14 \\
Mrk 78 &    E &  0.66 &  -239 & 1169 &    263 &     263 &       668 &  2.83 \\
Mrk 78 &    F &  1.32 &  -408 & 2339 &    449 &     449 &       430 &  4.94 \\
Mrk 78 &    G &  1.58 &  -226 & 2789 &    248 &     248 &       699 &  9.05 \\
Mrk 78 &    H &  1.12 &  -484 & 1980 &    533 &     533 &       357 &  3.54 \\
Mrk 78 &    I &  2.03 &  -257 & 1579 &    995 &     995 &       150 &  1.60 \\
Mrk 78 &    J &  1.47 &  -796 & 2609 &    875 &     874 &       182 &  3.05 \\
Mrk 78 &    K &  2.59 &  -165 & 2013 &    640 &     640 &       284 &  3.19 \\
\hline
NGC 1068 &    A &  7.32 &  -553 & 663 &    863 &     863 &        18 &  0.72 \\
NGC 1068 &    B &  5.08 &   216 & 411 &    432 &     429 &        41 &  0.82 \\
NGC 1068 &    C &  3.05 &   664 & 246 &   1328 &    1331 &        10 &  0.18 \\
NGC 1068 &    D &  3.50 &   392 & 283 &    784 &     780 &        23 &  0.35 \\
NGC 1068 &    E &  1.52 & -1292 & 138 &   2016 &    2025 &         5 &  0.07 \\
NGC 1068 &    F &  2.59 &  -834 & 235 &   1301 &    1302 &        11 &  0.18 \\
NGC 1068 &    G &  1.00 & -2218 &  90 &   3461 &    3455 &         1 &  0.03 \\
NGC 1068 &    H &  2.64 &   659 & 217 &   1430 &    1426 &         9 &  0.15 \\
NGC 1068 &    I & -1.42 &   485 & 117 &   1053 &    1051 &        15 &  0.11 \\
NGC 1068 &    J & -3.30 &   -67 & 306 &    116 &     115 &        79 &  1.31 \\
NGC 1068 &    K & -3.35 &  -304 & 304 &    475 &     472 &        40 &  0.57 \\
NGC 1068 &    L & -4.01 &  -645 & 364 &   1006 &    1008 &        15 &  0.35 \\
\hline
NGC 4151 &    A &  2.79 &   213 & 229 &    485 &     482 &        24 &  0.42 \\
NGC 4151 &    B &  2.08 &   229 & 170 &    522 &     518 &        23 &  0.30 \\
NGC 4151 &    C &  2.54 &   207 & 208 &    473 &     474 &        25 &  0.39 \\
NGC 4151 &    D &  1.05 &   416 & 151 &    446 &     444 &        28 &  0.31 \\
NGC 4151 &    E &  1.18 &  -386 & 170 &    413 &     414 &        30 &  0.36 \\
NGC 4151 &    F &  0.66 &   244 &  95 &    261 &     260 &        47 &  0.25 \\
NGC 4151 &    G & -0.63 & -1139 &  90 &   1219 &    1217 &         6 &  0.08 \\
NGC 4151 &    H & -1.42 &  -589 & 204 &    630 &     629 &        18 &  0.31 \\
NGC 4151 &    I & -1.98 &  -341 & 162 &    778 &     779 &        13 &  0.20 \\
NGC 4151 &    J & -2.79 &  -371 & 229 &    845 &     847 &        11 &  0.26 \\
NGC 4151 &    K & -3.50 &  -245 & 287 &    559 &     559 &        19 &  0.46 \\ \enddata
\tablecomments{The columns list (1) the galaxy name, (2) the assigned name for the outflow knot as identified in \othree images and gas kinematics (3) the observed distances (in arcsec), (4) observed velocities, (5) true/de-projected distances (in pc), (6) true/de-projected velocities using the parameters obtained for outflow bicone geometries via kinematics models, (7) velocities using the radiative-gravity model, (8) the calculated launch distance of these knots, and (9) the total time traveled for each knot from its launch distance to its true distance with the modeled velocity.}
\end{deluxetable*}

\subsection{Outflow Extents \& Launching Sites} \label{subsec:extents_launch}

The velocity profiles shown in Figure~\ref{fig:rad_drib_model} display the current distances and space velocities of bright NLR knots, along with their tracks over time, considering only the forces of AGN radiation pressure and the gravitational potential of the host galaxy and SMBH. The launch distances are indicated by tracing their motions back to zero velocity. The outflow knots labeled in alphabetical letters in Figure~\ref{fig:rad_driv_model} were chosen such that they are bright, easily distinguishable, and cover the maximum extents and ranges of velocities of the NLR outflows for each target using the \othree images (Figure~\ref{fig:O3}) and \othree kinematics (Figure~\ref{fig:kinematics}, and Appendix~\ref{app_kinematics}).
The observed velocity and position of each knot in Figure~\ref{fig:rad_driv_model} and Table~\ref{tab:rad_drive} represent an average of multiple points that comprise of comparable adjacent velocities measured in (Figure~\ref{fig:kinematics}, and Appendix~\ref{app_kinematics}). Similarly, the uncertainties (error bars in Figure~\ref{fig:rad_driv_model}) were calculated from the spread in data points in each defined knot. Additionally, the full range of launch radii accommodating each data points (Figure~\ref{fig:kinematics}, and Appendix~\ref{app_kinematics}) is shown in  \ref{fig:corr_lum}. The full range of launch radii for the five additional sources were obtained from radiative driving models provided in their respective literature, where  the maximum launch distance corresponds to an observed velocity.
Given the previously stated evidence for in-situ acceleration of NLR clouds from reservoirs of cold molecular gas, the launch distances indicate potential sites of star-formation gas that are being evacuated or blown away due to radiation pressure, revealing an important form of AGN feedback.
We discuss the interpretation of the observed kinematics and the results of the radiative driving models for the four individual targets below.

\paragraph{Mrk~3:} As seen in Figure~\ref{fig:rad_driv_model} (top left), we find outflows with de-projected velocities ranging from 1500 \kms~at 70~pc to 500 \kms~at $\sim$400~pc from the nucleus. 
There is no evidence for radially outflowing gas at larger radii, although the rotating gas on larger scales shows turbulence - as seen by the high velocity dispersion (FWHM) in Figure~\ref{fig:kinematics}. This turbulent/disturbed gas has been traced to $\sim$10\arcsec~(2.8~kpc) using ground-based spectroscopy \citep{Gnilka2020}. All of the observed outflowing knots are radiatively launched within $\sim$80~pc, and as close as within 10~pc of the nucleus. Radiative driving is possible at larger distances, and outflows can be produced up to $\sim$250 pc - beyond which the gas quickly stalls in the plane of the galaxy and cannot be driven radially outwards. 

\paragraph{Mrk~78:} The de-projected velocities in Mrk~78 show outflows spanning over $\sim$1 kpc to maximum distances of $\sim$2.5 kpc. There are likely outflows within the inner 1 kpc that are difficult to isolate due to the distance of Mrk~78 and the spatial resolution limit of HST. The true outflow velocities range from a maximum of $\sim$1000 \kms~for the gas launched closer to the nucleus, down to $\sim$200 \kms~for some of the most extended outflows. We also observe a few high velocity clouds that were launched relatively close to the nucleus and traveled long distances without decelerating appreciably (see knots I and J in the top right panel of Figure~\ref{fig:rad_driv_model}).
Almost all of the ionized outflows seem to originate outside $\sim$100 pc of the nucleus, which is likely due to our inability to resolve outflows in the innermost regions. 
The maximum launch radius affiliated with the bright emission line knots is $\sim$700~pc, although outflows can be produced by radiation pressure up to distances of $\sim$1.5~kpc from the AGN, which commensurate with the effective radius of the bulge (see \S\ref{subsec:galfit}).

\paragraph{NGC~1068:} As discussed in \S \ref{app_kinematics}, NGC~1068 presents the most diverse ionized gas kinematics in the sample, with a wide range of de-projected outflow velocities ranging from $\sim$200 \kms~to 3500~\kms. However, the majority of the observed knots have velocities $<$ 2000 \kms.
We find the majority of the outflows lie within $<$300~pc; however, a few knots are identified either at the edge or beyond the nominal bicone - as modeled in \cite{Das2007}. For example, knot A (at $\sim$500~pc) exhibits high outflow-like velocities and FWHM, but is located a little further than the expected extent of outflows.
While this could be a case of gas perturbation caused by the outflows running into the feeding lanes (in conjunction with the projection effects of the gas velocities in the line-of-sight), we derive the velocity profiles of knot A assuming that it is produced by radiative driving. 

All of the outflows are launched across the entire extent of the NLR (up to $R_{Launch}$ of $\sim$80 pc), and no outflows can be launched at distances $>$ $\sim$100~pc, according to our model.
The maximum launch radii and outflowing extent is only a fraction of the bulge of NGC~1068 ($r_{e}$ $\sim$ 500 pc), and much smaller than the galaxy disk, indicating that most of the star-forming gas in the circumnuclear regions might be unaffected by the AGN winds.

\paragraph{NGC~4151:} We find the maximum extent of the outflows is $\sim$300~pc from the nucleus, with true velocities ranging from $\sim$200--800~\kms~for the majority of the knots, and a few knots ranging up to $\sim$1200~\kms. Although the extents and velocities of the outflows in NGC~4151 are much smaller than in NGC~1068, the outflow launch distances are comparable. Based on our radiative driving model, the identified knots were launched from within $\sim$8-50~pc of the nucleus, with the radiative launching of outflows possible up to $\sim$100~pc. 

\subsection{Correlation with AGN luminosities} \label{subsec:corr_lum}

\begin{figure*}[ht!]
\begin{center}
\includegraphics[width=0.75\textwidth]{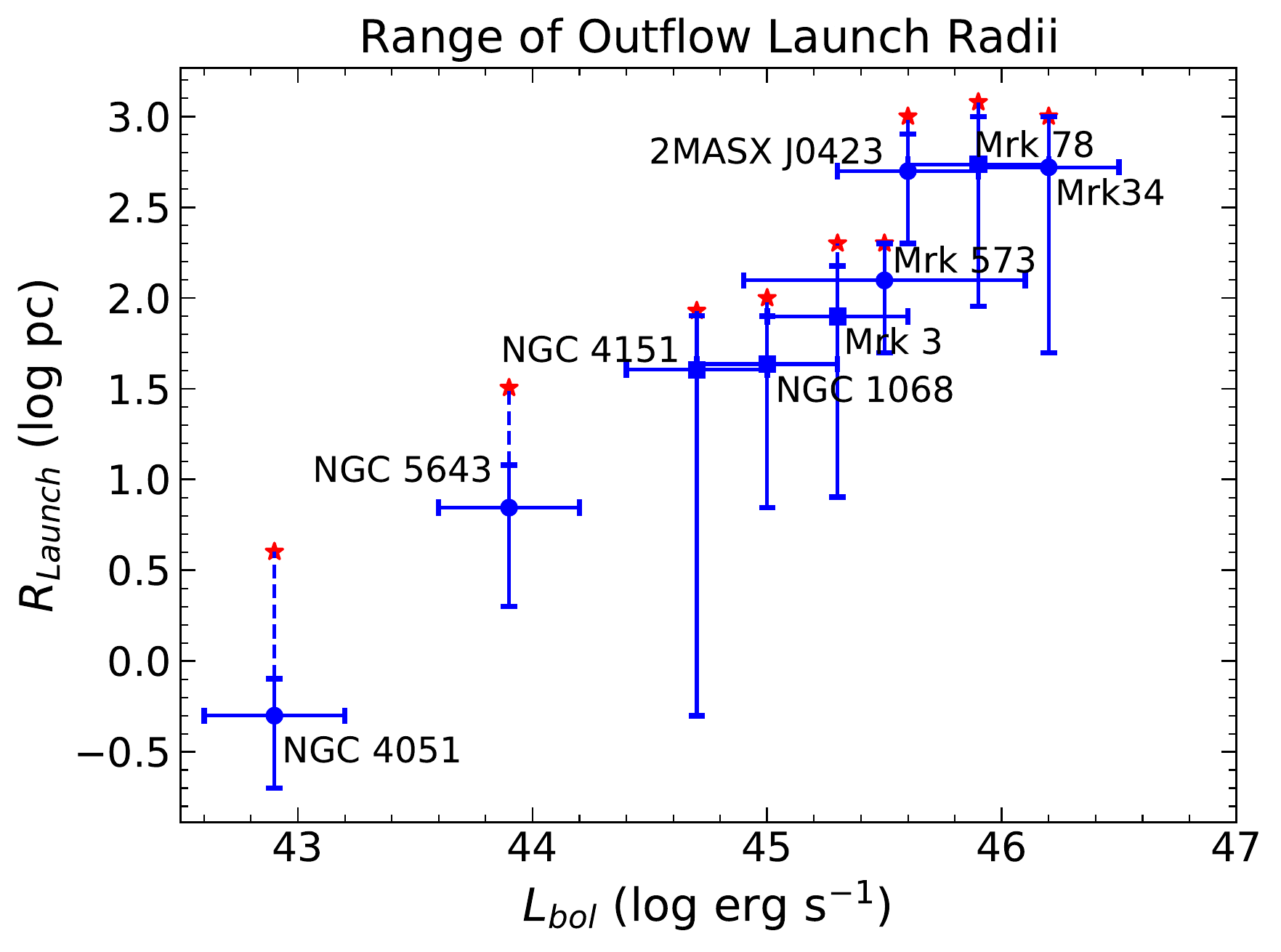}
\end{center}
\vspace{-1em}
\caption{Full range of the launch radii for different AGN using the radiative-gravity models determined for the observed gas kinematics. A force multiplier of $\mathcal{M}$ = 500 was used for the targets in this work and in \citetalias{Meena2021}. For the remaining targets, we adopted the models from previous studies, which are NGC 5643 \citep{Garcia2021}, Mrk 573 \citep{Fischer2017} and JMASX J0432 \citep{Fischer2019}. For Mrk~34, we revised the radiative driving model using the galaxy's enclosed mass profile, provided in \cite{Trindade2021}, to match our observed kinematics in \cite{Revalski2018b}. The positions of the blue squares (targets in this work) and circles (the additional five sources from the literature) on the y-axis represent the mean values of the calculated launch radii for each target. While these mean values do not pose physical significance their positions is helpful to visualize the trend between luminosity and launch distance.
The vertical bars in the launch radii (y-axis) correspond to the minimum and maximum launch distances of the spatially-resolved outflows from HST STIS observations. The uncertainties in the bolometric luminosities are $\pm$0.3 dex for all targets (references provided in Table~\ref{tab:sample}), except Mrk 573, which is $\pm$0.6 dex \citep{Kraemer2009}. The red $\ast$ represent the model turnover distance for each target beyond, which the outflows cannot be radiatively driven (see \S \ref{subsec:corr_TO}). It is to be noted that the correlation seen in this Figure is derived from  our assumption that the only AGN radiation pressure and gravity define the origin and dynamics of NLR outflows.}
\label{fig:corr_lum}
\end{figure*}

The full range of launch distances for the observed NLR outflows in each target are shown in Figure~\ref{fig:corr_lum}, based on a force multipler of $\mathcal{M}$ = 500. As expected from equation \ref{eq:vel_pro}, the launch distances increase with increasing luminosity. This trend is due to the assumption of radiative driving, and is not an independent correlation.
We note that for the nearby, lower-luminosity galaxies, the minimum launch radius is very close to the nucleus, indicating that the NLR clouds for these AGN are being produced as close as $\sim$1~pc from the SMBH. For more distant, higher-luminosity galaxies, the minimum launch distances are tens to hundreds of parsecs from the nucleus. This may be due to our inability to resolve outflows launched from smaller radii, or this may indicate outflows launched inside the resolution limit do not survive their travel to larger, spatially-resolved radii.

Figure~\ref{fig:corr_lum} also shows our modeled turnover radii, which is the maximum distance from the SMBH where the outflows can be radiatively launched. Beyond this distance, the gravitational field of the galaxy takes over and hinders any further driving of outflows.
In \citetalias{Meena2021}, we noted that this turnover distance (referred to there as the maximum $R_{Launch}$) is strongly dependent on the bolometric luminosity ($L_{bol}$) of the AGN (see Equation~\ref{eq:vel_pro}). We confirm this relation with our larger sample, shown in Figure~\ref{fig:corr_lum}. However, it is important to note again that this trend is expected, as L$_ {bol}$ is an input variable in the radiative-gravity model. In particular, a strong correlation with luminosity is expected for all galaxies with comparable bulge mass distributions. The dispersion in this relationship can be attributed to differences in the host galaxy radial mass profiles. For example, the large launch radii in 2MASX J0423 ($\sim$10 kpc) correspond to the relatively lower mass of the host galaxy \citep{Fischer2019}, as compared to our sample.
Given the assumption of radiative driving, higher luminosity AGN are capable of clearing out the reservoirs of star-forming gas from the centers of their host galaxies at larger distances compared to their lower luminosity counterparts.

\subsection{Comparison with Turnover Radii} \label{subsec:corr_TO}

\begin{figure*}[ht!]
\begin{center}
\includegraphics[width=0.480\textwidth]{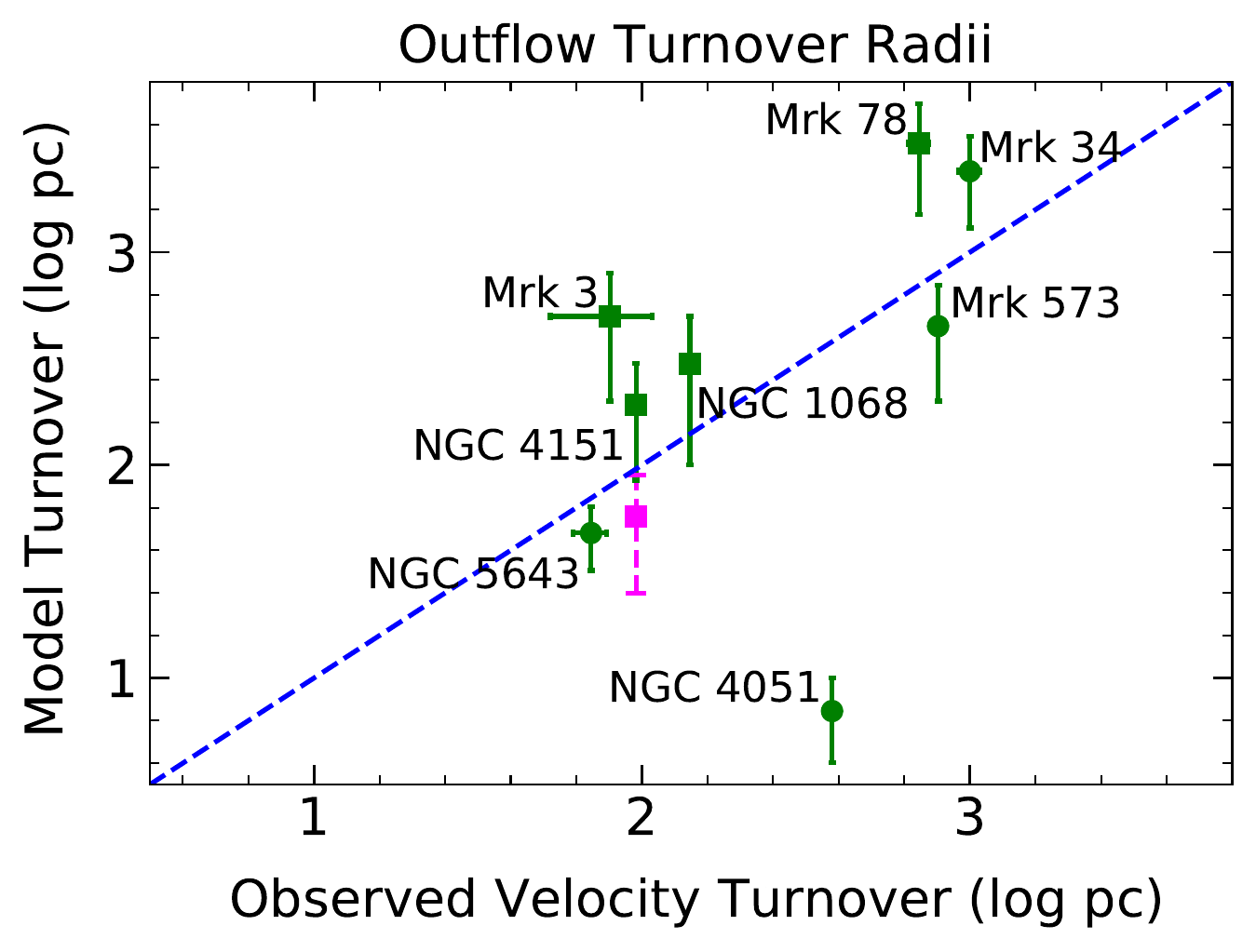}\hspace{3ex}
\includegraphics[width=0.49\textwidth]{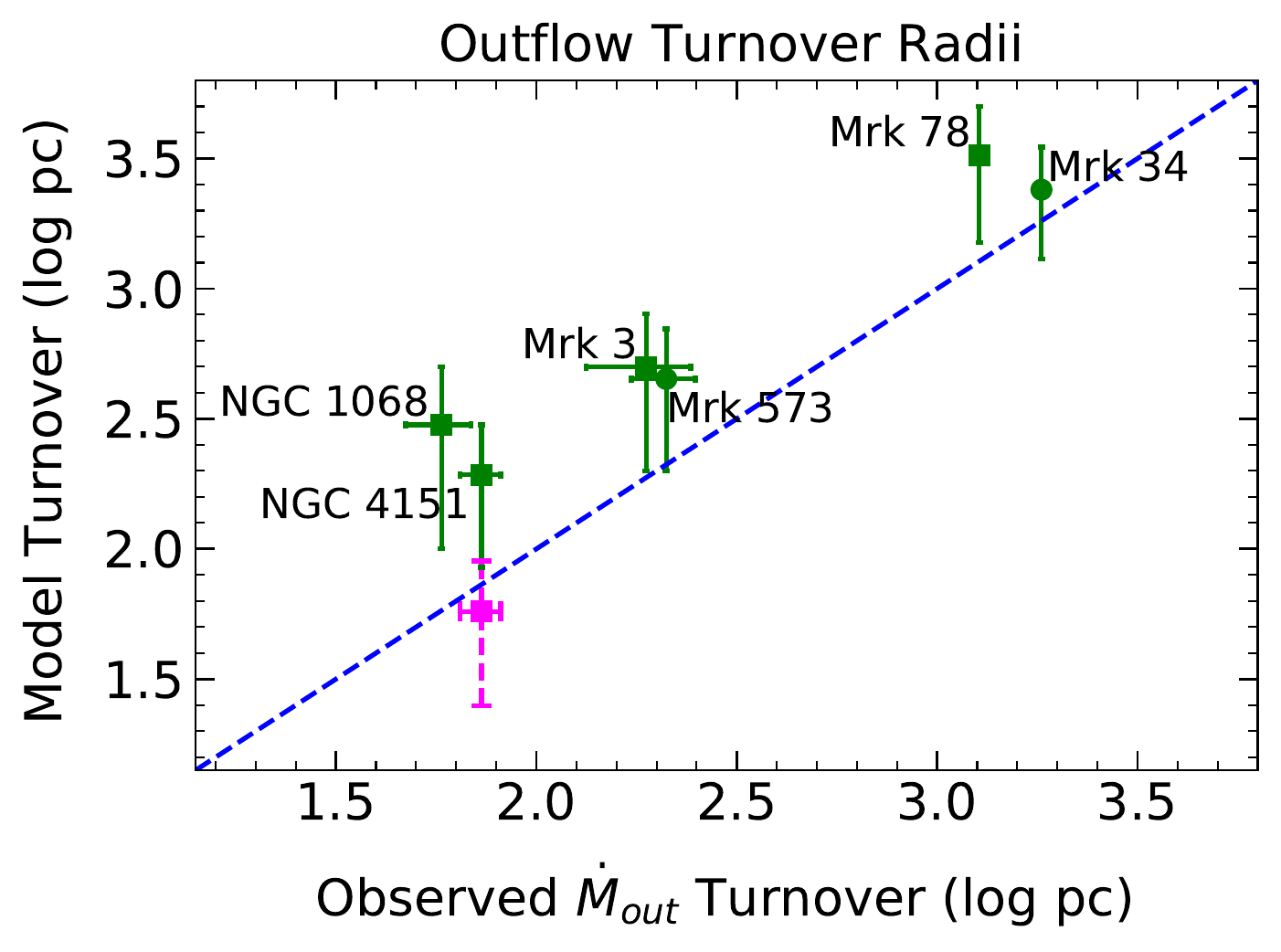}
\end{center}
\vspace{-1em}
\caption{A comparison of the modeled turnover radii based on radiative-gravity formalism and observed outflow quantities. Left: Comparison with the velocity turnover distances \citep{Fischer2013, Meena2021}. Right: Comparison with the mass outflow rate turnovers, i.e. distances at which the mass outflow rates peak) as presented in \cite{Revalski2021}. 
The data in green show the results for model turnover as determined in \S \ref{subsec:radiative_driving} and Figure~\ref{fig:rad_driv_model}, based on the AGN bolometric luminosities provided in Table~\ref{tab:sample}. The targets in this work and the literature are represented in squares and circles respectively. The vertical error bars correspond to $\mathcal{M}$ = 500 (for lower limit) and $\mathcal{M}$ = 3000 (for upper limit) for all targets except NGC 5643 ($\mathcal{M}$ = 200-400). The magenta dashed point is for NGC~4151 if we were to adopt log($L_{bol}$) = 44.14 erg s$^{-1}$ \citep{Kraemer2006,Crenshaw2015} in our radiative-gravity model.The horizontal uncertainties are associated with the bin size/radial distance for the velocity/mass outflow rate measurements for each target \citep{Revalski2021}. The blue dashed line corresponds to unity, where the model turnover radii equal the observed turnover radii.}
\label{fig:compare_TO}
\end{figure*}

Finally, we compare our outflow dynamical models with previously determined spatially-resolved kinematics, mass outflow rate measurements, and their observed turnover distances. 
Previous observational studies have determined that the velocities of the NLR clouds often show an observed increase to a maximum velocity at an empirical ``turnover radius'' ($r_{t}$, corrected for projection effects with a kinematic model), followed by a decrease, often to the galaxy's systemic or rotational velocity (\citealt{Crenshaw2000a,Crenshaw2000b,Fischer2013}, and references therein).
Values for $r_{t}$ were retrieved from \cite{Fischer2013} for each target and from \citetalias{Meena2021} for NGC 4051 (see also \S \ref{app_kinematics}). The values of $r_{t}$ were determined in these studies by fitting a biconical geometry to the observed kinematics.
Based on the observed kinematics, the biconical outflow model, and the mass distribution of ionized gas, we have determined spatially-resolved mass outflow rates for our sample, which also show a turnover in the observed values \citep{Crenshaw2015,Revalski2018a,Revalski2018b,Revalski2021}.
The observed mass outflow rate (${\dot{M}_{out}}$) turnover radii were obtained from the spatially-resolved measurements as provided in Figure~11 and Table~A1 of \cite{Revalski2021}, where mass outflow turnovers corresponds to the distance with a peak value of ${\dot{M}_{out}}$. 2MASX J0423 is excluded from this comparison, as we do not have mass outflow rate measurements for the outflows in this galaxy. Similarly, there are not yet spatially-resolved mass outflow rate measurements for NGC 4051, and those available for NGC~5643 \citep{Garcia2021} are based on lower spatial resolution data and do not employ photoionization models.

Figure~\ref{fig:compare_TO} (left) shows the comparison between our modeled turnover radii and the observed velocity turnovers. The range in model turnover radii represents the values associated with $\mathcal{M}$ = 500 for the lower limit and $\mathcal{M}$ = 3000 for the upper limit. The uncertainties in the velocity turnover radii are given as the spatial resolution of extracted spectra/velocity points.
We find that most of our modeled turnover radii (plus those from previous studies) are either equal to, or greater than, observed velocity turnovers. In other words, we do not see any increase in the outflow velocities beyond the model turnover - as expected from the radiative-gravity model. A force multiplier of $\mathcal{M}$ $\approx$ 500 for the model turnovers provides a reasonably good match to the observed values for NGC 5643, NGC~4151, and Mrk~34, while a slightly lower force multiplier ($\mathcal{M}$ $\sim$ 150-180) would be required for Mrk~3 and Mrk~78. We have adopted a radiative driving model for Mrk 573 from \cite{Fischer2017}, which assumed a $\mathcal{M}$ = 3000 (upper limit) and $M/L$ = 5. It must be noted that, unlike this work, the $M/L$ were not determined using B-V colors of the different brightness components of Mrk~573. Therefore, while the radiative-gravity formalism provided in \cite{Fischer2017} reasonably matches the NLR kinematics, the adopted force multiplier, as well as the model turnover, must be re-scaled to match this analysis.
Overall, based on the close similarities between modeled and observed turnover radii, we confirm that the dynamics of the NLRs in all of our targets (except NGC 4051) can be explained using only radiation pressure combined with gravitational deceleration. We do not require any additional forces in order to reproduce the observed outflow velocities.
This trend does, however, seem to break for the low mass, low luminosity AGN in NGC~4051, which has a much smaller modeled turnover ($\sim$ 2 pc) than the observed values (380 pc, \citetalias{Meena2021}). 

Finally, there is also a disparity in the modeled turnover radii for NGC~4151 based on the adopted bolometric luminosity. Our radiative-gravity model favors a $\mathcal{M}$ of 500 (considering $L_{bol}$ = 5 $\times$ 10$^{44}$ erg s$^{-1}$), but for a lower luminosity ($L_{bol}$ = 1.4 $\times$ 10$^{44}$ erg s$^{-1}$, \citealp{Kraemer2020}), we would require a higher $\mathcal{M}$ ($\sim$3000) to match the observed velocity turnover.

We find an even tighter relation between our modeled turnover and mass outflow rate turnover as seen in Figure~\ref{fig:compare_TO} (right).
The spatially-resolved mass outflow rates are calculated using outflow velocities and ionized gas masses, measured at each spatial distance from the SMBH (\citealp{Crenshaw2015, Revalski2018a, Revalski2018b, Revalski2021}). Therefore, mass outflow rates indicate the amount of ionized gas, with the measured outflow velocities at a given distance (r) from the central SMBH. As discussed in \cite{Revalski2021}, the outflow rates increase at larger distances into the NLR, as both outflow velocities and ionized mass increase. However, at a certain turnover distance, gravity dominates the radiation pressure and no more mass can be launched or accelerated. Therefore, while some outflowing ionized gas (that may have been accelerated at r $<$ modeled turnover radius can be observed beyond this distance, an overall deceleration is expected at larger radii. 
The outflow rate turnover radii for Mrk~3, Mrk~78 and Mrk~34 are slightly higher than the observed velocity turnover radii due to the presence of higher ionized gas mass increasing the total mass outflow rate at each spatial location. On the other hand, Mrk 573 and NGC~1068 have relatively lower gas masses than the other AGN with comparable luminosities (see \citealp{Revalski2021}), explaining the smaller mass outflow turnover distances. Nonetheless, for all of the targets, the observed mass outflow turnover is either equal to or smaller than our model turnover values, indicating no further acceleration of mass outflows beyond these radii, which is in agreement with the in-situ acceleration and radiative-gravity formalism.

The excellent agreement between the observed turnover distances and the lower limits of the model turnover indicates that $\mathcal{M} \approx$ 500 is the preferred value to use for these models, agreeing with recent investigations using photoionization models in \cite{Trindade2021} (see the Discussion in \S \ref{subsec:assum}). The ambiguity associated with NGC~4151 can also be observed in Figure~\ref{fig:compare_TO} (right), where the modeled and observed turnover radii are consistent for an adopted luminosity based on \othree flux and $\mathcal{M}$ = 500.
However, for a lower luminosity, a much higher $\mathcal{M}$ ($>$ 3000) is required (which is unlikely for the physical conditions of the NLR) to match both turnover distances, or there would have to be additional physical mechanisms besides radiation pressure accelerating the gas beyond the modeled turnover.

\section{DISCUSSION}\label{sec:discuss}

\subsection{Assumptions} \label{subsec:assum}
In exploring these models, it is important to clearly state the main assumptions included in the radiative-gravity formalism, the uncertainties associated with the model parameters, and their potential impact on the results.

\paragraph{Luminosity:} The bolometric luminosity of AGN used in Equation \ref{eq:vel_pro} is one of the biggest factors that affects the modeled values of the outflow launch radii shown in Figure~\ref{fig:rad_driv_model}.
As discussed in \S \ref{subsec:corr_lum}, higher luminosities can drive outflows from larger distances for a given force multiplier and enclosed mass distribution of the host galaxy. 
For example, as shown in Figure~\ref{fig:compare_TO}, for NGC~4151, an increase of $\sim$0.5 dex in the $L_{bol}$ would result in about a 3.5 times larger launch distance. For our model, we have adopted the bolometric luminosities calculated based on \othree luminosities using continuum subtracted \othree images. While most of our calculated $L_{bol}$ values are in agreement with the literature, for NGC~4151, we measured slightly higher values than our previously adopted values (log(L$_{bol}$) = 43.9 erg s$^{-1}$, \citealp{Crenshaw2015,Revalski2021}). However, the AGN of NGC~4151 is highly variable, where the UV continuum flux has been found to vary by a factor of $\sim$50 over the course of $\sim$two decades \citep{Kraemer2006}. Similarly, using an SED based on WIYN High Resolution Infrared Camera (WHIRC) and XMM-Newton observations Merritt, Bentz, \& Crenshaw (in prep) found a drop in the bolometric luminosity of NGC 4151 by a factor of $\sim$ 8 between 2003 ($\sim$3 $\times$ 10$^{44}$ erg s$^{-1}$) and 2015 ($\sim$4 $\times$ 10$^{43}$ erg s$^{-1}$). We prefer [O~III] luminosity-based measurements because they reflect an average state over the light travel time of the NLR (hundreds to thousands of years). We also note that the uncertainty in the distance to an AGN also affect the luminosity, as discussed below.

\paragraph{Distance:} In this work, we used the galaxy distances based on HI 21cm redshift and H$_{0}$ = 70 km s$^{-1}$ Mpc$^{-1}$ for all of the targets except NGC 4051 and NGC~4151 (for which we used the distances based on Cepheid measurements) and NGC~1068 (for which we used a Tully-Fisher (TF) distance, to be consistent with the rotation curves (Figure~\ref{fig:rotation}) provided in the cited literature). See Table~\ref{tab:sample} for relevant references of the adopted distances.
The galaxy distances may vary based on the method of measurement (redshift, TF, Cepheid etc.) or even for similar methods using different studies and/or calibrations. For example, there are a number of TF distance measurements available for NGC~1068 spanning a range of 7.2-14.4 Mpc. Similarly, the redshift-based distance measurements carry the uncertainties associated with the peculiar velocities of the galaxies (particularly at z$<$0.02) and the chosen value of the Hubble constant. These uncertainties in the galaxy distances affect the bolometric luminosity measurements, and therefore alter the outflow launch distances and modeled turnover radii. The adopted distances also affect the enclosed mass profiles (Figure~\ref{fig:masses}), determined from surface brightness decomposition of the galaxy.
For NGC~1068, while we adopted a higher value of TF distance, a lower value would reduce the enclosed mass, as well as the circular velocities shown in Figure~\ref{fig:masses}, subsequently increasing the model turnover. Furthermore, a change in adopted distance would also correspond to a new spatial scale and the de-projection factor used to measure true outflow distances, which, by themselves, would change the turnover distance by a factor of two in NGC~1068 based on the available range of distances. 

\paragraph{Mass-to-Light Ratio:} We determined the $M/L$ values for each individual brightness component of a galaxy, using B-V colors and relationships given in Table~1 of \cite{Bell2001}. These calculations offer an uncertainty in the enclosed mass, up to a factor of two, based on variations in the Initial Mass Function \citep{Bell2001}. This may lead to an error in the modeled turnover radii of up to a factor of 1.5 for a given AGN luminosity. Furthermore, the B and V magnitudes in our surface brightness decomposition may be affected due to emission from the NLR, leading to further uncertainties in $M/L$. However, as discussed in \S \ref{subsec:mass_vel}, our overall mass profiles are comparable to previous mass measurements and rotation curves, with slight overestimations for Mrk~78 and NGC~1068, indicating a satisfactory constraint on our measured $M/L$ values.

\paragraph{Spherical Distribution:} We assume a spherical symmetry in our radial mass distribution calculations using the relationships provided in \cite{Terzi2005} based on surface brightness decomposition. While this approximation can be adopted for the galaxy bulges, it is not quite appropriate to use spherical symmetry for more elongated or axisymmetric components such as bars and disks. However, we find that
the ionized gas outflow radii are dominated by host galaxy bulges in our sample. The overall contribution of the disk (or disk + bar) corresponding to outflow radii ($<$1 kpc) is less than 15$\%$ for Mrk 3, $\sim$10$\%$ for NGC 1068 and less than 1$\%$ for NGC 4151. Similarly, commensurate with the outflow extent of Mrk 78 ($\sim$ 3 kpc), the disk contribution is less than 15$\%$ in our mass calculation.
Correcting for axisymmetric and/or anisotropic distributions for these components will slightly reduce the total enclosed mass profiles, but is left for a future work.

\paragraph{Force Multiplier:} As shown in Figures~\ref{fig:corr_lum}-\ref{fig:compare_TO}, the outflow launch radii from the SMBH, and the modeled turnover, are subject to the value of the force multiplier used in our radiative-gravity formalism.
The force multiplier (and imposed acceleration) depends on the ionization state of the gas, optical depth (column density), and additional sources of opacity (such as internal dust), as well as the amount of radiation available as we move deeper into the cloud \citep{Chelouche2001, Dopita2002, Crenshaw2003}.
\cite{Trindade2021} discuss the change in force multiplier as a function of column density for different ionization parameters, including both dusty and dust-free models. They estimate a column density weighted force multiplier of $\mathcal{M}$ = 1040 associated with an ionization parameter of $log~U$ = -2. This value is almost 3 times smaller than previously used ($\mathcal{M}$ = 3000) in \cite{Fischer2017}. Similar to \citetalias{Meena2021}, we employ a range of $\mathcal{M}$ in our models, to determine the change in modeled turnover radii based on different $\mathcal{M}$ values.
We find that by using a conservative estimate of $\mathcal{M}$ $\approx$ 500, our model turnover radii sufficiently match the observed turnover distances determined with spatially-resolved velocity and mass outflow rate profiles, within a factor of $\sim$1.5 (Figure~\ref{fig:compare_TO}) for the majority of the sample, except Mrk 573 and NGC 4051.
While using a $\mathcal{M}$ = 3000 improves the model for Mrk 573, a much higher force multiplier would be required to satisfy the observed velocities in NGC 4051. Although further investigation is needed to evaluate these discrepancies, both theoretical and observational measurements indicate that a low force multiplier of $\mathcal{M}$ $\approx$ 500 is the most appropriate value for the ionized NLR clouds that are emitting most of the \othree emission in nearby AGN. 

\paragraph{Confinement:} A longstanding problem with both BLR and NLR clouds is the issue of confinement. It is difficult to maintain the integrity of an \othree emitting cloud in the NLR for the calculated dynamical travel times and distances given in Table~\ref{tab:rad_drive} in the absence of a confinement mechanism, such a hot external medium or a magnetic field providing pressure to confine the cloud. A spherically symmetric ionized gas cloud can expand to about twice its original radius within $\sim$ 2.5 $\times$10$^{10}$ s ($\sim$1000 yr), due to isothermal expansion of the cloud, subsequently increasing its ionization parameter by a factor of $\sim$10 \citep{Trindade2021}. As discussed in \S4.3 of \cite{Trindade2021}, for the measured velocities of NLR gas in our sample, an \othree emitting cloud will travel only a few pc and will quickly expand into a highly ionized gas emitting primarily X-ray emission lines long before completing its full trajectory (as shown in Figure~\ref{fig:rad_driv_model}). Thus, we assume that a confinement mechanism is at work, and we leave it to future studies to test this assumption and investigate the source(s) of confinement.

\subsection{Origins \& Impacts of NLR Outflows} \label{subsec:outflow_origins}

Following the discussion in \cite{Fischer2017}, we compare our models with observations of cold and warm molecular gas, and discuss possible connections between these gas reservoirs and the NLR outflows, including the potential of outflows for negative feedback by removing gas that would otherwise form stars. In general, the cold molecular gas (T $\leq$100 K) at the centers of galaxies is traced with sub-millimeter lines (particularly different CO transitions) using interferometric observations from telescopes including Atacama Large Millimeter Array (ALMA) and IRAM's Plateau de Bure Interferometer (PdBI) or Northern Extended Millimeter Array (NOEMA). The warm molecular gas near AGN is detected through the ro-vibrational H$_{2}$ emission lines (T $\approx$ 1000 -- 2000 K) observed in the near infrared using large ground- and spaced-based telescopes.

\paragraph{Mrk~3:} Using Gemini Near-Infrared Integral Field Spectrometer (NIFS) observations, \cite{Gnilka2020} provide
kinematic maps of H$_{2}$ emission in the nuclear regions of Mrk~3, which show warm molecular outflows travelling with velocities up to $\sim$600 \kms~inside $\pm$1\farcs5 ($\sim$400 pc) of the nucleus.
The majority of the H$_{2}$ emission is outflowing at typically lower velocities than in coincident or adjacent regions of ionized gas emitting [S~III]. Most of the H2 gas is concentrated inside of 0\farcs5 (~135 pc), similar to the launch distances of the ionized gas, indicating similar origins and a transition from cold (indicated by the dust lanes) to warm to ionized gas.
Nonetheless, these launching radii (or evacuation sites) are still much smaller than size of the host galaxy bulge ($>$2 kpc) or disk ($>$ 5 kpc) indicating that only a fraction of the host galaxy is being truly evacuated by the AGN feedback. While there is no evidence of gas evacuation at these large radii, using ground-based spectroscopy, \cite{Gnilka2020} discuss the presence of disturbed/turbulent gas up to $\sim$3.2 kpc, which may also halt the star-formation activities at larger radii in Mrk~3.

\paragraph{Mrk~78:} Outflows can be produced by radiation pressure up to a distance of $\sim$1.5 kpc from the AGN, commensurate with the effective radius of the bulge in Mrk~78 (as determined in \S \ref{subsec:galfit}). Therefore, the large launch radii in Mrk~78 suggest a case of strong negative feedback from AGN radiation pressure, where the gas in a significant fraction of the galaxy is being ionized and evacuated, thereby potentially impeding star-formation as well as (eventually) black hole accretion.
We note that several of the [O~III] knots (eg. F,H,I,J,K) follow the nuclear dust lane (see Figure~\ref{fig:O3}) and also coincide with shocked emission found by \cite{Fornasini2022}. This work found that the gas in these regions is heavily influenced by shocks versus photo-ionization, which dominates the inner regions of NLR. The coincidence between shock emission and observed high-velocity outflow knots may indicate a different origin for these knots than the launch radii calculated solely from radiation pressure. Nonetheless, these regions still lie inside our calculated launch distances and do not affect the overall modeled turnover radii.
Currently, there is no existing information regarding cold/warm molecular gas in the nuclear/circumnulear regions of Mrk~78. Future observations of these gas phases may provide more information regarding the origins of outflows as well as total mass ejections into the host galaxy.

\paragraph{NGC~1068:} Using high angular resolution observations with ALMA, \cite{Garcia-Burillo2016} map the CO(6-5) molecular lines and the 432 $\mu$m dust continuum emission in the circumnuclear disk (CND) of NGC~1068, and propose the radius of the putative torus R$_{torus}$ = 3.5$\pm$0.5 pc. The torus is apparently connected with the CND via streams of molecular gas as further reported in \cite{Garcia-Burillo2019}. Using intensity maps and kinematics of various CO emission lines, they also show the presence of an outflowing molecular ring in the CND, with a diameter $\sim$400 pc, which coincides with the edge of the ionized gas outflows in NGC~1068. This outflowing/expanding ring is also observed in the warm molecular gas in H$_{2}$ emission using Gemini-NIFS \citep{Barbosa2014, May2017}.
More importantly, both of these observations reveal a clear deficit of cold gas (as seen in the CO(2-1) maps in Figure 3, \citealp{Garcia-Burillo2019}) as well as warm gas (Figure~10, \citealp{Barbosa2014}) in the inner 100 --130 pc. These gas-deficient regions (or cavities), however, are pervaded by the ionizing bicone, and particularly coincide with our modeled launch radii of the ionized gas outflows. 
This coincidence between outflow launch radii and lack of molecular gas at those distances suggest that the ionized outflows originated in-situ from the molecular reservoirs in the central $\sim$100 pc, which is now nearly cleared out from interaction with AGN winds, as also claimed by the above authors.

\paragraph{NGC~4151:} \cite{Storchi-Bergmann2009} use Gemini-NIFS K band observations to present maps of 2.1218 $\mu$m H$_{2}$ emitting gas, all of which seem to lie outside the ionized gas bicone. They propose that this emitting gas is related to a large molecular gas reservoir that was a possible feeding source to the AGN, which was depleted by the AGN feedback in the regions occupied by the ionizing bicone. More recently, \cite{May2020} find bullets of outflowing warm molecular gas inside the ionized gas bicone in the 1.7480 $\mu$m line and the suggest existence of an expanding molecular cavity structure. Similar to NGC~1068, the size of this cavity (and their interpretation of the nuclear gas dynamics) is mostly in agreement with our proposed launch radii, further reinforcing the in-situ origin of ionized gas outflows and ongoing removal of molecular gas in the NLR.
Once again, these launch radii and extents of the ionized gas outflows much smaller than the bulge and disk sizes, which indicates little AGN feedback from NLR outflows in the bulk of the bulge.

\subsection{NLR Dynamics \& Outflow Driving Mechanisms} \label{subsec:outflow_mech} 

The originally proposed radiation-gravity (including drag) formalism in \cite{Das2007}, which included two of the current authors, was not successful in explaining the NLR kinematics in NGC~1068.
However, we now recognize that the incompatibility of that model was due to the basic assumption that the outflows are launched from a single point in the NLR. If so, the clouds launched closed to the nucleus (at $\sim$ 1 pc) would achieve terminal velocity and would not slow down under the gravitational influence of the host galaxy. While clouds launched further away from the nucleus can eventually turn down towards zero velocity, they can't achieve the maximum velocities. This pattern can also be seen in Figure~\ref{fig:rad_driv_model}. However, by including our recent understanding of ``in-situ''  acceleration, we 
can confirm that the outflows can be launched at multiple radii from the SMBH and throughout the NLR, providing the observed velocities.
Furthermore, the correlation between model turnover distances, where gravity from the enclosed mass begins to overcome radiation pressure, and previously determined turnovers from kinematic models of the observed velocities, provides strong evidence that only radiation pressure and gravity are needed to explain the NLR kinematics in these AGN.
Additionally, based on the turnover radii comparison, we find that for the AGN with $L_{bol}$ $>$ 10$^{44}$ erg sec$^{-1}$, our previous estimated value of the force multiplier (close to 500) as used in \citetalias{Meena2021}, works well to bring the model and observed turnovers into agreement. As discussed in \S \ref{subsec:assum}, the adopted results for Mrk 573 (from \citealt{Fischer2017}) do not account for the accurate host galaxy mass measurements as well as the factor of $\sim$1.4 (mean atomic weight of proton), which explains the use of higher $\mathcal{M}$ to match the outflow velocity turnover distances.
Overall, we find that AGN radiation pressure is the dominant mechanism responsible for the origins of outflows, and radiation and gravity are the two major forces that define the dynamics of NLRs in most galaxies in our presented sample, including NGC~1068.

However, this argument may not be true for lower luminosity AGN. As seen in Figure \ref{fig:compare_TO}, for NGC 4051 ($L_{bol}$ $=$ 10$^{42.9}$) the velocity turnover is much higher than our modeled turnover, even for $\mathcal{M}$ $=$ 3000. NGC 4051 has more precise distance measurement using Cepheid variable stars. Similarly, the host galaxy mass was consistent with the stellar (and baryonic) mass measurements provided in \cite{Robinson2021}. This leaves us with two scenarios that could be the cause of presented difference in our model: 
(1) Radiation pressure may not be the only and/or main force behind the driving and acceleration of NLR outflows in NGC 4051; and (2) We severely overestimated the observed turnover distance in \citetalias{Meena2021}, because for this low-luminosity AGN the radiation driving model indicates that the turnover is at $\sim$4 pc, which is inside of the central resolution element of $\sim$8 pc and is therefore undetectable.
Various other lower luminosity AGN have been claimed to have significantly extended NLRs and outflows, including the low-ionization nuclear emission line regions (LINERS), which have low luminosity as well as much smaller Eddington ratio \citep{Walsh2008,Rodriguez-Ardila2017,Hermosa2022,Cazzoli2022}. 
Given similar mass distribution as the galaxies in our sample, these low luminosity AGN would have much smaller launch radii according to our radiative acceleration plus gravitation deceleration model. 
Further investigations across a wider range of AGN luminosities are required to test the consistency of our proposed model and determine if radiation pressure is universally the principal driving source for NLR outflows, or if there are additional physical process needed to produce the observed large-scale outflows.

\section{SUMMARY AND CONCLUSIONS} \label{sec:conclusion}
We have investigated the driving mechanisms powering NLR outflows in a sample of four nearby Seyfert galaxies using spectroscopy and imaging from HST and APO. We determined the rotational and outflow kinematics of the \othree ionized gas using our multi-Gaussian fitting routine to accurately identify the number of meaningful velocity components. We mapped the large-scale rotation curves of the host galaxies using long-slit spectroscopy, and compared them with previous rotation curves to separate the outflow and rotation components, and determine the extents of the outflows. 
We tested whether AGN radiation pressure and host galaxy plus SMBH gravitational potential are the fundamental mechanisms that govern the dynamics of NLR outflows. Our main results are summarized below.
\begin{enumerate}

\item We measured the radial stellar mass distributions of the host galaxy components using surface brightness decomposition of space- and ground-based images and $M/L$ ratios from B$-$V colors. We provide the first mass measurements for the host galaxies of Mrk~3 and Mrk~78. For Mrk~3 and NGC~4151, the circular velocities based on the enclosed mass as a function of radii match with those of the observed rotational kinematics. For Mrk~78 and NGC~1068, these circular velocities are higher than the observed rotation by a maximum of 50 \kms, indicating a possible overestimation in our mass measurements.
 
\item We developed velocity profiles of the ionized gas outflows identified by their \othree emission using an analytical model based on acceleration induced by AGN radiation pressure and deceleration from the gravitational potential of the host galaxy and SMBH. These velocity profiles allow us to trace the launch distances of the observed outflows, and we find that gas is launched at multiple radii from the AGN, which is consistent with in-situ acceleration of the gas.

\item Unlike NGC 4051 \citepalias{Meena2021} where all of the outflows were launched within $\sim$2 pc of the nucleus, the higher luminosity AGN exhibit outflows that are produced at tens to hundreds of pc from the nucleus. We confirm that the maximum outflow launch distances strongly depend on the AGN bolometric luminosity, and higher luminosity AGN are capable of evacuating the gas reservoirs by launching outflows at up to bulge scales.

\item The maximum outflow launch radii correspond to the model turnover radii where gravitational deceleration begins to overcome radiative acceleration, and the velocities start to decrease. We compared our model turnover radii with the observed values from the kinematics, and find they are approximately equal. This agreement confirms our hypothesis that radiation pressure and gravity are the two primary mechanisms responsible for the kinematics of NLR outflows in our sample for AGN with $L_{bol}$ $>$ 10$^{44}$ erg sec$^{-1}$. 

\item Given various uncertainties associated with distances, luminosities, and $M/L$ measurements, we conclude that the radiative driving of NLR outflows may be characterized using an approximate value for the force multiplier of $\mathcal{M}$ = 500, which is significantly lower than values of $\sim$3000 used in some previous studies that are more applicable to the physical conditions at the ionized face of a NLR cloud.

\item We find that the consistency between the model and observed turnover distances breaks at the lower luminosity AGN NGC 4051 (provided in \citetalias{Meena2021}), which indicates that either the velocity turnover is inside of the resolution element of the observations, or there may be additional physical processes that can accelerate outflows further away than our modeled launch radii.

\item Finally, we also compare the cold and warm molecular gas distribution in Mrk~3, NGC~1068, NGC~4151 with those ionized gas outflows launch radii and extents, which indicates possible evolutionary connections between different gas phases and gas evacuation due to AGN ionization and driving. However, further investigations using high-angular resolution observations of the molecular gas distribution are required to explore similar scenarios in Mrk 78.
\end{enumerate}

\acknowledgements
The authors would like to thank the anonymous referee for their constructive feedback, which improved the clarity of this manuscript. B.M. also thanks Dr. Misty Bentz for her helpful suggestions.
This research has made use of NASA's Astrophysics Data System. This research has made use of the NASA/IPAC Extragalactic Database (NED), which is operated by the Jet Propulsion Laboratory, California Institute of Technology, under contract with the National Aeronautics and Space Administration. IRAF is distributed by the National Optical Astronomy Observatories, which are operated by the Association of Universities for Research in Astronomy, Inc., under cooperative agreement with the National Science Foundation
This research has made use of the VizieR catalogue access tool, CDS, Strasbourg, France (DOI : 10.26093/cds/vizier). The original description of the VizieR service was published in 2000, A\&AS 143, 23.

\facilities{APO (ARCTIC), \textit{HST} (STIS, ACS, WFPC2)}

\software{IRAF \citep{Tody1986, Tody1993}, MultiNest \citep{Feroz2019}, SAOImage DS9 \citep{ds92000}, Mathematica \citep{Mathematica}, Python (\citealp{VanRossum2009}, \url{https://www.python.org}), Astropy \citep{astropy:2013}, Interactive Data Language (IDL, \url{https://www.harrisgeospatial.com/Software-Technology/IDL}), \textsc{GALFIT} \citep{Peng2002}, DiskFit \citep{Sellwood..Spekkens2015, Kuzio2012soft, Peters2017}}

\bibliography{refs.bib}{}
\bibliographystyle{aasjournal}


\appendix
\restartappendixnumbering

\section{Spectral Fitting} \label{subsec:app_spec_fit}

We fit multiple Gaussian profiles to the observed \othree\llothree~ emission lines extracted at each spatial location along the slits using our automated multi-component Gaussian fitting routine. This fitting routine, first presented in \cite{Fischer2017}, utilizes a multimodal nested sampling algorithm called MultiNest \citep{Feroz2008, Feroz2009, Feroz2019, Buchner2014}. The algorithm adopts Bayesian statistics to determine the most meaningful number of Gaussians to fit to the asymmetric and bumpy \othree line profiles, and provide the velocity centroid, full width at half maximum (FWHM), and integrated flux of each kinematic component. We employed the same routine to measure the ionized gas kinematics in earlier studies
(\citealp{Fischer2017, Gnilka2020, Revalski2021, Meena2021}), and a detailed description is given in the Appendix of \cite{Fischer2017}.

\begin{figure*}[ht!]
\centering
\includegraphics[width=0.49\textwidth]{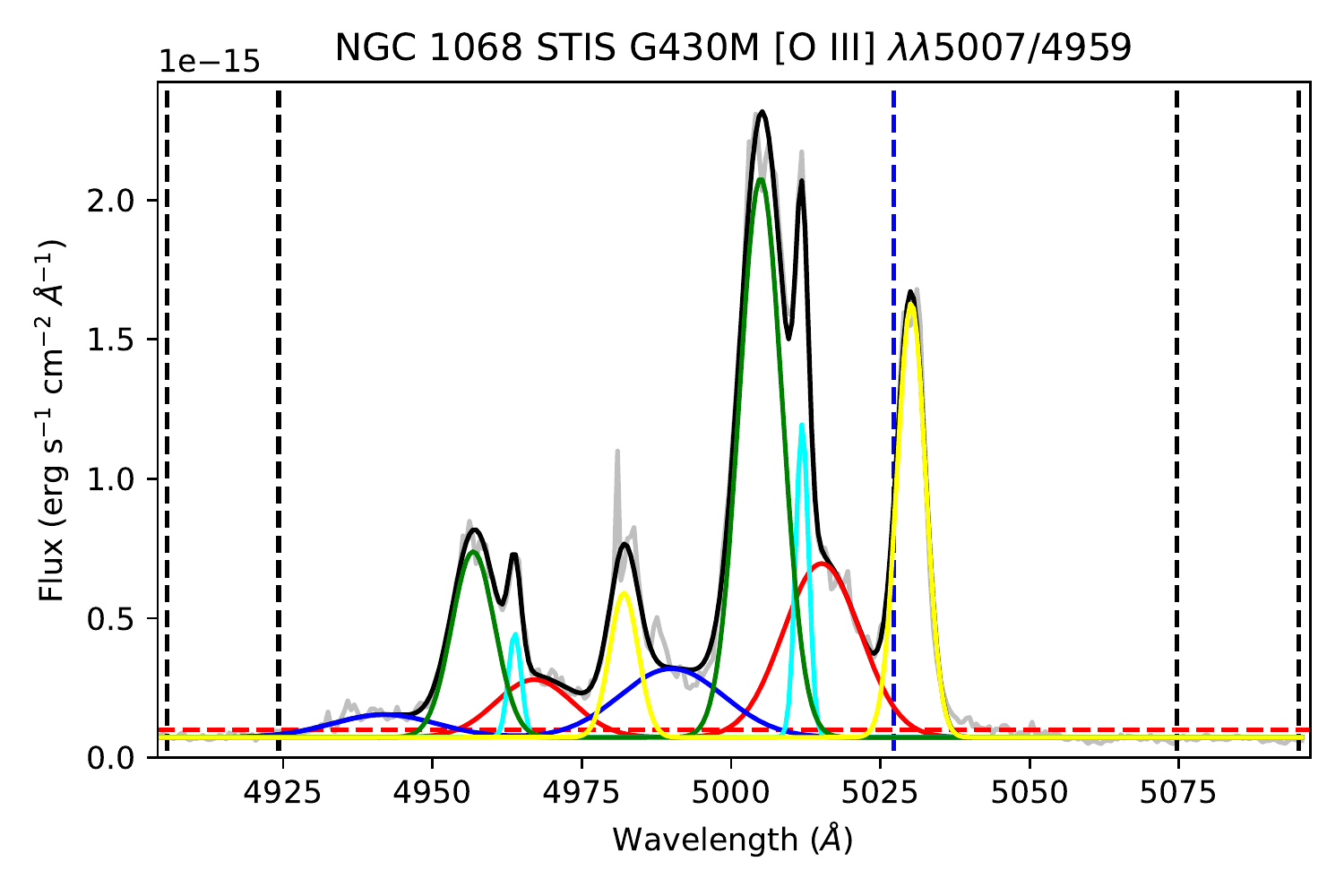}
\includegraphics[width=0.49\textwidth]{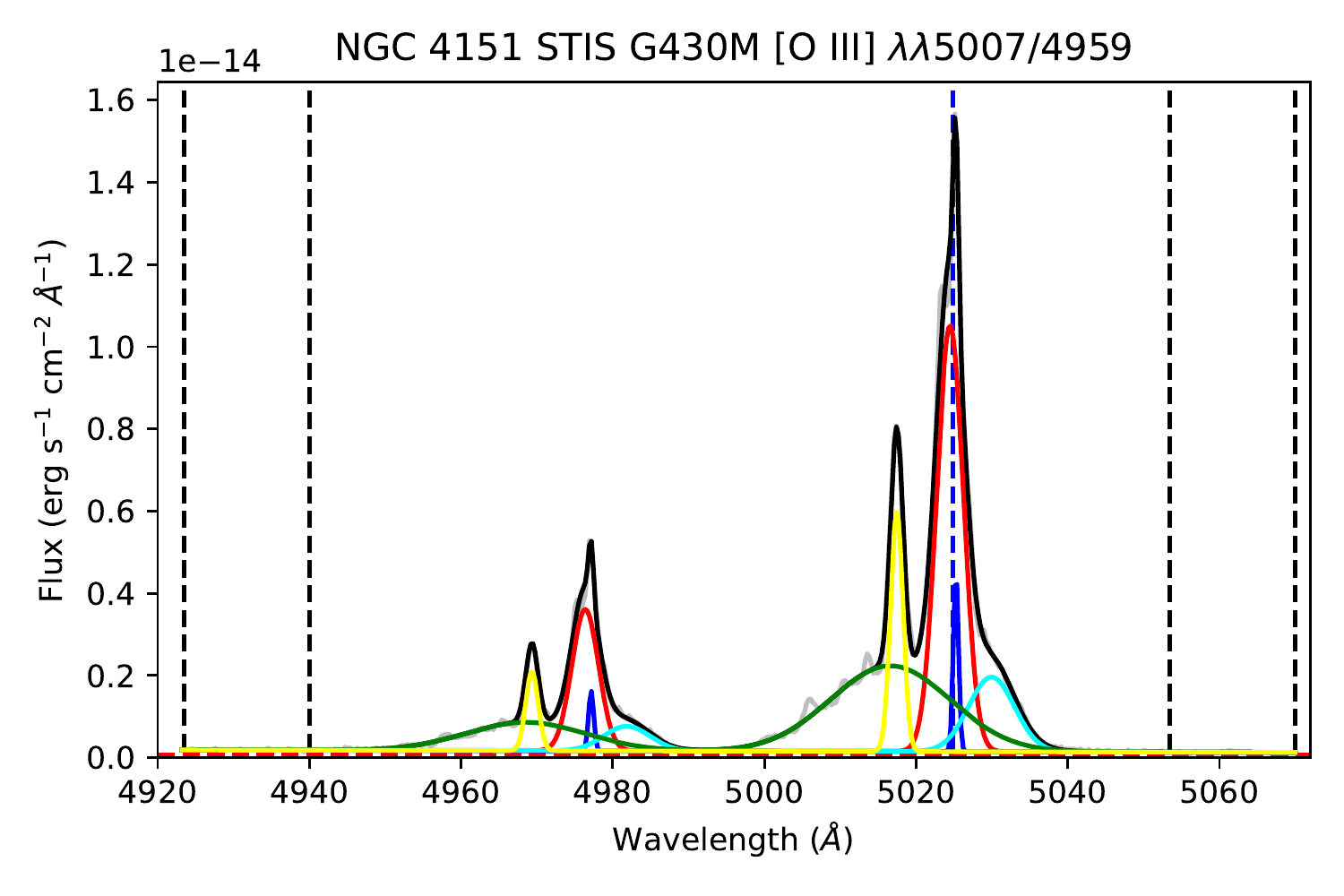}
\caption{Examples of multi-Gaussian fitting of the \othree\llothree~lines observed using HST STIS G430M slits. Left: Spectrum extracted $\sim$1\farcs4 north of the nucleus for NGC~1068. Right: Spectrum extracted at $\sim$0\farcs25 south of the nucleus for NGC~4151. The data are shown in gray, and the total multi-Gaussian fit is shown with a black solid line. The continuum regions are outlined using vertical black dashed lines, and the $3\sigma$ S/N limit is represented by the horizontal dashed red line. The vertical blue dashed line corresponds to the systemic redshift of the galaxy.}
\label{fig:fit}
\end{figure*}

\begin{figure*}[ht!]
\centering
\includegraphics[width=0.49\textwidth]{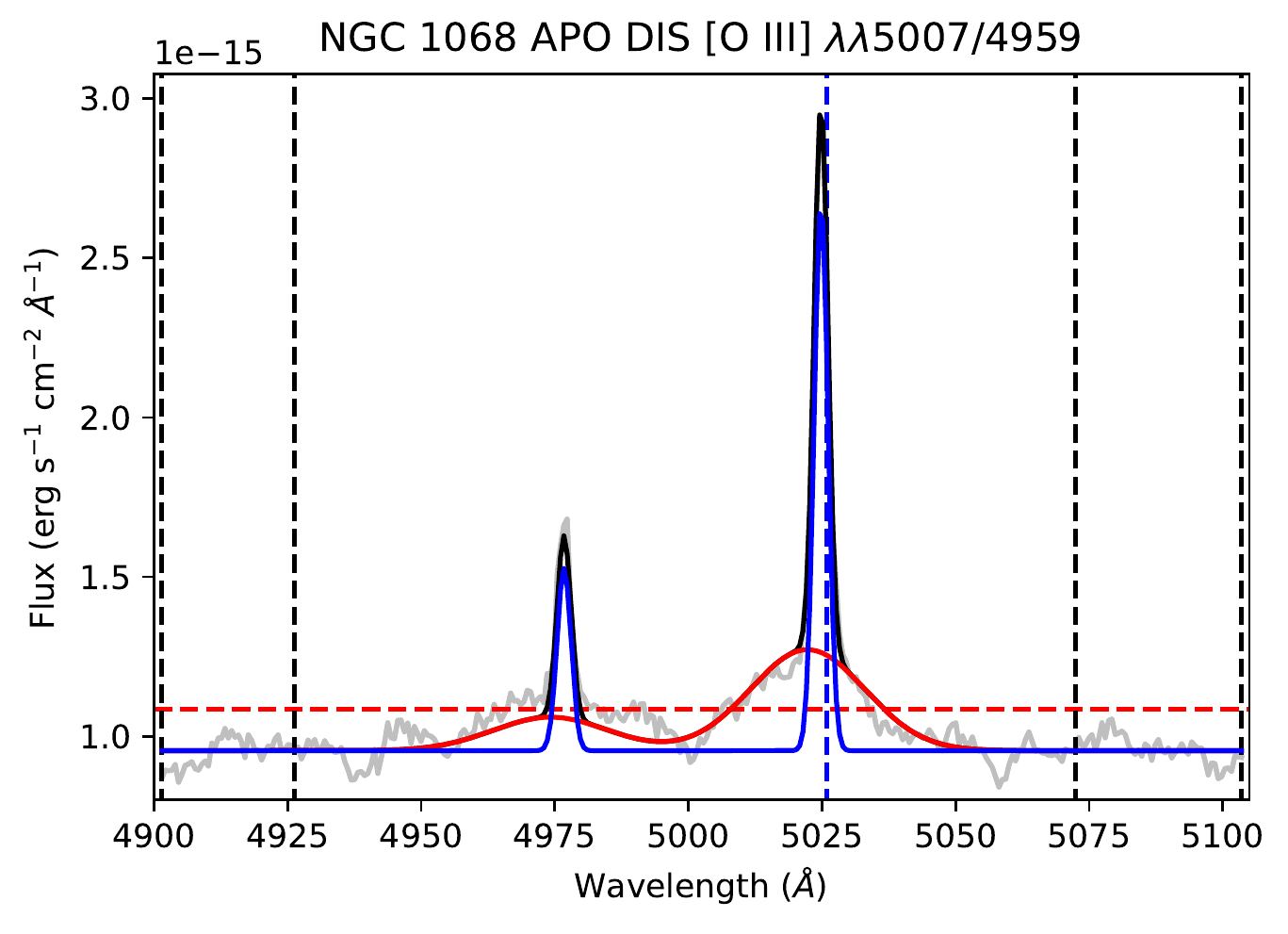}
\includegraphics[width=0.49\textwidth]{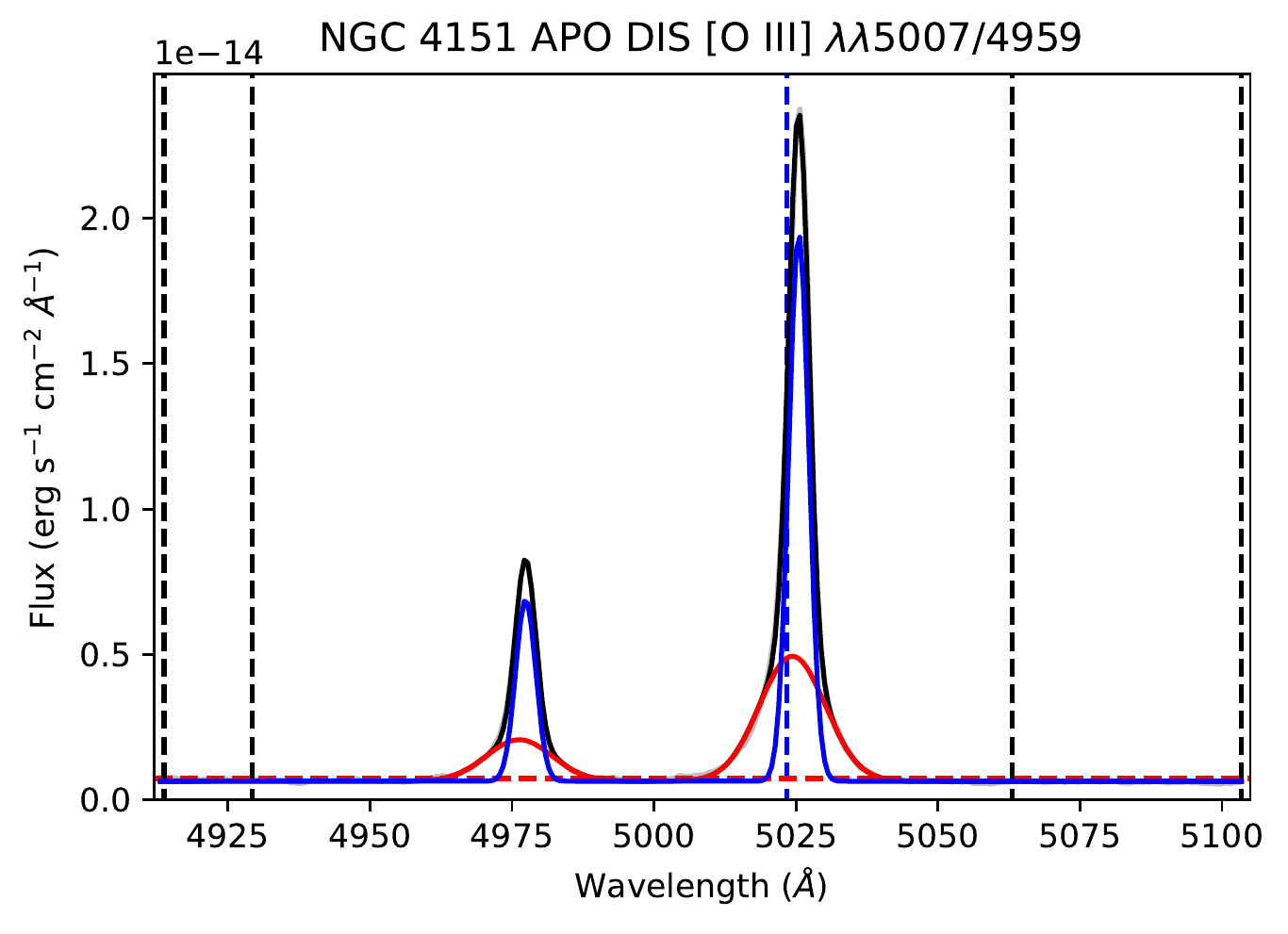}
\caption{Examples of two-component Gaussian fits to the \othree\llothree~lines observed using APO DIS B1200 slits. 
Left: Spectrum extracted at $\sim$11\arcsec north of the nucleus of NGC~1068. Right: Spectrum extracted at $\sim$6\arcsec north of the nucleus of NGC~4151. The broad red wing is associated with the scattered light contamination, while the narrow blue line fits correspond to the emission from the rotating gas. The data, continuum regions and systemic redshifts are shown in the same manner as Figure~\ref{fig:fit}.}
\label{fig:apofit}
\end{figure*}

To achieve the best signal-to-noise ratio (S/N), we extracted the G430M spectra binned every two pixels (each pixel is $\sim$0\farcs05) in the cross-dispersion direction, which matches the resolution of $\sim$0\farcs1~pix$^{-1}$. The low-dispersion G430L spectrum for Mrk~3 was fit for a bin size of 0\farcs05 to accommodate the lower-spectral resolution, allowing detection of distinct emission line peaks to acquire smooth velocity profile.
To achieve the best signal-to-noise ratio (S/N), we extracted the G430M spectra in binned over every two pixels in the cross-dispersion directions, which gives the spatial resolution of $\sim$0\farcs1~pix$^{-1}$.
We optimized the fitting process by fixing the relative height ratio of $\lambda5007$ and $\lambda4959$ to their theoretical value of 3.01 and their wavelength separation to 47.9 \AA\ in the rest frame of the source \citep{Osterbrock_Ferland2006}.
We provided further constraints on the fit by limiting the minimum height of the Gaussian to $3\sigma$ above the background noise, where the standard deviation ($\sigma$) was calculated from the average flux measured in the continuum regions without any strong emission or absorption features.
The minimum width of the Gaussian was also restricted to the spectral resolution (FWHM of the line-spread function or LSF) of the instrument. Examples of multi-Gaussian fits to the \othree lines observed in NGC~1068 and NGC~4151 are shown in Figure \ref{fig:fit}. We identified up to three kinematic components in Mrk~3 and Mrk~78, and a maximum of five components in NGC~1068 and NGC~4151. For the DIS spectra of NGC~1068 and NGC~4151, we employed the same fitting routine to apply two Gaussian profiles to the emission lines extracted from the regions outside of the central 20\arcsec~and 9\arcsec, respectively, to avoid dewar contamination. One of the Gaussians corresponds to the broad wing from the scattered light contamination, and the other Gaussian fits the strong narrow emission tracing the host galaxy rotation. Examples of the fits including scattered light treatment to the extended \othree emission lines extracted from DIS B1200 spectra of NGC~1068 and NGC~4151 are shown in Figure~\ref{fig:apofit}.

\section{HST-STIS Kinematics}\label{app_kinematics}

The \othree velocity map using the same fitting routine for the G430L dataset is also presented in \cite{Gnilka2020}. However, in that work, the fitting process was constrained to fit a maximum of 2 Gaussian components, potentially associated with an outflow and a rotational components. In this work, the automated routine could fit up to 6 components, and was able to identify a maximum of 3 significant kinematic components. Nevertheless, the velocity variation remains similar except for a few  points presenting a smoother variation in the velocity field close to the nucleus. Although the single available long-slit does not cover the entire NLR seen in the \othree image in Figure~\ref{fig:O3}, it provides significant coverage. 
The majority of the kinematics is dominated by outflowing gas, as seen by the high velocity and/or high FWHM points in Figure~\ref{fig:kinematics} (top left panel). The observed outflow velocities reach up to $\sim$600 \kms~(redshift) and $\sim$ $-$700 \kms~(blueshift) at projected distances of $\sim$0\farcs2 SW and NE of the nucleus, respectively. A similar symmetry is noted $\sim$0\farcs4 SW and NE of the nucleus, with velocities of $\sim$ $-$500 \kms~(blueshift) and $\sim$400 \kms~(redshift) in the respective directions.  A gradual decrease in the radial velocity is seen in both directions, similar to that identified in \cite{Fischer2013}.
While it is difficult to distinguish the rotational components within the inner $\sim$2\arcsec, a transition from outflow to rotation (velocity $<$ 200 \kms) can be noticed in the SW at $\sim$1\farcs4 (after knot E). Beyond these radii, the ionized gas follows a rotational pattern at the ends of the backward ``S'' shape as seen in the \othree image (Figure~\ref{fig:O3}), and noted in \cite{Gnilka2020}. It is reasonable to conclude that we trace the full extent and peak velocities of the outflows using the single slit in Mrk~3 that we use to derive the radiative-driving model in \S \ref{subsec:radiative_driving}. Finally we selected data-points clustered together with comparable velocities to match the locations of the knots of emission (in assigned alphabetical letters) with those seen in the \othree image (see Figure~\ref{fig:kinematics}.
Due to low S/N and oversampling of the low-dispersion spectra for Mrk~3, the single-component outliers near $\sim$1\arcsec~to the NE suggest that those high-velocity outflows should be considered with caution.

\subsection{Mrk~78}

\begin{figure*}[ht!]
\centering
\includegraphics[width=0.32\textwidth]{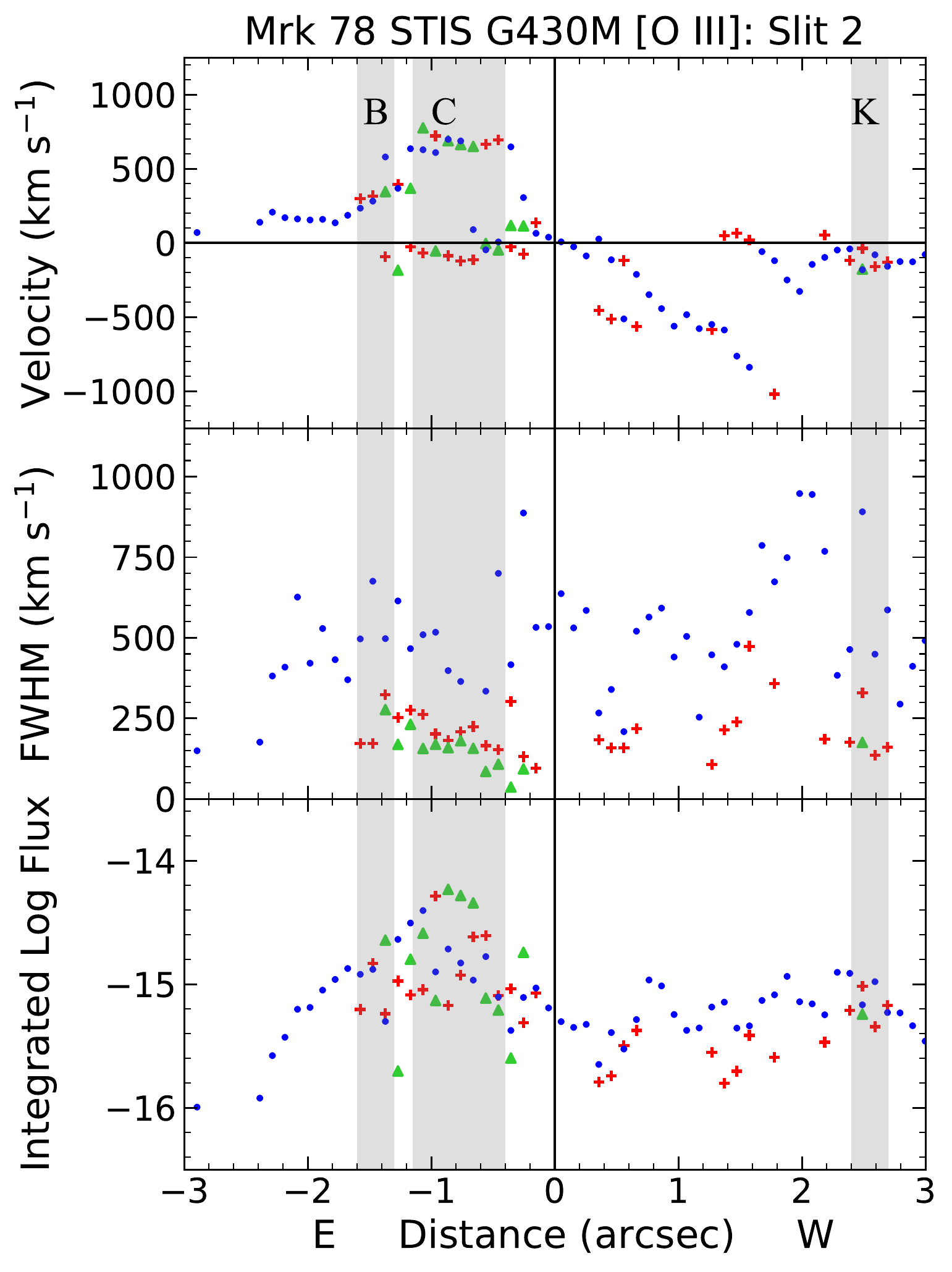}\hspace{1ex}
\includegraphics[width=0.32\textwidth]{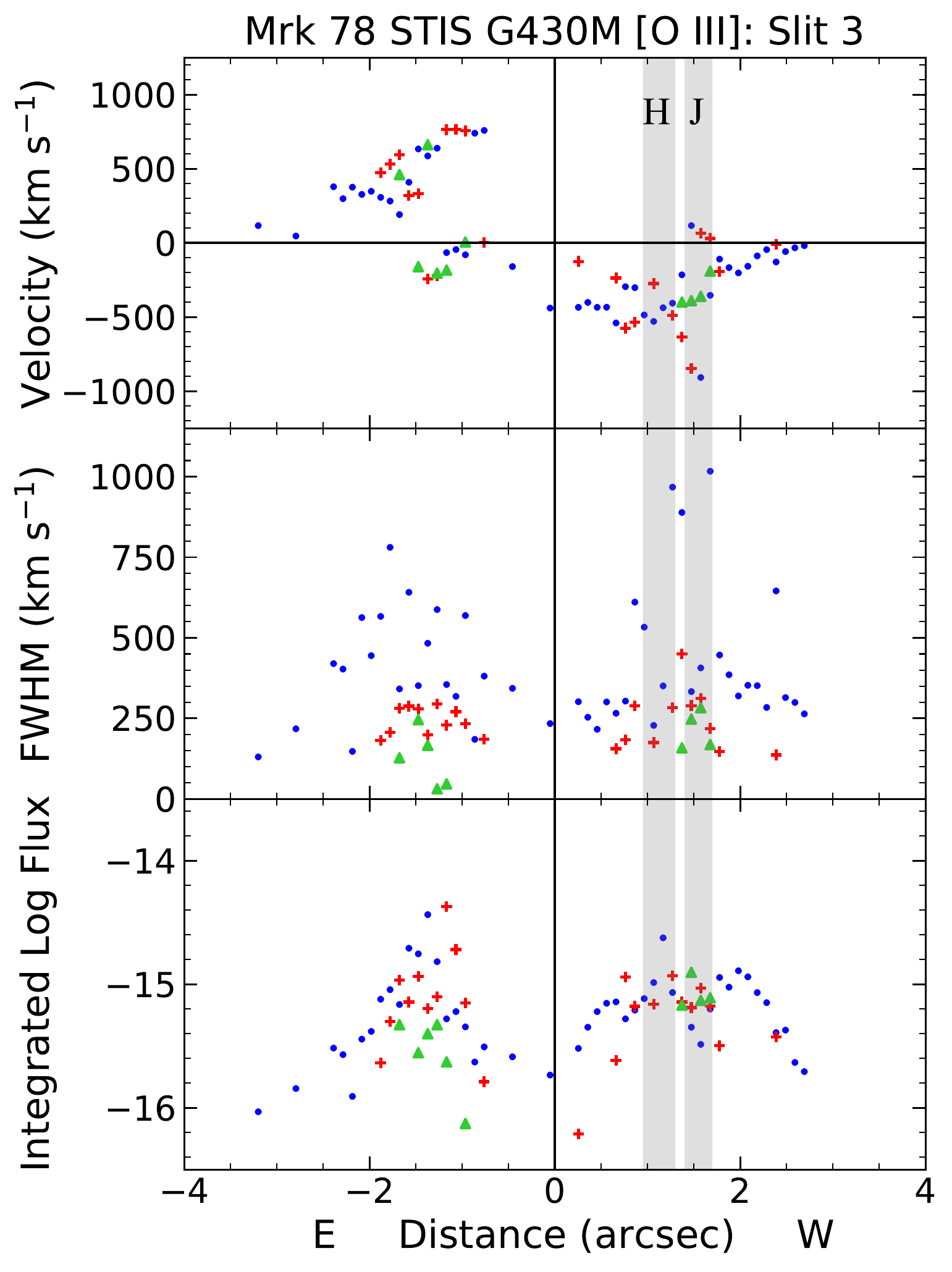}\hspace{1ex}
\includegraphics[width=0.32\textwidth]{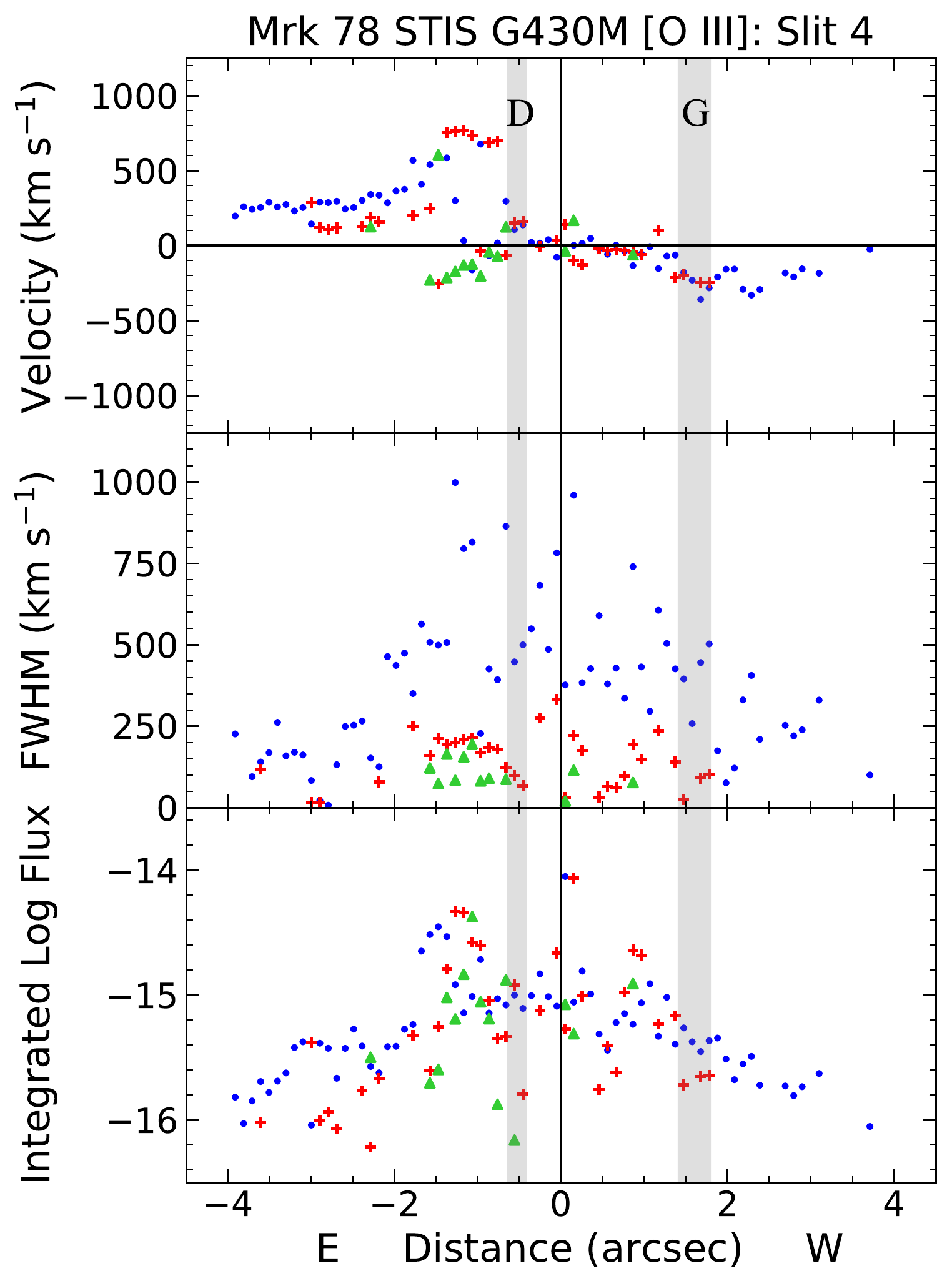}
\caption{The HST-STIS kinematics of the G430M slits as shown in Figure \ref{fig:O3}.}
\label{fig:mrk78stis_vel}
\end{figure*}

The radial velocity, FWHM, and flux maps for Mrk~78 (Figures~\ref{fig:kinematics} and \ref{fig:mrk78stis_vel}) follow the same pattern as presented in \cite{Fischer2011}. However, here we binned the spectra to a resolution element of $\sim$0\farcs1 in order to achieve higher S/N, which yielded a slightly more inhomogeneous set of patchy data points in the velocity fields than the continuous variations presented in the previous work. The clumpiness of the NRL in Mrk~78 can also be observed in the \othree image (Figure~\ref{fig:O3}). \cite{Revalski2021} provides \halpha kinematics using the STIS G750M slits at the same location as the G430M slits used in this work. The \othree kinematics strongly resemble the \halpha kinematics with a slight exception of fewer \othree points close to the center in slit C (see Figure~\ref{fig:mrk78stis_vel}), which is caused by the reddening from the dust lane going through the nucleus, as also noted in \cite{Fischer2011}. The kinematics are dominated by outflows up to $\pm$3\arcsec~from the nucleus, with a strong redshift to the East and a strong blueshift to the West. The velocities peak up to $\sim$1\farcs5 in both directions with velocities varying from 1000 \kms~to 500 \kms~in different slits, before slowly declining to systemic velocity of the galaxy beyond 2\arcsec.  The majority of the velocities still have high FWHM ($>$500 \kms) out to 3\arcsec, suggesting the presence of ``kinematically disturbed'' gas - a term coined by \cite{Fischer2018} for the gas that appears to be rotating based on their low velocity centroids but has high dispersion (FWHM $>$ 250 \kms) indicating some form of AGN influence.
Slit 4 (PA = 61.5\arcdeg) shows a decrease in velocities (to 200 \kms) as well as FWHM (to 250 \kms) at close to 3\arcsec, displaying a possible transition to rotating ionized gas.
Finally, similar to Mrk~3, we determine the velocities of the distinct knots of emission as identified in the \othree image (Figure~\ref{fig:O3}) by selecting a group of points clustered together in the observed velocity distribution. Some of the knots spread across multiple slits, however, for consistency, we assigned each specific knot to a single slit. The kinematics distribution of a few of the knots as identified in slit 1 are shown in the top-right panel of Figure~\ref{fig:kinematics} and the kinematics of the rest of slits (and labeled emitting knots) are provided in Figure~\ref{fig:mrk78stis_vel}.

\subsection{NGC~1068}

\begin{figure*}[ht!]
\centering
\includegraphics[width=0.32\textwidth]{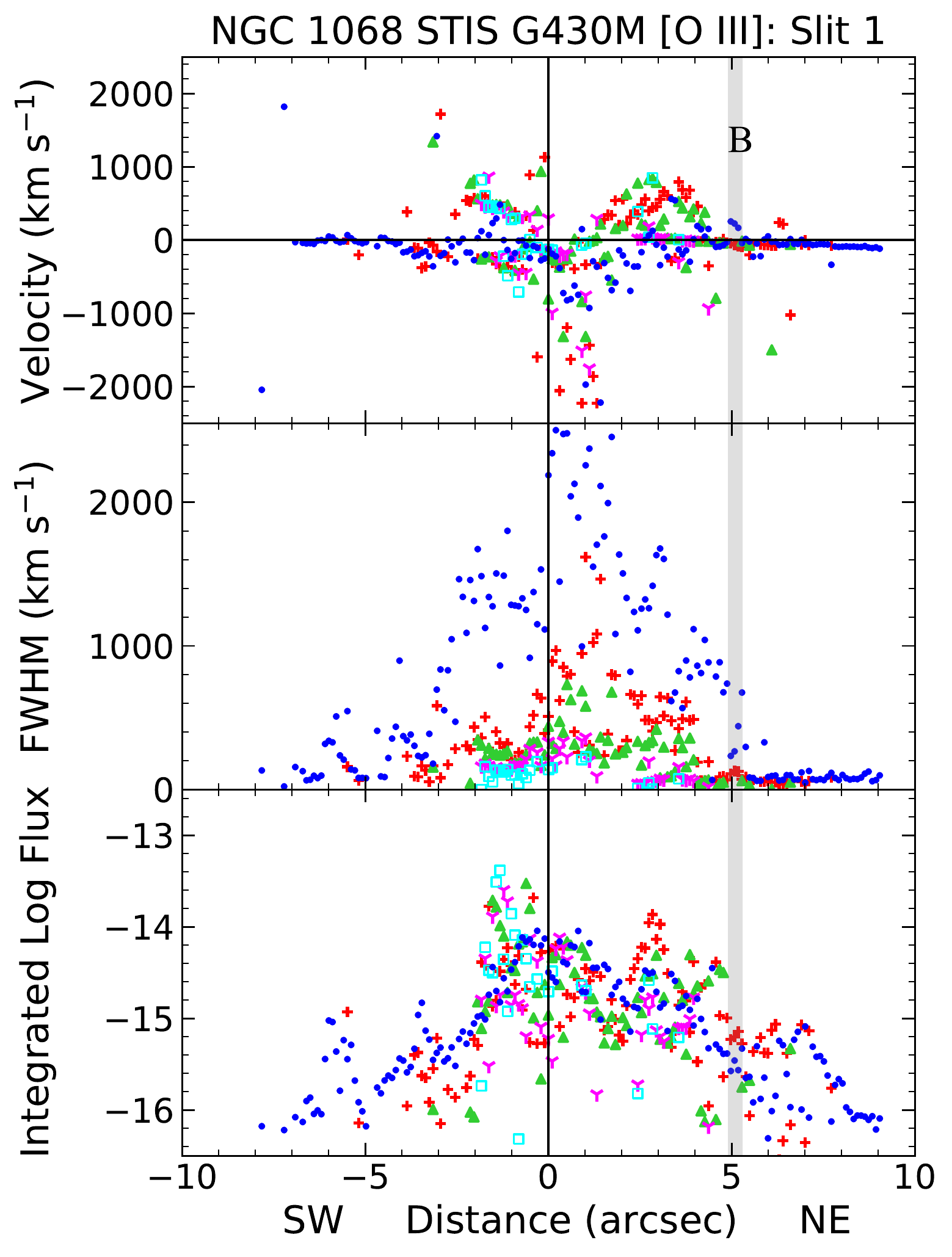}\hspace{1ex}
\includegraphics[width=0.32\textwidth]{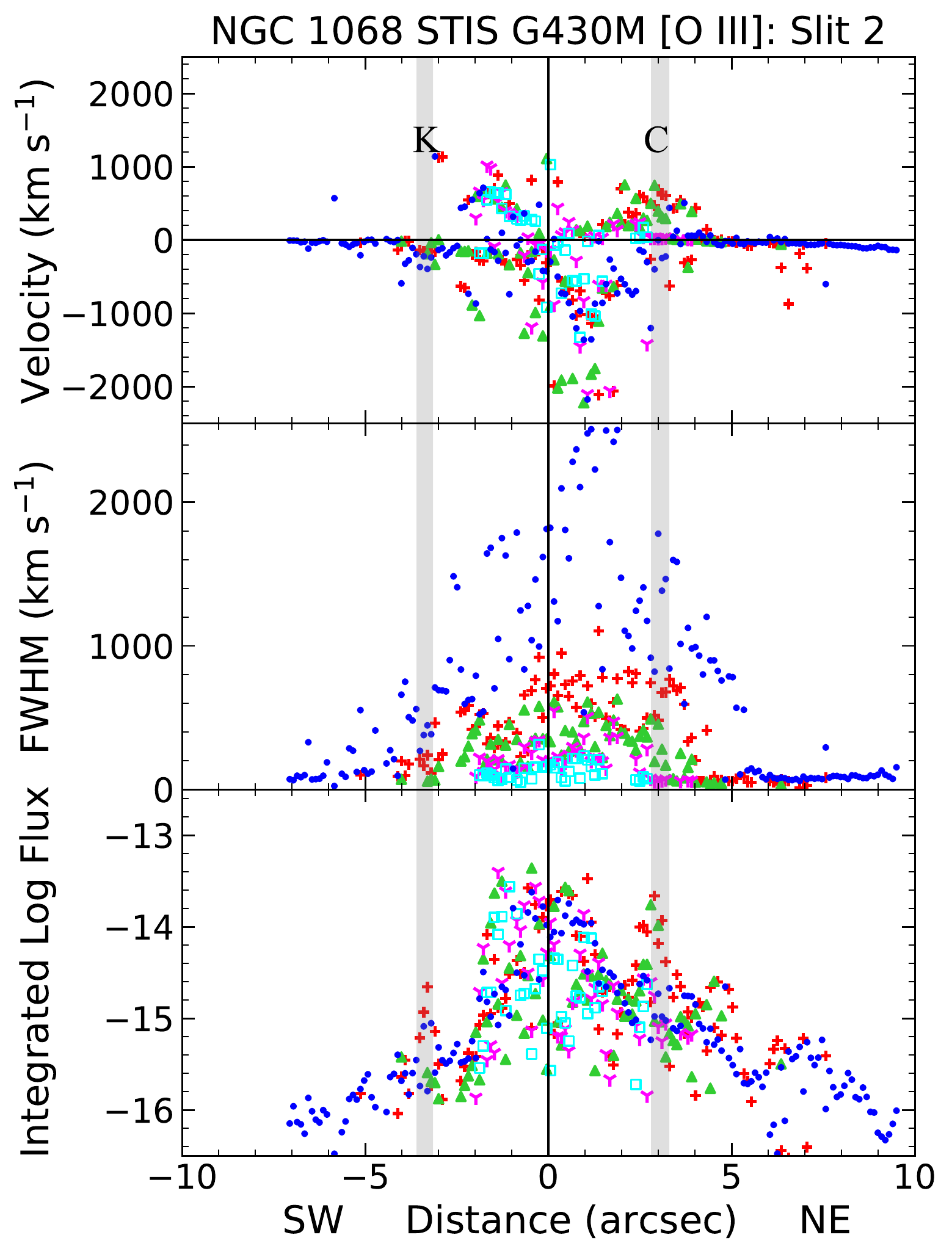}\hspace{1ex}
\includegraphics[width=0.32\textwidth]{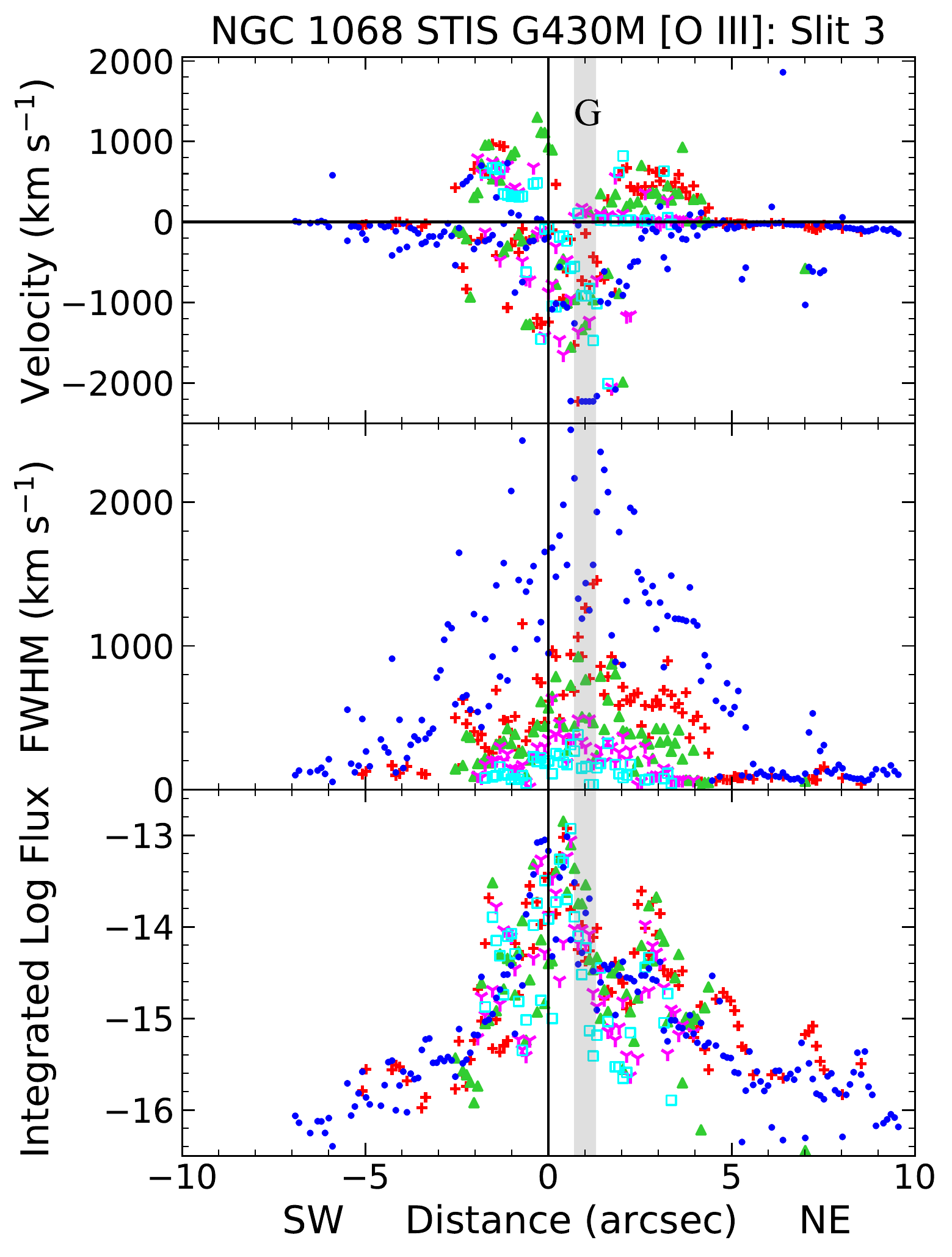}
\caption{The HST-STIS kinematics of the G430M slits as shown in Figure \ref{fig:O3}.}
\label{fig:ngc1068stis_vel}
\end{figure*}

\begin{figure*}[ht!]
\centering
\includegraphics[width=0.32\textwidth]{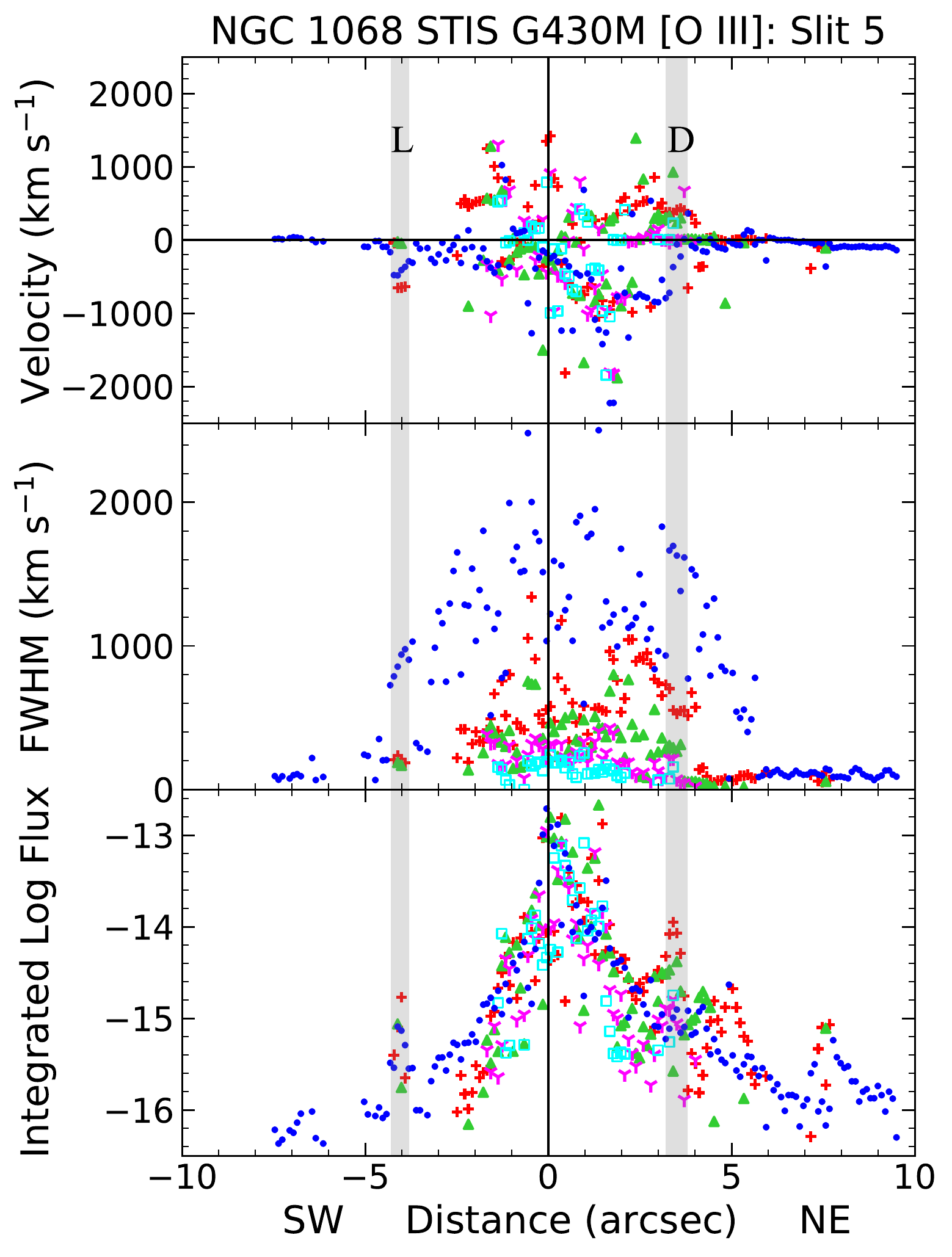}\hspace{1ex}
\includegraphics[width=0.32\textwidth]{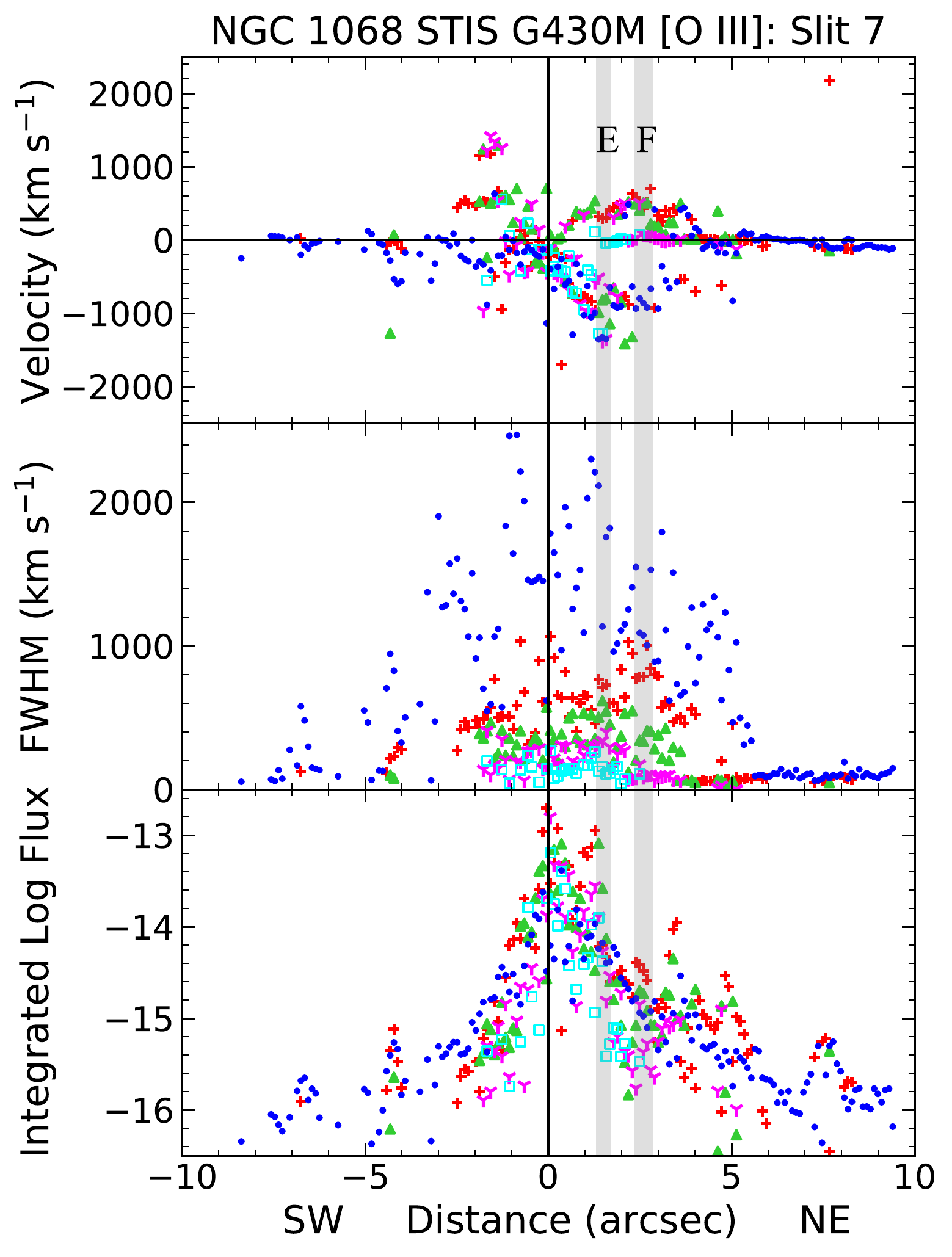}\hspace{1ex}
\includegraphics[width=0.32\textwidth]{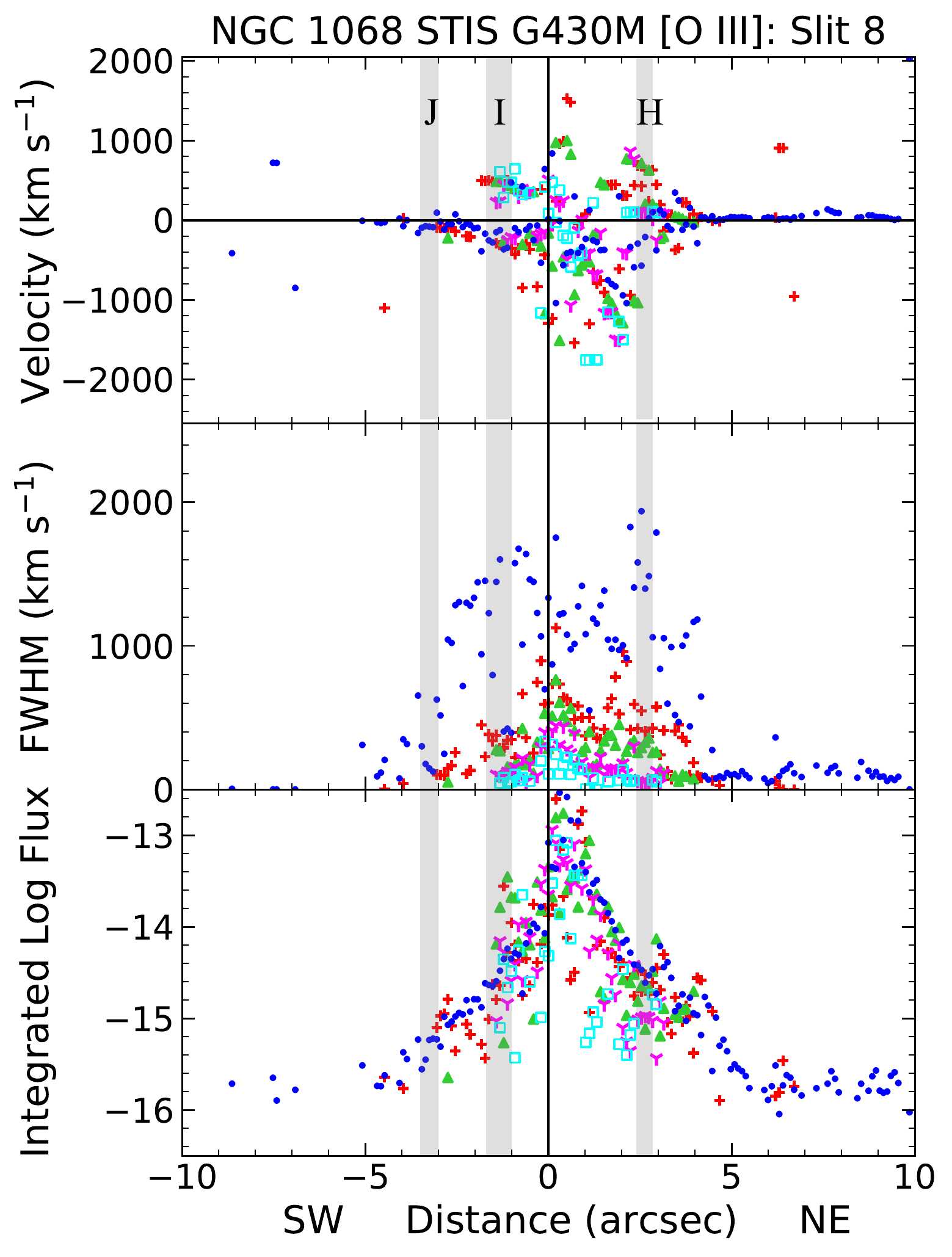}
\caption{continue Figure \ref{fig:ngc1068stis_vel}}
\label{fig:ngc1068stis_vel2}
\end{figure*}

The long-slit observations of NGC~1068 present highly complex kinematics where up to 5 components were required to fit a single emission line (see Figure~\ref{fig:fit}). 
In \cite{Das2006} only a maximum of 3 components were employed, using a visual inspection of distinct peaks in the \othree lines as compared to our Bayesian method. While they were able to determine an overall radial variation and the kinematic structures of the NLR gas, they lack the fine granularity in the velocity maps due to the enforced simplicity to the spectral fits. For example, in the measurements of slit 4, \cite{Das2006} did not identify the points with velocity gradient of $\sim$1000 - 2000 \kms~shown in the bottom-left panel of Figure~\ref{fig:kinematics} at a distance from 2\arcsec~to 0\farcs5 SW of the nucleus. We find the intrinsic widths (FHWM corrected for instrument LSF) of kinematic components up to $\sim$2500 \kms, while the maximum FWHM of only $\sim$1500 \kms~is reported in \cite{Das2006}. 
This difference in the velocity dispersions is due to the fact that the previous method fit the $\lambda$5007 emission independent of the $\lambda$4959 companion line, and therefore failed to detect the wide, low-flux blueshifted components that overlap with the red wing of $\lambda$4959 but are visible on the blueshifted side.
The automated multi-Gaussian routine fits both of the \llothree~lines simultaneously and when it detects any blueshifted wing, it fits a component to both lines with a fixed flux ratio of 3.01 for $\lambda$5007/$\lambda$4959. Finally, the majority of the points close to the systemic velocity of the galaxy (defined as zero \kms) as seen in the SW were also undetected for slits 4-8 in \cite{Das2006}. We also found more kinematic points associated with the knot A (Figure~\ref{fig:O3}) in slit 4 (at $\sim$7\arcsec~to the NE) illustrating a strong divergence in the kinematics (higher velocities and FWHM) from the surrounding gas, which predominantly carries a rotational pattern (see Figure~\ref{fig:kinematics}). Additionally, we found a larger number of points with blueshifted velocities $\sim$2000 \kms~at $\sim$1\arcsec - 2\arcsec~in the NE. Similar to \cite{Das2006}, we observe both redshifted and blueshifted outflows SW and NE of the center, with a general increase in the velocities followed by a slow decline to the systemic velocity close to 4\arcsec~is - seen in the overall kinematics for all slits (see Figures~\ref{fig:ngc1068stis_vel} and \ref{fig:ngc1068stis_vel2}). The peak redshifted velocities are $\sim$1000 \kms~($\sim$1200 \kms~in slit 7) at $\sim$1\farcs5 in SW and $\sim$800 \kms~at $\sim$2\arcsec - 3\arcsec~in SW. The lower number of blueshifted points in the SW indicates extinction due to dust in the plane of the host galaxy, as also noted in \cite{Das2006}. The widths of all kinematic components peak at or close to the nucleus with a gradual decline in both directions. Interestingly, the highest FWHM component starts to show a rotational pattern at about 4\arcsec~, as seen by the low velocities near systemic in the kinematic maps.

In a few instances, the observed mean velocity for the disturbed gas deviates from a rotational pattern to slightly higher velocities ($\sim$ 200 - 400 \kms, see knots J, K and L in Figures~\ref{fig:ngc1068stis_vel} and \ref{fig:ngc1068stis_vel2}) suggesting outflows. Similarly, knot A (Figure~\ref{fig:kinematics}) shows velocities up to $\sim$600 \kms~while the surrounding gas illustrates a clear rotational pattern with low velocities and FWHM. These aberrations are associated with distinct bumps (increases) in the radial flux distribution, which may correspond to renewed outflows from gas reservoirs such as nuclear dust spirals (see \citealp{Fischer2017}).

\subsection{NGC~4151}

\begin{figure*}[h]
\centering
\includegraphics[width=0.32\textwidth]{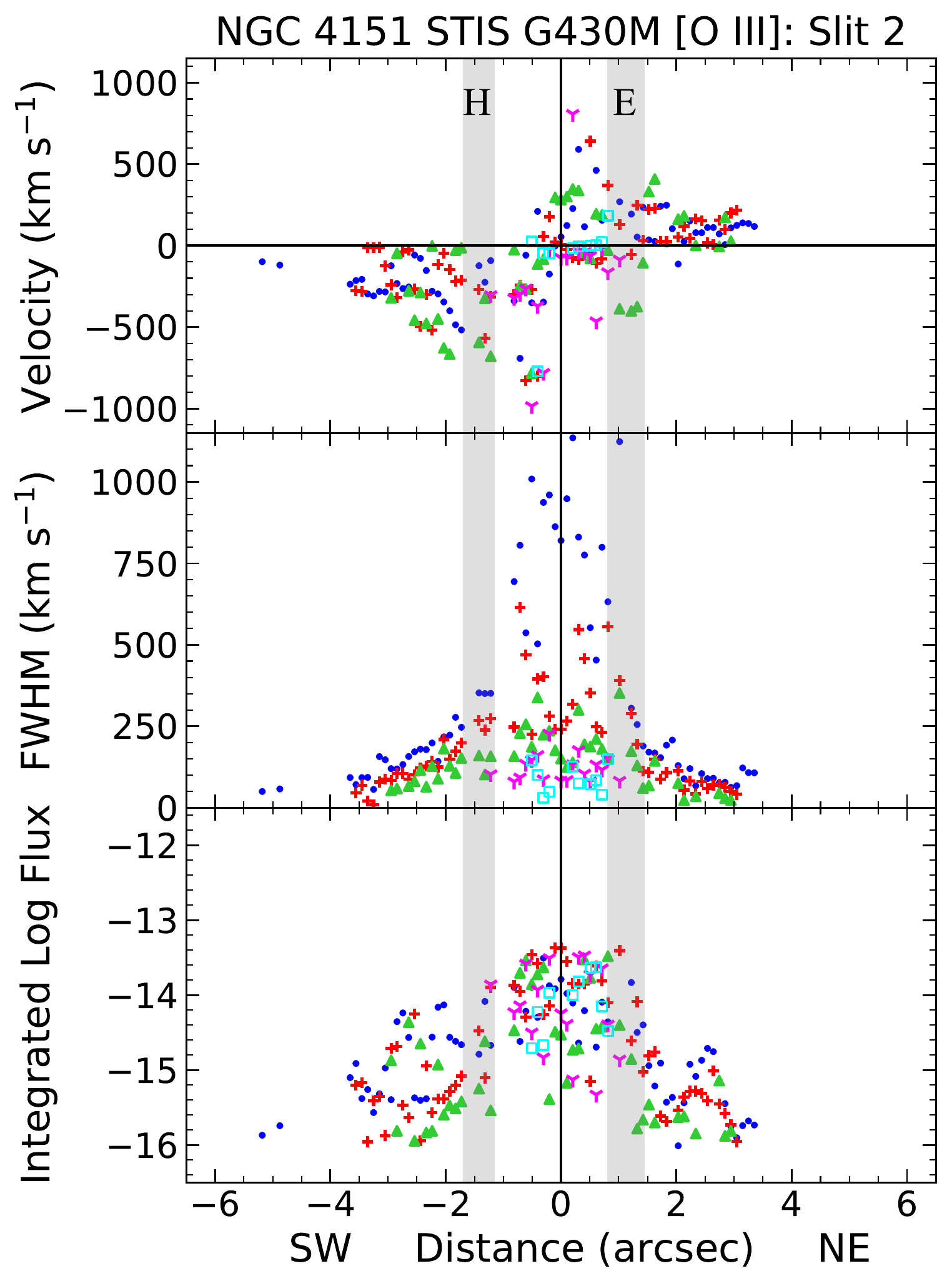}\hspace{1ex}
\includegraphics[width=0.32\textwidth]{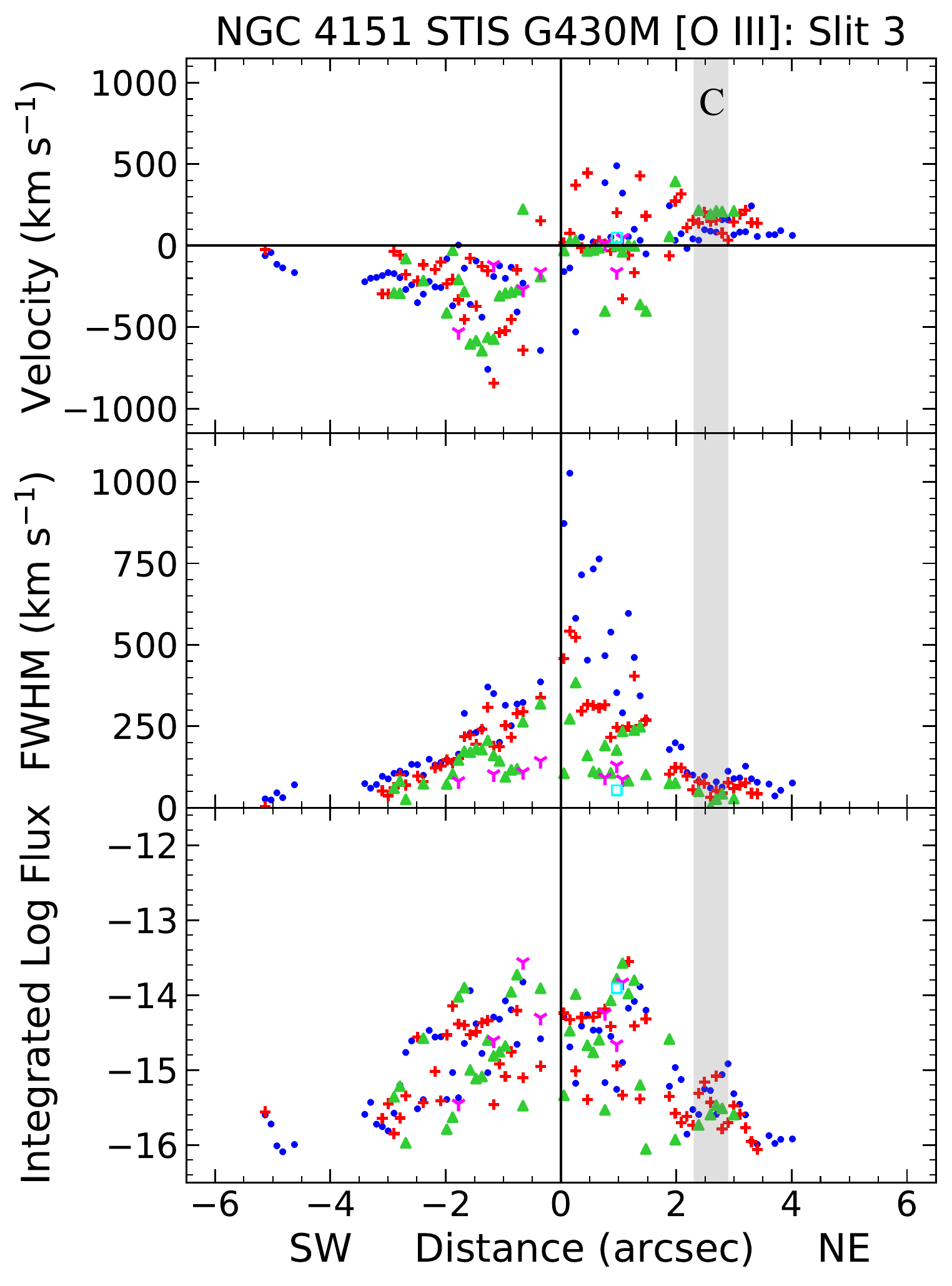}\hspace{1ex}
\includegraphics[width=0.32\textwidth]{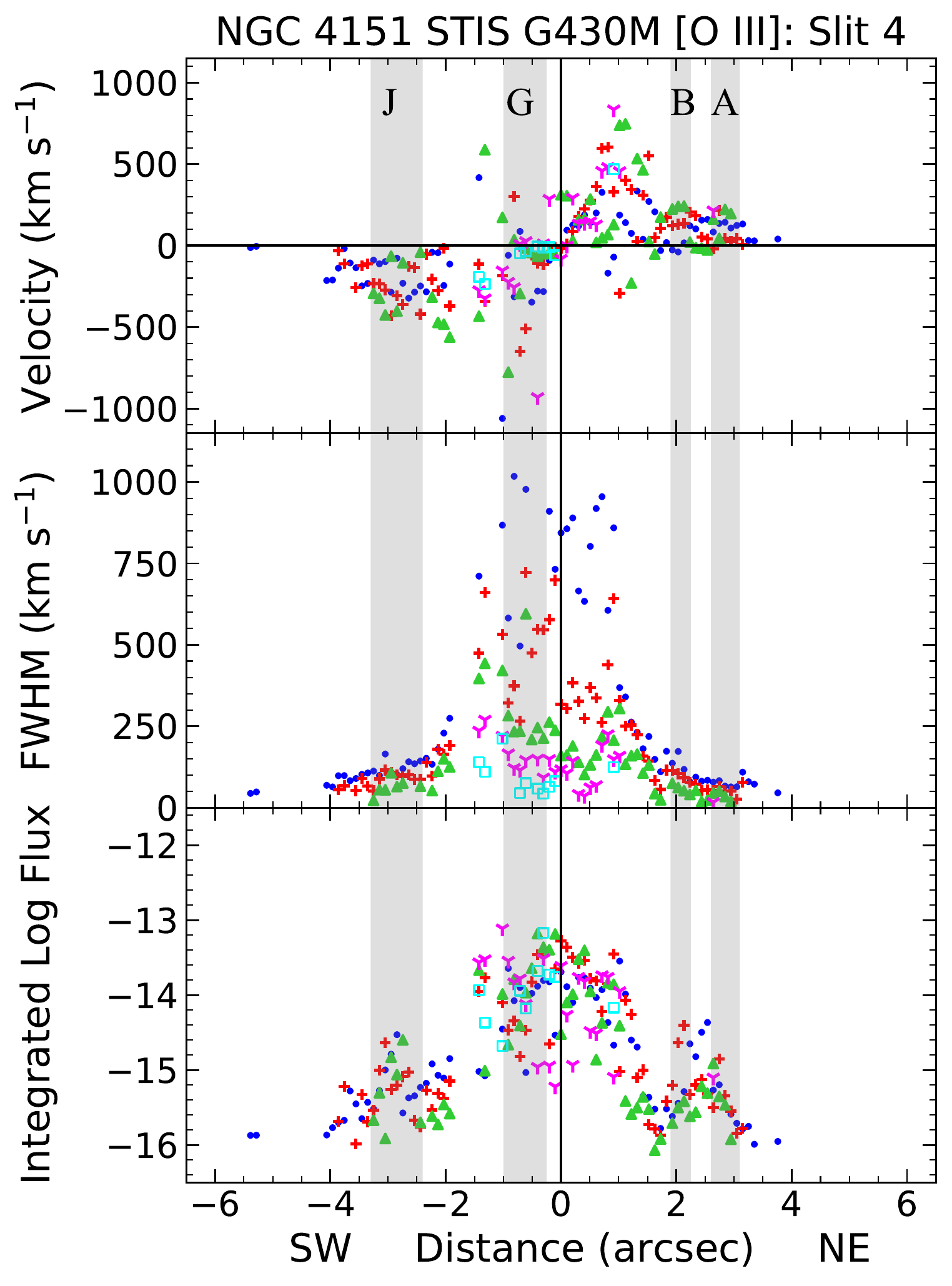}
\caption{The HST-STIS kinematics of the G430M slits as shown in Figure \ref{fig:O3}.}
\label{fig:ngc4151stis_vel}
\end{figure*}

\begin{figure*}[ht!]
\centering
\includegraphics[width=0.45\textwidth]{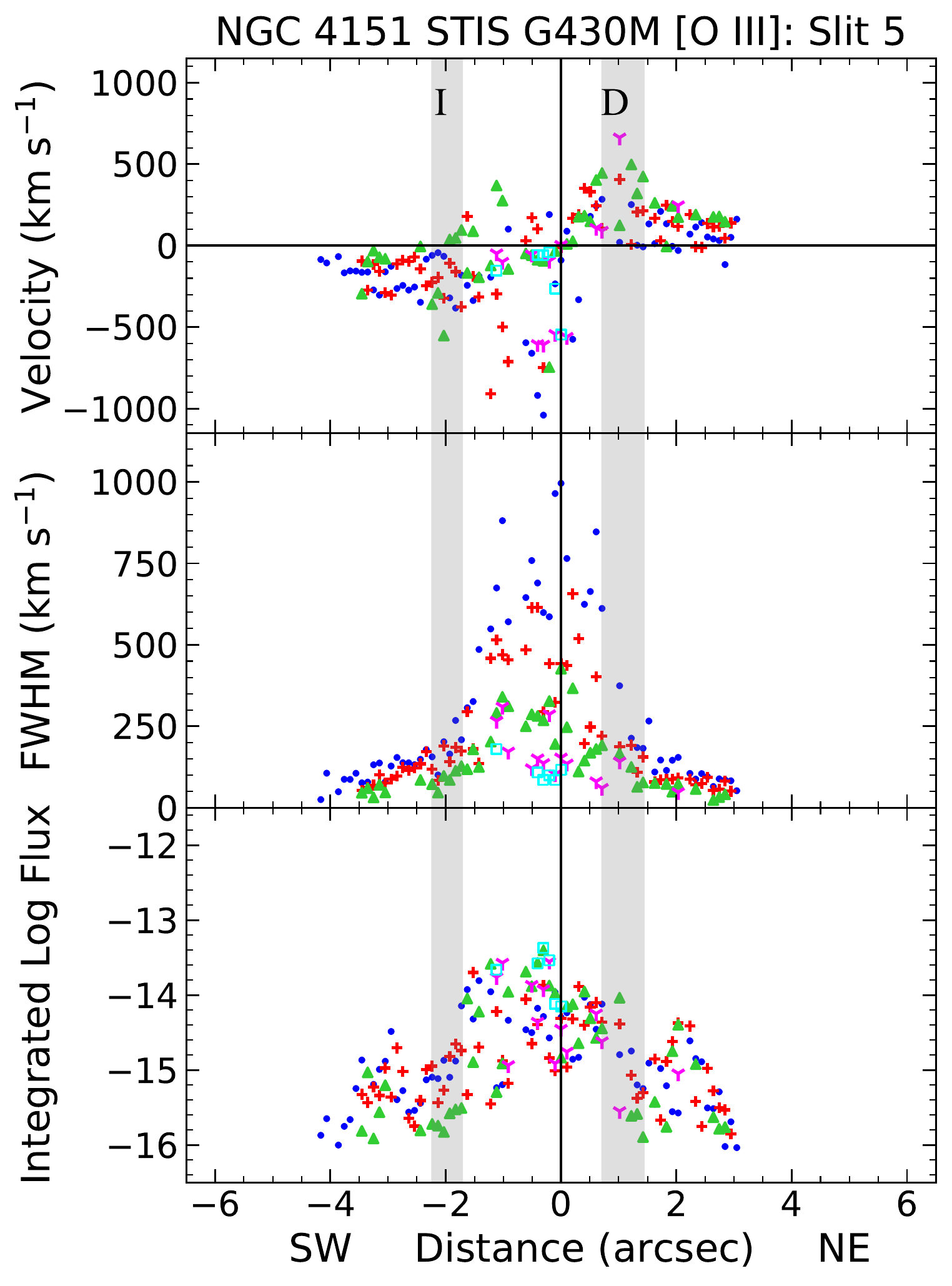}\hspace{2ex}
\caption{continue Figure \ref{fig:ngc4151stis_vel}}
\label{fig:ngc4151stis_vel2}
\end{figure*}

Similar to NGC~1068, we identified up to 5 components in the NLR gas kinematics of NGC~4151 (see bottom-right panel in Figure~\ref{fig:kinematics}, Figure~\ref{fig:ngc4151stis_vel} and \ref{fig:ngc4151stis_vel2}), likely as a result of the close proximities and large apparent brightnesses of these two Seyfert galaxies. Our kinematic maps are almost identical to those presented in \cite{Das2005} using only 3 component fits based on visual inspections. However, the kinematics were over-sampled in \cite{Das2005} as they fit spectra extracted for each pixel ($\sim$0\farcs05), while we employ two-pixel binning to match the STIS resolution element. Regardless of the spatial sampling, our automated fitting routine is able to recover the same kinematic components found by \cite{Das2005} at nearly all radii. This consistency indicates that the emission line profiles can be accurately decomposed without manual selection of the kinematic components.
The kinematics are mostly blueshifted in the SW and redshifted in the NE of the nucleus, with an exception of a few scattered points within $\sim$1\arcsec. We observe an increase in the radial velocities to a projected distance of $\sim$1\arcsec, with peak velocities $\sim$ 800 \kms~ before gradually declining to $\sim$ 100 \kms~at $\sim$4\arcsec. Unlike the other targets, we see a steeper drop in the FWHM variation away from the center, where we observe higher FWHM ($>$250 \kms) corresponding to high mean velocities ($\sim$500) within $\pm$2\arcsec~and low FHWM ($<$250 \kms) for the points outside these radii. At these larger distances (2\arcsec-4\arcsec) two sets of kinematics can be noticed: one with velocities 100 - 300 \kms, most likely associated with the far side of the outflow bicone (as described in \citealp{Das2005}), and another with with velocities less than 100 \kms, possibly from the underlying rotation (see \S \ref{subsec:rotation}). 
The points outside 4\arcsec~radii are associated with the host galaxy rotation based on their low velocities and low FWHM. There is no sign of ``disturbed" gas in the observed kinematic fields. 

Detailed discussions of the NLR gas kinematics for all of these targets are also available in \cite{Ruiz2001, Gnilka2020} for Mrk~3, \cite{Fischer2011} and \cite{Revalski2021} for Mrk~78, \cite{Cecil2002, Das2006} for NGC~1068 and \cite{Das2005} for NGC~4151.

\end{document}